\documentclass[11pt]{amsart}
\usepackage[english]{babel}
\usepackage{graphicx,amsmath,amssymb}
\usepackage{bbm}
\usepackage[numbers,sort&compress]{natbib}

\usepackage{fullpage}

\usepackage{tikz}
\usetikzlibrary{positioning,shapes}
\usepackage{pgfplots}
\usepackage{standalone}

\newenvironment{probleme}[1]{\vskip 0.2 cm\noindent{\bf {\sc
#1}}}{\vspace*{0.2 cm}}
\newcommand{\inputProblem}{\newline\noindent{\bf Input.~}}
\newcommand{\outputProblem}{\newline\noindent{\bf Output.~}}

\newenvironment{outdent}
{\begin{list}{}{\leftmargin-2cm\rightmargin\leftmargin}\centering\item\relax}
{\end{list}\ignorespacesafterend}

\theoremstyle{plain}
\newtheorem{theo}{Theorem}
\newtheorem{prop}[theo]{Proposition}

\newtheorem{lem}[theo]{Lemma}

\theoremstyle{remark}

\newtheorem{rem}{Remark}

\def\R{\mathbb{R}}
\def\Z{\mathbb{Z}}
\def\P{\mathbb{P}}
\def\bbM{\mathbb{M}}
\def\E{\mathbb{E}}
\newcommand{\ind}{\mathbbm{1}}

\usepackage{url}

\usepackage{xspace}
\newcommand{\SP}{\textsc{Shortest Path Problem}\xspace}
\newcommand{\RCSP}{\textsc{Resource Constrained Shortest Path Problem}\xspace}

\newcommand{\MRCSP}{\textsc{Monoid Resource Constrained Shortest Path Problem}\xspace}

\newcommand{\rplus}{\oplus}

\newcommand{\bigrplus}{\bigoplus}

\newcommand{\rleq}{\leqslant}
\newcommand{\rgeq}{\geqslant}
\newcommand{\nrleq}{\nleqslant}

\newcommand{\rset}{\mathcal{R}}
\newcommand{\rcost}{c}

\newcommand{\rmeas}{\rho}
\newcommand{\rmeasset}{\mathcal{S}}

\newcommand{\re}{x}

\newcommand{\NP}{$\mathcal{NP}$}
\newcommand{\leqst}{\leq_{\mathrm{st}}} 

\newcommand{\meet}{\wedge} 
\newcommand{\bigmeet}{\bigwedge}
\newcommand{\join}{\vee} 
\newcommand{\bigjoin}{\bigvee}
\newcommand{\meetst}{\meet_{\mathrm{st}}} 

\newcommand{\cvar}{\mathrm{CVaR}}

\usepackage{color}

\begin{document}

\title{Algorithms for non-linear and stochastic resource constrained shortest path.}

\author{Axel Parmentier}

\address{A. Parmentier, \'Ecole Nationale des Ponts et Chauss\'ees, CERMICS, 6-8 avenue Blaise Pascal, Cit\'e Descartes, 77455 Marne-la-Vall\'ee, Cedex 2, France}
\email{axel.parmentier@cermics.enpc.fr}


\begin{abstract}

Resource constrained shortest path problems are usually solved thanks to a smart enumeration of all the non-dominated paths. Recent improvements of these enumeration algorithms rely on the use of bounds on path resources to discard partial solutions. 
The quality of the bounds determines the performance of the algorithm. The main contribution of this paper is to introduce a standard procedure to generate bounds on paths resources in a general setting which covers most resource constrained shortest path problems, among which stochastic versions. 

In that purpose, we introduce a generalization of the resource constrained shortest path problem where the resources are taken in a monoid. The resource of a path is the monoid sum of the resources of its arcs. The problem consists in finding a path whose resource minimizes a non-decreasing cost function of the path resource among the paths that respect a given constraint. 
Enumeration algorithms are generalized to this framework. We use lattice theory to provide polynomial procedures to find good quality bounds. These procedures solve a generalization of the algebraic path problem, where arc resources belong to a lattice ordered monoid. The practical efficiency of the approach is proved through an extensive numerical study on some deterministic and stochastic resource constrained shortest path problems.



\end{abstract}

\maketitle

\section{Introduction} 
\label{sec:introduction}

\subsection{Problem statement} 
\label{sub:problem_statement}


Several methods have recently been developed to increase the efficiency of the shortest path problem solvers \cite{bast2014route}. A common feature to all these methods is the use of lower bounds on costs of paths to discard partial paths in an enumeration of all the paths. In this paper, we exploit the properties of lattices to extend this lower bound idea to algorithms for a generic version of the resource constrained shortest path problem where arc resources belong to a lattice ordered monoid. The lattice ordered monoid point of view covers a wide range of applications, among which stochastic versions of the resource constrained shortest path problem.

Let $(\rset,\rleq)$ and $(\rmeasset,\leq)$ be two partially ordered sets. A map $\rmeas : \rset \rightarrow \rmeasset$ is \emph{isotone} if $x \rleq y$ implies $\rmeas(x) \leq \rmeas(y)$. Let $(\rset,\rplus)$ be a set endowed with a law of composition. $(\rset,\rplus)$ is a \emph{monoid} if $\rplus$ is associative and admits a neutral element in $\rset$. A partial order $\rleq$ is a \emph{compatible order} on $(\rset,\rplus)$ if all translations $y \mapsto x \rplus y$ and $y \mapsto y \rplus x$ are isotone. An \emph{ordered monoid} $(\rset,\rplus,\rleq)$ is a monoid endowed with a compatible order. A partially ordered set $(\rset,\rleq)$ is a lattice if any pair of elements $(\re,\tilde{\re})$ of $\rset^{2}$ admits a greatest lower bound $\re \meet \tilde{\re}$ or \emph{meet} and a least upper bound $\re \join \tilde{\re}$. A \emph{lattice ordered monoid} $(\rset,\rplus,\rleq)$ is an ordered monoid such that $\rleq$ induces a lattice structure.

Let $(\rset,\rplus,\rleq)$ be a lattice ordered monoid. Consider the following problem.
\begin{probleme}\MRCSP
\inputProblem{A digraph $D=(V,A)$, two vertices $o,d \in V$, a collection $(x_{a}) \in \rset^{A}$, and two isotone mappings $\rcost: \rset \rightarrow \mathbb{R}$ and $\rmeas : \rset \rightarrow \{0,1\}$.}
 \outputProblem{An $o$-$d$ path $P$ such that $\rmeas\left(\bigrplus_{a \in P}x_{a}\right) =0$ and with minimum $\rcost\left(\bigrplus_{a \in P}x_{a}\right)$.}
 \end{probleme} 

$\rset$ is the \emph{set of resources}. The \emph{resource} of a path $P$ is $\bigrplus_{a \in P}x_{a}$ and we denote it $\re_{P}$, the \emph{cost} of $P$ is $\rcost\left(\bigrplus_{a \in P}x_{a}\right)$, and $P$ is \emph{feasible} if and only if $\rmeas\left(\bigrplus_{a \in P}x_{a}\right)$ is a equal to $0$. The function $\rmeas$ is later referred as the \emph{infeasibility function}. 

The \MRCSP is \NP-hard as it contains the usual \RCSP, which is obtained by using the set $\rset = \mathbb{R}^{2}$, the cost $\rcost(x^{1},x^{2}) = x^{1}$, and the function $\rmeas((x^{1},x^{2}))$ equal to $1$ if and only if $x^{2}>M$ for a given $M\in \mathbb{R}$. Finally, we emphasize that $\rplus$ is possibly non commutative. 

This lattice ordered monoid framework has two main strengths. First, it enables to deal with stochasticity in path problems. And second, the lattice ordered monoid structure enables to define polynomial procedure to compute lower bounds on paths resources, and to use these bounds to speed-up solution algorithms. We now sketch the main ideas behind the treatment of stochastic path problems and the solution schemes.


\subsection{Application to stochastic path problems} 
\label{sub:application_to_stochastic_path_problem}

Suppose that for each arc $a$ we have a random variable $\xi_{a}$. A large class of stochastic path problems can be expressed as follows. Given an origin vertex $o$ and a destination $d$, find an $o$-$d$ path minimizing
$$\min_{P} \E\left[f\left(\sum_{a\in P}\xi_{a}\right)\right],$$
\noindent under the constraint
$$\P\left(\sum_{a\in P}\xi > \tau\right) \leq \alpha, $$
\noindent where $f$ is a non-decreasing function, and $\tau$ and $\alpha$ are constants. We use two main ideas to model such problems within the \MRCSP framework. First, a space of random variables endowed with the addition and the almost sure order is a lattice ordered monoid. And second, many if not most probability functionals that intervene in stochastic path problems are monotone with respect to this order and can therefore be modeled withing this framework. On our example, it suffices to define $\re_{a} = \xi_{a}$, $\rcost(\xi) = \E\left[f(\xi)\right]$, and $\rmeas(\xi) = \ind_{(\alpha,1]}(\P(\sum_{a\in P}\xi > \tau) )$, where we introduce the notation $\ind_{I}$ that we frequently use in the paper to denote the indicator function of a set $I$.

 Finally, the use of alternative stochastic orders enables to exploit assumptions such as the independence of the $\xi_{a}$ to improve the performance of our solution algorithms. 

\subsection{Solution scheme} 
\label{sub:solution_scheme}

We propose solution schemes for the \MRCSP when $\bigrplus_{a\in C}x_{a} \rgeq 0$ for each cycle $C$ in $D$. These schemes are in two steps. We start by computing for each vertex $v$ a lower bound $b_{v}$ on the resource $\re_{P}$ of all the $v$-$d$ paths. Then, we use these lower bounds to discard partial solutions in an enumeration of all the paths. 

Bounds $b_{v}$ are used in enumeration algorithms to compute lower bounds on the resource of any $o$-$d$ path starting by an $o$-$v$ path $P$. Indeed, given an $o$-$v$ path $P$ and a $v$-$d$ path $\tilde{P}$, $$\re_{P}\rplus b_{v} \rleq \re_{P} \rplus \re_{\tilde{P}} = \re_{P+\tilde{P}},$$ 
where $P+\tilde{P}$ denotes the path composed of $P$ followed by $\tilde{P}$. The resource $\re_{P}\rplus b_{v}$ is therefore a lower bound on the resource of any $o$-$d$ path starting by $P$. As $\rmeas$ and $\rcost$ are isotone, $\rmeas(\re_{P}\rplus b_{v}) = 1$ implies that there is no feasible $o$-$d$ path starting by $P$, and $\rcost(\re_{P}\rplus b_{v})$ is a lower bound on the cost of any $o$-$d$ path $P$ starting by $P$. The enumeration algorithm we propose therefore enumerates all the paths satisfying
\begin{equation}\label{eq:introDP}
	\rmeas(\re_{P} \rplus b_{v}) = 0 \quad \text{and} \quad \rcost(\re_{P} \rplus b_{v}) \leq c_{od}^{UB},
\end{equation}
where $v$ is the destination of $P$ and $c_{od}^{UB}$ is an upper bound on the cost of an optimal solution.

The practical efficiency of the approach relies on our ability to compute in a preprocessing a tight lower bound on the resource of all the $v$-$d$ paths. As $(\rset,\rleq)$ is a \emph{lattice}, the meet $b_{v}^{\mathrm{opt}} = \bigmeet_{P\in \mathcal{P}_{vd}}\re_{P}$, where $\mathcal{P}_{vd}$ denotes the set of $v$-$d$ paths, is the tightest lower bound. Indeed, by definition of the meet,
\begin{equation*}
	b_{v} \rleq \re_{P} \text{ for all $P$ in }\mathcal{P}_{vd} \quad \Leftrightarrow \quad b \rleq b_{v}^{\mathrm{opt}} = \bigmeet_{P\in \mathcal{P}_{vd}}\re_{P}.
\end{equation*}
Unfortunately, we have shown \cite{parmentier2016thesis} that, unless $\mathcal{P} = \mathcal{NP}$, there is no polynomial algorithm that enables to compute $b_{v}^{\mathrm{opt}}$ in polynomial time even on some simple lattice ordered monoid where resources are positive. However, we provide polynomial procedures that compute lower bounds on $b_{v}^{\mathrm{opt}}$. To that purpose, we show that the following equation always admits solutions, that its solutions are lower bounds on $b_{v}^{\mathrm{opt}}$, and we introduce polynomial procedures that compute its greatest solution.
\begin{equation}\label{eq:monoidDPEquation}
	\left\{\begin{array}{l}
	b_{d} = 0, \\
	b_{v} = b_{v} \meet \displaystyle \bigmeet_{(v,u)\in \delta^{+}(v)}(x_{(v,u)} \rplus b_{u}) \text{ for all } v\in V\backslash \{d\}.
	\end{array}
	\right.
\end{equation}
Note that Equation \eqref{eq:monoidDPEquation} can be interpreted as a generalization of the Ford-Bellman dynamic programming equation for the usual shortest path problem, where the minimum has been replaced by the meet operator $\meet$, and the sum by the operator $\rplus$. Our polynomial procedures are generalizations of the Ford-Bellman and of the Dijkstra algorithm for the usual shortest path problem.

\subsection{Contributions and plan} 
\label{sub:contributions_and_plan}

We can sum-up our contributions as follows:
\begin{itemize}
	\item We introduce a versatile algebraic framework for constrained shortest path problems. This framework notably enables to deal with non-linearity and stochasticity in the objective and in the constraints. 
	\item We provide polynomial procedures to compute the solution of the generalized dynamic programming equation \eqref{eq:introDP}. The problem of solving this equation when the resource set has the structure of idempotent semiring is known as the algebraic path problem and has received much attention. Our procedures extend well-known algorithms of this community to the more general setting of lattice ordered monoids. Besides, contrary to the algebraic path problem community, we do not interpret the solution of \eqref{eq:introDP} as the solution of the problem but as lower bounds that can then be used in an enumeration algorithm. This new interpretation together with the extension to lattice ordered monoids enable to apply these algebraic methods to a much wider range of paths problems. The lattice ordered monoid point of view is notably essential to model stochastic path problems.
	\item We generalize the usual enumeration algorithms for resource constrained shortest path problems to our framework. The use of bounds is these algorithms is easy thanks to the lattice ordered monoid structure.  
	\item Concerning stochastic path problems, we show that our framework can deal with most probability functionals of interest, among which the version independent risk measures. Besides, we can deal with a wide range of probability distributions for the random variables $\xi_{a}$, and we can solve approximated versions of problems with other distribution through sampling.
	\item We show the practical efficiency of our approach through extensive numerical experiments on deterministic and stochastic resource constrained shortest path problems. To the best of our knowledge, our algorithms are the first practically efficient ones for paths problems with probabilistic constraints. 
	\item Finally, we provide strategies to improve the performance of our enumeration algorithms on difficult problems thanks to a longer preprocessing.
\end{itemize}


After introducing some notions on digraphs and ordered algebraic structures in Section \ref{sec:preliminaries}, we detail the connections between our framework and algorithms and those of the literature on usual, algebraic, resource constrained, and stochastic path problems in Section~\ref{sec:literature_review}. Sections \ref{sec:enumeration_algorithms} and \ref{sec:bounding_algorithms} introduce respectively our enumeration and bounding algorithms for the \MRCSP. Section \ref{sec:numerical_experiments_on_usual_resource_constrained_shortest_path_problem} tests the numerical performance of our algorithms on instances of the usual resource constrained shortest path problem. In Section \ref{sec:stochastic_path_problems}, we explain how to model stochastic path problems within our framework, and test numerically the performance of our algorithm on some non-constrained and constrained stochastic path problems. We conclude the paper with techniques to improve the performance of our algorithms on difficult instances in Section \ref{sec:what_to_do_on_difficult_problems}.

\section{Preliminaries} 
\label{sec:preliminaries}

\subsection{Digraphs} 
\label{sub:digraphs}

A \emph{digraph} $D$ is a pair $(V,A)$, where $V$ is the set of \emph{vertices} and $A$ is the set of \emph{arcs} of $D$. An \emph{arc} $a$ links a \emph{tail} vertex to a \emph{head} vertex. An arc $a$ is \emph{incoming} to (resp. outgoing from) $v$ if $v$ is the \emph{head} (resp. the tail) of $v$. The set of arcs incoming to (resp. outgoing from) $v$ is denoted $\delta^{-}(v)$ (resp. $\delta^{+}(v)$).  A \emph{path} is a sequence of arcs $a_{1},\ldots,a_{k}$ such that for each $i \in \{1,\ldots,k-1\}$, the head vertex of $a_{i}$ is the  tail vertex of $a_{i+1}$. Note that with this definition, paths can contain multiple copies of an arc or of a vertex. A path $P$ is said to be \emph{elementary} if it contains at most one copy of each vertex. The \emph{origin} of a path it the tail of its first arc and its \emph{destination} is the head of its last arc. Given two vertices $o$ and $d$ in $V$, an $o$-$d$ path $P$ is a path with origin $o$ and destination $d$. Finally, a \emph{cycle} is a path whose origin is identical to its destination.

Given two paths $P$ and $Q$ such that $P$ ends in the origin of $Q$, we denote $P+Q$ the path made of $P$ followed by $Q$. Given two vertices $u$ and $v$, we denote $\mathcal{P}_{uv}$ the set of $u$-$v$ paths.

\subsection{Lattice ordered monoids and algebraic structures} 
\label{sub:lattice_ordered_monoids_and_algebraic_structures}

We now introduce additional properties on lattices. When a subset $S$ of a partially ordered set admits a greatest lower bound (resp.~a least upper bound), we again call it its meet (resp.~its join), and denote it $\bigmeet S$ or $\bigmeet_{\re \in S}\re$ (resp.~$\bigjoin S$ or $\bigjoin_{\re\in S}\re$). Any finite subset of a lattice admits a meet and a join. A lattice is \emph{complete} if any subset $S\subseteq \rset$ admits a meet and a join. It is conditionally complete if any bounded subset $S \subseteq \rset$ admits a meet and a join. 
\begin{align}\label{eq:latticesConditionnallyComplete}
& \text{All the lattices we consider in this paper are conditionally complete.} 
\end{align}

If $\rset$ is conditionally complete, then $\rset\cup\{-\infty,+\infty\}$ is complete, where $-\infty$ (resp.~$+\infty$) is smaller (resp.~greater) than any element in $\rset$. The lattice $\rset\cup\{-\infty,+\infty\}$ is a \emph{completion} of $\rset$. 
We sometimes need our lattices to be complete to be able to define some quantities needed in the paper. When such quantities are defined, we mention that they may belong to the completion $\rset\cup\{-\infty,+\infty\}$ of $\rset$.

We denote $0$ the neutral element of the operator $\rplus$ of a lattice ordered monoid $(\rset,\rplus,\rleq)$. A resource $\re$ is \emph{positive} if $\re>0$. A lattice ordered monoid $(\rset,\rplus,\rleq)$ is a lattice ordered group if $(\rset,\rplus)$ is a group. A lattice ordered group is \emph{Archimedean} if, for each $\re,\tilde{\re} \in \rset$ such that $\re>0$, there exists $n$ in $\Z_{+}$ such that $n\re \rgeq \tilde{\re}$, where $n\re = \underbrace{\re\rplus\cdots\rplus\re}_{n\text{ times}}$.

\section{Literature review} 
\label{sec:literature_review}

As an algebraic framework for non-linear and stochastic resource constrained shortest path problems, our work is at the cross-road of several branches of the literature. First, our algorithms are naturally interpreted as generalizations of the usual shortest path problem algorithms. Second, our bounding algorithms are generalizations to lattice ordered monoids of those of the algebraic path problem community. Third, our framework can be seen as a versatile alternative to other resource constrained shortest path framework with enhanced version of the enumeration algorithms. Finally, the restriction of our algorithms to stochastic path problems compare favorably to the existing literature on the topic.

\subsection{Usual shortest path problem} 
\label{sub:usual_shortest_path_problem}

Variants of the \SP have been thoroughly studied during the last six decades. As we already mentioned, there are two main types of algorithms for the \SP. 
The first ones, such as Dijkstra's algorithm or Ford-Bellman algorithm, compute the shortest path between one vertex and all the other ones. In that sense their output is a solution of the dynamic programming equation \eqref{eq:monoidDPEquation} where $\rplus$ is the usual sum on $\R$ and the meet $\meet$ is the minimum. The standard algorithm to solve it is Dijkstra's algorithm \cite{dijkstra1959note} when arcs costs are non-negative, and Ford-Bellman \cite{ford1956maximal,bellman1958routing} when they can be negative. We generalize both of them to compute solutions of \eqref{eq:monoidDPEquation}.

The second ones are enumeration algorithms and use bounds to discard paths in an enumeration of all the paths. The typical example of enumeration algorithm is A$^{*}$ algorithm \cite{hart1968formal}, that we generalize to our setting. These algorithms are called goal directed algorithms as they compute the shortest path only between a given pair of origin and destination vertices. When good bounds are used, enumeration algorithms are faster than polynomial algorithms. \citet{bast2014route} survey the rich literature developed in the last few years on the topic in the context of online route planning systems. However, the goals of these recent contributions are orthogonal to our ones: their objective is to be able to compute quickly the solution of an easy path problem between any $o$-$d$ pair, while we want to compute the solution of a  difficult problem between one given $o$-$d$ pair.

\subsection{Bounding algorithms and algebraic path problems} 
\label{sub:bounding_algorithms_and_resource_constrained_shortest_path_problem}


Equation \eqref{eq:monoidDPEquation} is not the first generalization of the usual dynamic programming equation. Indeed, when $(\rset,\meet,\rplus)$ is an idempotent semiring, the problem of solving Equation \eqref{eq:monoidDPEquation} is called the \emph{algebraic path problem}. The literature on the topic considers idempotent semirings of various generality \cite{aho1974design,backhouse1975regular,carre1971algebra,zimmermann1981linear,fink1992survey,gondran2008graphs,lehmann1977algebraic,roy1959transitivite,mohri2002semiring} and is surveyed in \cite{fink1992survey}. Our framework generalizes the algebraic path problem:  the idempotent semiring structure is stronger than the lattice ordered monoid one. Indeed, if $(\rset,+,\times)$ is an idempotent semiring, then $(\rset,\times, \rleq_{+})$ is a lattice ordered monoid, where $\rleq_{+}$ is the idempotent semiring canonical order: $\re \rleq_{+}\tilde{\re}$ if and only if $\re + \tilde{\re} = \re$. Its meet operator is $+$. In the other direction, if $(\rset,\rplus,\rleq)$ is a lattice ordered monoid with meet operator $\meet$, then $(\rset,\meet,\rplus)$ is an idempotent semiring if and only if $\rplus$ \emph{distributes} with respect to $\meet$, i.e.
$$a \rplus (b \meet c) = (a \rplus b) \meet (a \rplus c)  \quad \text{and}\quad (a \meet b) \rplus c = (a\rplus b) \meet (a \rplus c).$$
\noindent If $b_{v}^{\dagger}$ is not necessarily equal to $b_{v}^{\mathrm{opt}}$ on lattice ordered monoids, these two quantities are equal on idempotent semirings.

Two types of algorithms have been developed for the algebraic path problem. The first ones generalize respectively the Ford-Bellman \citep{cousot1977abstract,cousot1979constructive} and the Dijkstra \citep{mohri2002semiring} algorithm. Our algorithms can be seen as generalization to the lattice ordered monoid setting of these algorithms. Our generalized Dijkstra algorithm is in particular very similar to the one developed by \citet{mohri2002semiring}. Therefore, we can use results from the algebraic path problem community to obtain stronger versions of our convergence theorems when $\rplus$ \emph{distributes} with respect to $\meet$. The second type of algorithms for the algebraic path problem consider Equation \eqref{eq:monoidDPEquation} as a system of linear equations in the idempotent semiring $(\rset,\rplus,\rleq)$, and generalize the Gauss-Seidel and the Gauss-Jordan algorithms to that setting \citep{gondran2008graphs,zimmermann1981linear}. Unfortunately, these algorithms do not generalize well to the lattice ordered monoid setting. 


Finally, we present in Section \ref{sec:what_to_do_on_difficult_problems} a technique to improve the quality of bounds that builds a lower envelope on the set $\{\re \in \rset| \re_{P} \rleq \re$ for some $P$ in $ \mathcal{P}_{vd}\}$. Techniques to build such a lower envelope have recently been proposed \citep{amato2016efficiently,apinis2013combine} in the context of static program analysis. However, their process for building the lower envelope is orthogonal to our one.

\subsection{Enumeration algorithms and resource constrained shortest path problems} 
\label{sub:enumeration_algorithms_and_resource_constrained_shortest_path_problems}

\citet{irnich2005shortest} provide a resource constrained shortest path framework and survey the exact and heuristic approaches to resource constrained shortest path problems. Their resource constrained shortest path framework is based on the notion of resource extension functions \cite{irnich2008resource}: there is one such function $\zeta_{a}$ for each arc $a$, and the resource of a path formed of the arcs $a_{1},\ldots,a_{k}$ is $\zeta_{a_{k}}\circ \ldots \circ \zeta_{a_{1}}(0)$. The main difference between their framework and our one is the associativity of $\rplus$. This makes our framework slightly less versatile, but enables us to compute lower bounds on paths resources and to use them to speed up the resolution. Besides, most problems of the literature can be modeled within our framework, as we argue in Chapter 3 of \cite{parmentier2016thesis}.

There are three main types of exact solution schemes to solve resource constrained shortest path problems: constraint programming, branch and bound, and enumeration algorithms. Constraint programming approaches \cite{rousseau2004solving,fahle2002constraint,gualandi2009constraint,de2001combining,junker1999framework} combine specifically designed search, domain reduction, and propagation algorithms. Concerning Branch-and-Bound algorithms, specific branching patterns have been developed by branch and bound algorithms for resource constrained shortest path problems: they branch on cycles, on arcs, and on resources \cite{irnich2008resource,righini2009decremental,beasley1989algorithm,borndorfer2001scheduling}. Finally, our work enters in the field of enumeration algorithms, and we now detail the literature on that topic.

In their survey on resource constrained shortest path problems, \citet{irnich2005shortest} describe the enumeration algorithms of the literature as variants of a generic enumeration algorithm. This generic enumeration algorithm has many similarities with the one we propose in Section \ref{sec:enumeration_algorithms}. To obtain a practical algorithm from the generic enumeration algorithm, one must choose define a key, some bounds, and a dominance rule. The key defines which paths are processed first. The bounds and the dominance rules are what enable to discard paths. The dominance rule is an order on the set of resources that enables to discard paths whose resources are dominated. \citet{desrochers1988generalized} and \cite{powell1998generalized} provide specific keys for routing problems with time windows, but these strategies apply to many resource constrained shortest path problems. \citet{irnich2008resource} provides general techniques to define resource extension functions leading to good dominance rules, and techniques to handle path with identical resources \citep{irnich2006shortest}. The remaining of the techniques are problem-specific \cite{beasley1989algorithm,feillet2004exact,kohl19992,larsen1999parallelization,irnich2006shortest,ioachim1998dynamic}. Finally, we note that variants of the enumeration algorithm based on the $k$-shortest path problem \citep{eppstein1998finding} have been proposed \cite{handler1980dual,beasley1989algorithm,santos2007improved}. These variants and problem-specific dominance rules can be used within our setting when appropriated.

Several techniques have been proposed to compute bounds when $(\rset,\rplus,\rleq)$ is $\R^{n}$ endowed with its product order and sum. Some contributions solve a usual shortest path problem for each component of the resource \cite{joksch1966shortest,desrochers1988generalized,dumitrescu2003improved,lozano2013exact}. Another branch of the literature uses Lagrangian relaxation on an integer formulation of the problem to obtain lower bounds \cite{handler1980dual,santos2007improved,carlyle2008lagrangian,dumitrescu2003improved}. These methods require the absence of negative cost cycles, in order to be able to solve the Langrangian relaxation using a shortest path problem. The case with negative cost cycles is considered by \citet{feillet2004exact}. When $(\rset,\rplus,\rleq)$ is $\R^{n}$, these Lagrangian techniques provide bounds that are typically tighter than our ones, but that require longer computations along the enumeration algorithm. 

The strength of our framework lies in our bounding algorithms, that enable to use lower bounds to discard path and good keys in non-linear and stochastic settings.

\subsection{Stochastic path problems} 
\label{sub:stochastic_paths_problems}

There are two types of stochastic path problems: offline problems, where the entire path is chosen a priori, and online problems, whose solution is a policy that updates the path used given the realization of uncertainty on the first arcs. Given that the solution of the \MRCSP in an $o$-$d$ path, it enables to model offline stochastic path problems. However, interestingly, the lower bounds computed in our framework provide an optimal policy \cite{parmentier2016thesis} for the well-studied online stochastic on time arrival problem \citep{fu2001adaptive,fu1998expected,hall1986fastest,fan2006optimal,nie2006arriving,samaranayake2012tractable,sabran2014precomputation,flajolet2014robust}. We now review the approaches to the different offline stochastic path problems.

Stochastic shortest path problems without constraints have been extensively studied since the seminal work of \citet{frank1969shortest}. Models differ by the type of distribution they use for arc random variables, and by the probability functional they optimize. The objective of a first line of paper is to find a path maximizing the probability of on time arrival, or analogously, a path with minimum quantile of given order. Approaches have been developed for both continuous \citep{frank1969shortest,chen2005path,nikolova2006stochastic,nikolova2010high} and discrete \citep{mirchandani1976shortest} distributions. \citet{chen2013finding} describe an efficient labeling algorithm to deal with normal distributions on the arcs. This algorithm is not so far from our label correcting algorithm applied with the lattice ordered monoid presented in Section 5.1.2 of \cite{parmentier2016thesis} when restricted to $\rmeas(\cdot) = \P(\cdot > \tau)$. A second line of papers defines a shortest path as a path minimizing the expectation of a cost function \citep{loui1983optimal}.
Dynamic programming can be used when cost functions are affine or exponential \citep{eiger1985path}.
\citet{murthy1996relaxation,murthy1998stochastic} present an efficient labeling algorithm when arc distributions are normal and cost functions are piecewise-linear and concave. All these objective functions are isotone with respect to the stochastic orders we use and can therefore be modeled within the \MRCSP framework. Besides, we show in Section \ref{sec:stochastic_path_problems} that we can also model version independent risk measures. Concerning the distributions, our approach can handle all the distributions used in the literature, among which discrete, normal, and gamma distributions for independent random variables, and scenario based distributions for non-independent random variables. In this paper, we focus on discrete independent distributions and scenario based distributions. Lattice ordered monoids for other distributions can be found in Chapter 5 of \cite{parmentier2016thesis}. The specificity of our solution approach is the use of lower bounds for stochastic orders.

Our approach cannot deal with positive linear combinations of means and variances in the objective \cite{sivakumar1994variance,nikolova2006stochastic,nikolova2010high}, as these objective functions are not isotone with respect to stochastic order.

The problem of finding a minimum cost path for deterministic arc costs under stochastic resource constraints have been introduced in \cite{kosuch2010stochastic}, and a solution algorithm based on linear programming is derived. We are not aware of other approaches specifically dedicated to stochastic constraints in path problems. However, such problems have been considered in the context of column generation. A wide range of stochastic versions of the traveling salesman and the vehicle routing problems have been studied in the last decades. When such problems are solved by column generation, the pricing subproblem is a stochastic resource constrained shortest path problem. Uncertainty in customer presence \cite{jaillet1988priori,jaillet1988probabilistic}, in demand \citep{sungur2008robust,gounaris2013robust,bertsimas1996new,bertsimas1992vehicle}, and in travel time \citep{jaillet2016routing,chang2009stochastic,jula2006truck,li2010vehicle,mazmanyan2009stochastic,russell2008vehicle,tacs2014vehicle,adulyasak2015models} have been considered. Using the modeling techniques we introduce in Chapter 3 of \cite{parmentier2016thesis}, most probability functionals considered in this literature can be dealt with using the lattice ordered monoid and the probability functionals presented in this chapter. However, we underline that, in the context of vehicle routing problems, the graph is often complete. In that case, the bounds provided by our approach are likely to be of poor quality, and thus our solution approach may not suit to the specific structure of these problems on complete graphs.



\section{Enumeration algorithms} 
\label{sec:enumeration_algorithms}

\subsection{Generic enumeration algorithms} 
\label{sub:generic_enumeration_algorithms}
In this section, we give three algorithms for the \MRCSP: the generalized A$^{*}$, the label dominance, and the label correcting algorithms. These three algorithms share the same structure. They enumerate all the paths in the graph using tests to discard partial paths. They differ only by the tests they use to discard paths, and by the keys they use to determine in which order the paths are processed. We therefore give a generic algorithm, and define the algorithms used in practice as specializations of this generic algorithm. Later in this section, we sometimes call optimal path an optimal solution of the \MRCSP.

We now describe the \emph{generic enumeration algorithm}. A list $L$ of partial paths $P$, and an upper bound $c_{od}^{UB}$ on the cost of an optimal solution are maintained. Initially, $L$ contains the empty path at the origin $o$, and $c_{od}^{UB} = +\infty$. While $L$ is not empty, the following operations are repeated.

\begin{enumerate}
	\item Extract a path $P$ \emph{of minimum ``key''} from $L$. Let $v$ be the destination of $P$. \label{step:key}
	\item If $v = d$ and $P$ is feasible and better than the current solution, i.e.~$\rmeas(\re_{P}) = 0$, and $\rcost(\re_{P}) < c_{od}^{UB}$, then update $c_{od}^{UB}$ to $\rcost(\re_{P})$.
	\item Else if \emph{``test'' returns ``yes''}, extend $P$: for each arc $a$ outgoing from $v$, add $P+a$ to $L$. \label{step:test}
\end{enumerate}

We obtain our different algorithms by specifying the key and the test respectively in the first and in the last step. We now introduce several keys and test. 

Our first key and test rely on lower bounds $b_{v}$ on the resources of all the $v$-$d$ paths. Section \ref{sec:bounding_algorithms} provides polynomial algorithms to compute such bounds in a preprocessing. Given an $o$-$v$ path $P$, the \emph{lower bound test} is expressed as follows.

\smallskip
	(Low) \begin{minipage}{\dimexpr\textwidth-2cm} \emph{Does $P$ satisfy $\rmeas(\re_{P}\rplus b_{v}) = 0$ and $\rcost(\re_{P}\rplus b_{v}) \rleq c_{od}^{UB}$?}
\end{minipage}
\smallskip

\noindent The isotony of $\rmeas$ and $\rcost$ implies that a subpath $P$ of an optimal path satisfies this test. 

An $o$-$v$ path $P$ \emph{dominates} an $o$-$v$ path $\tilde P$ if  $\re_{P} \rleq \re_{\tilde P}$. The \emph{dominance test} maintains along the algorithm a list $M_{v}$ of non-dominated $o$-$v$ paths for each vertex $v$, and is expressed as follows.

\smallskip
	(Dom) \begin{minipage}{\dimexpr\textwidth-2cm} \emph{Is $P$ non-dominated by any path in $M_{v}$?}
\end{minipage}
\smallskip

\noindent 
If the answer is yes, then before extending $P$, we remove from $M_{v}$ and $L$ all the paths in $M_{v}$ dominated by $P$, and add $P$ to $M_{v}$. 
The rationale behind this test is that, given the way $M_{v}$ is built, there is an optimal path whose subpaths are all non-dominated.

The aim of keys is to extend first the most promising paths. As an optimal solution is a feasible solution of minimum cost, we use as keys estimations of the cost of an optimal path. When bounds $b_{v}$ are computed, given an $o$-$v$ path $P$, the quantity $\rcost(\re_{P}\rplus b_{v})$ provides a lower bound on the cost of any $o$-$d$ path starting by $P$, and is therefore a good estimator of how promising $P$ is. When no bounds are computed, we use $\rcost(\re_{P})$, the cost of the partial solution considered.

We can now describe our enumeration algorithms.  The \emph{generalized A$^{*}$} is obtained from the generic algorithm by using the lower bound $\rcost(\re_{P}\rplus b_{v})$ as key in Step \ref{step:key}, and extending a path $P$ in Step \ref{step:test} only if it satisfies the lower bound test (Low). When $(\rset,\rplus,\rleq) = (\R,+,\leq)$, $\rmeas = 0$, and $\rcost(x) = x$, it corresponds to the usual A$^{*}$ algorithm. The \emph{label dominance} algorithm is obtained from the generic algorithm by using the partial path cost $\rho(c_{P})$ as key in Step \ref{step:key}, and extending an $o$-$v$ path $P$ in Step \ref{step:test} only if it satisfies the dominance test (Dom). The \emph{label correcting} algorithm is obtained from the generic algorithm by using the lower bounds $\rcost(\re_{P}\rplus b_{v})$ as key in Step \ref{step:key}, and extending an $o$-$v$ path $P$ in Step \ref{step:test} only if it satisfies both the lower bound test (Low) and the dominance test (Dom).

\begin{table}
	\begin{tabular}{p{4.2cm}cccc}
	\hline
	Algorithm & Test & Key & Pre-processing & Maintained \\
	&&& structures & structures \\
	\hline
	Generalized A$^{*}$ (A$^{*}$) & (Low) & $\rcost(\re_{P} \rplus b_{v})$ & $b_{v}$ & $L$, $c_{od}^{UB}$ \\
	Label dominance (dom.) & (Dom) &$ \rcost(\re_{P})$ & --- & $L$, $c_{od}^{UB}$, $M_{v}$\\
	Label correcting (cor.) & (Dom), (Low) &$ \rcost(\re_{P}\rplus b_{v})$ & $b_{v}$ & $L$, $c_{od}^{UB}$, $M_{v}$\\
	\hline
	\end{tabular}
	\caption{\MRCSP algorithms.}
	\label{tab:mrcspAlgo}
\end{table}

The properties of these algorithms are summed up in Table \ref{tab:mrcspAlgo}. We underline fact that bounds $b_{v}$ must be computed in a preprocessing when the generalized A$^{*}$ and the label correcting algorithms are used, and that the label correcting and the label dominance algorithm need to maintain lists of non-dominated paths $M_{v}$. 

Alternative combinations of our tests and keys are possible but less interesting. Indeed, our three algorithms reach trade-offs between the pre-processing time, and the quality of the keys and tests. Once bounds $b_{v}$ have been computed, performing the lower bound test (Low), and computing $c(\re_{P}\rplus b_{v})$ require a fix number of operations $\rplus$, $\rleq$, and $\meet$ of the lattice ordered monoid. Thus, if time has been spent computing bounds $b_{v}$ in a preprocessing, it is always better to use both the lower bound test and the key $c(\re_{P}\rplus b_{v})$. The alternative is to use the label dominance algorithm to avoid spending time in the preprocessing. 

The label dominance algorithm corresponds to the standard algorithm for the resource constrained shortest path problem \cite{irnich2005shortest} in the resource extension framework. The strength of the \MRCSP framework is that it enables to introduce the generalized A$^{*}$ and the label correcting algorithms, which both rely on the use of bounds.

When it comes to practical performances, it is well known that when lower bounds $b_{v}$ can be computed, the label correcting algorithm outperforms the label dominance algorithm on usual resource constrained shortest path problems \cite{dumitrescu2003improved}. The relative performance of the generalized A$^{*}$ and of the label correcting algorithm depends on the problem considered. Indeed, the dominance test enables to discard more paths, but its complexity is linear in the size of $M_{v}$, and it may slow the algorithm if dominance is rare and lists $M_{v}$ become large. This is notably the case of some problems with numerous or stochastic constraints. Finally, the numerical experiments in Sections \ref{sec:numerical_experiments_on_usual_resource_constrained_shortest_path_problem} and \ref{sec:stochastic_path_problems} show that the label dominance algorithm tends to perform well on easy problems, while the generalized A$^{*}$ algorithm tends to perform better on difficult problems with many deterministic constraints or one stochastic constraint.





\subsection{Convergence of the algorithms} 
\label{sub:convergence_of_the_algorithms}


We now prove the convergence of our enumeration algorithms when $x_{C} \rgeq 0$ for all the cycles $C$ of $D$. When this assumption is not satisfied, these algorithms must be adapted to discard non-elementary paths, using for instance the techniques developed by \citet{feillet2004exact}. 

\subsubsection{Generalized A$^{*}$ algorithm} 
\label{ssub:generalized_Astar}
The following assumptions will be used to define some settings under which ones the generalized A$^{*}$ algorithm converges:
\begin{align}
&\parbox{15.5cm}{For all $a$, $b < \bigjoin \rset$ and $\re > 0$ in $\rset$, there exists and $n \in \Z_{+}$ such that $nq\rplus a \nrleq b$.}\label{eq:weakArchimedeanProperty} \\
& \parbox{15.5cm}{There exists a feasible $o$-$d$ path $P$ such that $\displaystyle c^{-1}\left((-\infty,c(\re_{P})]\right) \cap \rho^{-1}(0)$ is upper-bounded by a resource $\displaystyle\re_{M} < \bigjoin\rset$.} \label{eq:costUpperBoundedCondition}
\end{align}

In Assumptions \eqref{eq:weakArchimedeanProperty} and \eqref{eq:costUpperBoundedCondition}, $\bigjoin \rset$ may be in the completion of $\rset$. Assumption \eqref{eq:weakArchimedeanProperty} is a weaker version for ordered monoid of the Archimedean property. 
It is satisfied by all the lattice ordered monoids considered in this paper. $\R^{2}$ endowed with its product sum and order is an example of lattice ordered group that is not Archimedean but which satisfies Assumption~\eqref{eq:weakArchimedeanProperty}. Finally, Assumption \eqref{eq:costUpperBoundedCondition} says that the set of resources of potentially optimal solutions is bounded. Indeed, an optimal path must be feasible and of cost non-greater than $\rcost(\re_{P})$: the feasible resources belong to $\rmeas^{-1}(0)$, and $c^{-1}\left((-\infty,c(\re_{P})]\right)$ is the set of resources whose cost is non-greater than the cost of path $P$.
\begin{theo}\label{theo:labelSettingConvergence}
Suppose that at least one of the following conditions is satisfied.
\begin{enumerate}
 	\item[(a)] $D$ is acyclic.
 	\item[(b)] Assumptions \eqref{eq:weakArchimedeanProperty} and \eqref{eq:costUpperBoundedCondition} are satisfied, $\re_{a}$ is positive for each arc $a$, and $b_{v} \rgeq 0$ for each vertex $v$.
 	\item[(c)] Assumptions \eqref{eq:weakArchimedeanProperty} and \eqref{eq:costUpperBoundedCondition} are satisfied, $\rplus$ is commutative, and $\bigrplus_{a \in C}\re_{a}$ is positive for any cycle $C$ in $D$.
 \end{enumerate}
 \noindent Then the generalized A$^{*}$ algorithm converges after a finite number of iterations, and at the end, if  $c_{od}^{UB}$ is finite, then it is the cost of an optimal solution of the \MRCSP. Otherwise, the problem admits no feasible solutions. 
\end{theo}

We note that in case (b), the hypothesis that $b_{v}\rgeq 0$ is not restrictive, as $\re_{a}> 0$ implies that $\re_{P} \rgeq 0$ for all paths $P$. Finally, Case (c) is notably satisfied when $(\rset,\rplus,\rleq)$ is an Archimedean lattice ordered group and cycle resources are positive. Indeed, Theorem 10.19 in \cite{blyth2005lattices} ensures that any Archimedean lattice ordered group is commutative, and Assumption \eqref{eq:weakArchimedeanProperty} is a consequence of the Archimedean property in ordered groups.

\begin{lem}\label{lem:upperBoundInGeneralizedAstar}
Let $P$ be an $o$-$d$ path satisfying $\rmeas(\re_{P}) = 0$. Then at a given step of the generalized A$^{*}$ algorithm, at least one of the following statements is satisfied:
\begin{itemize}
	\item there is a subpath $P'$ of $P$ in $L$,
	\item $c_{od}^{UB} \leq \rcost(\re_{P})$.
\end{itemize}
\end{lem}
Note that $P'$ can be equal to $P$.

\begin{rem}\label{rem:gapLowerBound}
	As $c(\re_{P'}\rplus b_{v}) \leq c(\re_{P})$, where $v$ is the destination of $P$, Lemma~\ref{lem:upperBoundInGeneralizedAstar} implies that, if we stop the algorithm before its convergence, the minimum $c(\re_{P'}\rplus b_{v})$ for $P' \in L$ provides a lower bound on the cost of an optimal path.
\end{rem}

\begin{proof}
We start with preliminary results. Paths are added to $L$ only due to extension of paths in Step~3. A path $Q$ can therefore be in $L$ only if its subpaths have been considered, removed from $L$, and extended by the algorithm. Thus, at a given step of the generalized A$^{*}$ algorithm, for each path $Q$ with origin $o$, exactly one of the following statements is satisfied:
\begin{itemize}
	\item $Q$ has been considered by the generalized A$^{*}$ algorithm,
	\item a subpath $Q'$ of $Q$ is in $L$,
	\item a strict subpath $Q'$ of $Q$ has not been extended by the algorithm when considered. 
\end{itemize}
Besides, if a feasible $o$-$d$ path $Q$ has already been considered, Step 2 of the algorithm implies $c_{od}^{UB} \leq \rcost(\re_{Q})$.

We now prove Lemma \ref{lem:upperBoundInGeneralizedAstar}. Suppose that none of the statements of Lemma \ref{lem:upperBoundInGeneralizedAstar} are satisfied. As $P$ is a feasible $o$-$d$ path, the two results above imply that a subpath $P'$ of $P$ has not been extended by the algorithm when considered. Let $P'$ be this subpath, and $v'$ be its destination. As $c_{od}^{UB}$ decreases along the algorithm, the hypothesis implies that $c_{od}^{UB} > \rcost(\re_{P})$ when $P'$ is considered. As $b_{v'}$ is a lower bound on the resource of all $v'$-$d$ paths, we have $\re_{P'} \rplus b_{v'} \rleq \re_{P}$. By monotonicity of $\rmeas$ and feasibility of $P$, we have $\rmeas(\re_{P'} \rplus b_{v'}) = 0$. By monotonicity of $\rcost$, we have $\rcost(\re_{P'} \rplus b_{v'}) \leq \rcost(\re_{P}) < c_{od}^{UB}$ when $P'$ is considered. The two last inequalities imply that $P'$ satisfies the lower bound test, which contradicts the fact that $P'$ has not been extended, and we obtain the lemma.
\end{proof}

\begin{lem}\label{lem:pathsNotBoundedByQm}
Under Assumption \eqref{eq:costUpperBoundedCondition}, if an $o$-$v$ path $Q$ such that $\re_{Q}\rplus b_{v} \nrleq \re_{M}$ is considered by the algorithm, it does not satisfy the lower bound test (Low).
\end{lem}
\begin{proof}

Suppose that Assumption \eqref{eq:costUpperBoundedCondition} is satisfied, and that $Q$ is considered by the algorithm. If $\rmeas(\re_{Q}\rplus b_{v}) = 1$, path $Q$ does not satisfy the lower bound test and we obtain the result. We now prove that, if $\rmeas(\re_{Q}\rplus b_{v}) = 0$, we have $c_{od}^{UB} < c(\re_{Q} \rplus b_{v})$ when $Q$ is considered by the algorithm, which then implies that $Q$ does not satisfy the lower bound test, which gives the lemma. Suppose that it is not the case. We place ourselves at the step when $Q$ is considered. As a consequence, $Q$ minimizes $\rcost(\re_{Q} \rplus b_{v})$ among the paths in $L$. Let $P$ and $\re_{M}$ be as in Assumption \eqref{eq:costUpperBoundedCondition}. By definition of $P$ and $\re_{M}$, the hypothesis $\re_{Q}\rplus b_{v} \nrleq \re_{M}$ implies $c(\re_{P}) < c(\re_{Q} \rplus b_{v}) \leq c_{od}^{UB}$ when $Q$ is considered. Lemma \ref{lem:upperBoundInGeneralizedAstar} implies that there is a subpath $P'$ of $P$ in $L$ when $Q$ is considered. By monotonicity of $c$ we have $\rcost(\re_{P'} \rplus b_{v'}) \rleq \rcost(\re_{P}) < \rcost(\re_{Q} \rplus b_{v})$. This contradicts the fact that $Q$ minimizes $\rcost(\re_{Q} \rplus b_{v})$ among the paths in $L$, and gives the lemma.
\end{proof}

\begin{lem}\label{lem:labelSettingConvergence}
Suppose that \emph{(a)}, \emph{(b)}, or \emph{(c)} is satisfied, then there is a finite number of paths in $D$ that satisfy the lower-bound test.
\end{lem}
\begin{proof}
In case (a), graph $D$ is acyclic and there is a finite number of paths. We now suppose that we are in case~(b) or~(c): let $\re_{M}$ be as in one Assumption \eqref{eq:costUpperBoundedCondition}. Lemma \ref{lem:pathsNotBoundedByQm} ensures that only $o$-$v$ paths $P$ such that $\re_{P} \rplus b_{v} \rleq q_{M}$ can satisfy the lower bound test. We show the lemma by proving that there is a finite number of such paths.
 
As we are in case (b) or (c), any elementary cycle $C$ in $D$ satisfies $\re_{C} > 0$. Thus, given an elementary cycle $C$, an elementary path $Q$, and a vertex $v$ in $P$, as the resource of $C$ is positive, Assumption \eqref{eq:weakArchimedeanProperty} implies that there exists an integer $n_{C,Q,v}$ such that $(n_{C,Q,v} \re_{C}) \rplus \re_{Q} \rplus b_{v} \nrleq \re_{M}$. As there is a finite number of elementary paths and a finite number of elementary cycles in $D$, we can define $n$ to be an integer such that
\begin{equation}\label{eq:defOfNinLem:labelSettingConvergence}
	(n \re_{C}) \rplus \re_{Q} \rplus b_{v} \nrleq \re_{M}
\end{equation}
for any elementary cycle $C$, elementary path $Q$, and bound $b_{v}$. Let $n_{C}$ be the number of elementary cycles in $D$.

The proof of case (b) relies on the following well-known result. 

{\em Any path in a directed graph can be decomposed in a sequence of elementary paths and elementary cycles.}

Suppose that we are in case (b), let $P$ be a path with at least $2 nn_{C}|V|$ arcs and consider such a decomposition. As an elementary path or an elementary cycle contains at most $|V|$ arcs, this decomposition contains at least $nn_{C}$ cycles, and thus at least $n$ copies of a given cycle~$C_{0}$. As the resource of all arcs are positive by hypothesis of case~(b), we have $\re_{P} \rgeq n\re_{C_{0}}$. We therefore have $\re_{P} \rplus b_{v} \rgeq n\re_{C_{0}} \rplus b_{v}$, and by applying Equation~\eqref{eq:defOfNinLem:labelSettingConvergence} with the empty path as $Q$, we obtain $\re_{P} \rplus b_{v} \nrleq \re_{M}$. As a consequence, only $o$-$v$ paths $P$ with fewer than $2nn_{C}|V|$ arcs can satisfy $\re_{P} \rplus b_{v} \rleq q_{m}$, and Lemma \ref{lem:pathsNotBoundedByQm} ensures that there is a finite number of paths that satisfy the lower bound test in case (b). 
 
We now consider case (c). As the monoid is supposed to be commutative, the resource of a path does not depend on the order of the sequence of its arcs, but only on its multiset of arcs. The proof of case (c) relies on the following well-known result. 

{\em The multiset of arcs of any path in a directed graph can be decomposed in the union of the sets of arcs of an elementary path and of several elementary cycles. }

Suppose that we are in case (c), let $P$ be a path with at least $n|V|(n_{C} + 1)$ arcs and consider such a decomposition where $Q$ denotes the elementary path. As by hypothesis of case~(c), the operator $\rplus$ is commutative, the resource of $P$ is entirely defined by the resource of its arcs, independently of their order. As an elementary path or an elementary cycle contains at most $|V|$ arcs, this decomposition contains at least $nn_{C}$ cycles, and thus at least $n$ copies of a given cycle $C_{0}$. As, by hypothesis of case (c), all cycles are positive, we have $\re_{P} \rplus b_{v} \rgeq n\re_{C_{0}} \rplus \re_{Q} \rplus b_{v}$. Equation \ref{eq:defOfNinLem:labelSettingConvergence} ensures that only paths with less than $n|V|(n_{C}+1)$ arcs can satisfy $\re_{P} \rplus b_{v} \rleq q_{m}$, and Lemma \ref{lem:pathsNotBoundedByQm} ensures that there is a finite number of paths that satisfy the lower bound test in case (c).  
\end{proof}

\begin{proof}[Proof of Theorem \ref{theo:labelSettingConvergence}] As any path inserted in $L$ is the extension of a previously considered path, a given path is considered at most once by the algorithm. Thus, Lemma \ref{lem:labelSettingConvergence} implies the convergence after a finite number of iterations as only paths satisfying the lower bound test can be extended by the algorithm.

At the end of the algorithm, list $L$ is empty and Lemma \ref{lem:upperBoundInGeneralizedAstar} ensures that $c_{od}^{UB}$ is a lower bound on the cost of any $o$-$d$ path satisfying $\rmeas(\re_{P}) = 0$. Besides, Step~2 of the algorithm ensures that if $c_{od}^{UB}$ is different from $+\infty$, then there is a path $P$ such that $c(\re_{P}) = c_{od}^{UB}$ and $\rmeas(\re_{P}) = 0$. This concludes the proof.
\end{proof}

\subsubsection{Label correcting and label dominance algorithms} 
\label{ssub:label_correcting_algorithm}

\begin{theo}\label{theo:labelCorrectingConvergence}
Suppose that the resource of any cycle $C$ in $D$ satisfies $\re_{C}\rgeq 0$, then the label correcting algorithm converges after a finite number of iterations, and at the end, if  $c_{od}^{UB}$ is finite, then  $c_{od}^{UB}$ is the cost of a non-dominated optimal solution of the \MRCSP. Otherwise, the problem admits no feasible solutions.
\end{theo}

\begin{theo}\label{theo:labelDominanceConvergence}
Suppose that the resource of any cycle $C$ in $D$ satisfies $\re_{C}\rgeq 0$, then the label dominance algorithm converges after a finite number of iterations, and at the end, if  $c_{od}^{UB}$ is finite, then  $c_{od}^{UB}$ is the cost of a non-dominated optimal solution of the \MRCSP. Otherwise, the problem admits no feasible solutions.
\end{theo}

Remark that the label correcting and the label dominance algorithms converge under weaker conditions than those required for the convergence of the generalized A$^{*}$ algorithm in Theorem \ref{theo:labelSettingConvergence}. 

\begin{lem}\label{lem:labelCorrectingConv}
Suppose that the resource of any $C$ in $D$ satisfies $\re_{C}\rgeq 0$, then if a path $P$ containing a cycle is considered by the label dominance or the label correcting algorithms, then it does not satisfy the dominance test (dom).
\end{lem}

\begin{proof}
As paths in $L$ are added by extension of paths previously in $L$, we only need to prove the result for paths ending by a cycle. Let $P$ be such a path, let $Q + C$ be its decomposition into a path and a cycle, and let $v$ be the common destination vertex of $P$ and $Q$. By hypothesis, we have $\re_{C} \rgeq 0$. As a consequence, $\re_{P} =  \re_{Q} \rplus \re_{C} \rgeq \re_{Q}$. As $P$ is processed, all its subpaths have been extended by the algorithm, and thus path $Q$ has necessarily been extended. This implies that either $Q$ or a path $Q'$ such that $\re_{Q'} < \re_{Q} \leq \re_{P} $ is in $M_{v}$, and thus $P$ is dominated by a path in $M_{v}$ and is therefore not extended.
\end{proof}

\begin{proof}[Proof of Theorem \ref{theo:labelCorrectingConvergence}]
As any path inserted in $L$ is the extension of a previously considered path, a given path is considered at most once by the algorithm. Thus, as there is only a finite number of acyclic paths in a graph, Lemma \ref{lem:labelCorrectingConv} ensures that the algorithm converges after a finite number of iterations.

Step~(b) of the algorithm ensures that $c_{od}^{UB}$ is non-smaller than the cost of an optimal solution of the \MRCSP. We now prove that at the end of the algorithm, $c_{od}^{UB}$ is equal to the cost of an optimal solution. Indeed, suppose that it is not the case. Let $P$ be an optimal solution. Let $\mathcal{L}$ be the set of all the paths that have been contained in $L$ along the algorithm, and for each vertex $v$ in $P$, let $P_{ov}$ be the subpath of $P$ starting  $v$. Let $v$ be the last vertex of $P$ such that there is an $o$-$v$ path $Q$ in $\mathcal{L}$ with $\re_{Q} \rleq \re_{P_{ov}}$. It exists because the empty path $P_{oo}$ is added to $L$ at the beginning of the algorithm, and is therefore in $\mathcal{L}$. Besides, it is not equal to $d$, as otherwise we would have $c_{od}^{UB} \leq \rcost(\re_{P})$. Among the $o$-$v$ paths dominating $P_{ov}$ in $\mathcal{L}$, let $Q$ be the first one generated by the algorithm. 
By definition of $Q$ and as any path that has been in $M_{v}$ along the algorithm is in $\mathcal{L}$, there is no path dominating $Q$ in $M_{v}$ when $Q$ is processed. 
As $c_{od}^{UB}$ decreases along the algorithm and by hypothesis, when $Q$ is processed we have $\rmeas(\re_{Q} \rplus b_{v}) \leq \rmeas(\re_{P_{ov}} \rplus b_{v}) \leq \rmeas(\re_{P}) = 0$ and $\rcost(\re_{Q} \rplus b_{v}) \leq \rcost(\re_{P_{ov}} \rplus b_{v}) \leq \rcost(\re_{P}) < c_{od}^{UB}$. As a consequence, $Q$ has been extended, and $Q + (v,w)$ is in $\mathcal{L}$, where $w$ be the vertex after $v$ in $P$. Besides, we have $\re_{Q + (v,w)} = \re_{Q} \rplus \re_{(v,w)}\rleq \re_{P_{ov}} \rplus \re_{(v,w)} = \re_{P_{ow}}$, which contradicts the definition of $v$.
\end{proof}

The proof of Theorem \ref{theo:labelDominanceConvergence} is analogous and can be found in \cite{parmentier2016thesis}.

\section{Bounding algorithms} 
\label{sec:bounding_algorithms}

We now come to the computation of lower bounds $b_{v}$ on the resource of any $v$-$d$ path $P$. We have already mentioned in the introduction that such lower bounds $b_{v}$ satisfy $b_{v}\rleq b_{v}^{\mathrm{opt}}$, where $b_{v}^{\mathrm{opt}}= \bigmeet_{P \in \mathcal{P}_{vd}}\re_{P}$, and $\mathcal{P}_{vd}$ is the set of $v$-$d$ paths. The bound $b_{v}^{\mathrm{opt}}$ is therefore the best lower bound. We prove in Proposition 4.11 of \cite{parmentier2016thesis} the following complexity results on the computation of $b_{v}^{\mathrm{opt}}$.

\begin{prop}\label{prop:bvoptComplexity}
Unless $\mathcal{P} = \mathcal{NP}$, there is no polynomial algorithm independent of $\rset$ that enables to compute $b_{v}^{\mathrm{opt}}$ even when restricted to a commutative monoid with positive resources.
\end{prop}
The proof is a reduction of the usual resource constrained shortest path problem.

Computing $b_{v}^{\mathrm{opt}}$ is therefore difficult in the general case. However, Theorem \ref{theo:relationsBetweenBounds} shows that when $\rplus$ distributes with respect to $\meet$, $b_{v}^{\mathrm{opt}}$ can be computed in polynomial time. The remaining of the section introduces polynomial procedures to compute lower bounds on $b_{v}^{\mathrm{opt}}$. 

\subsection{Extended Ford-Bellman algorithm} 
\label{sub:extended_ford_bellman_algorithm}
Let $\left(b_{v}^{n}\right)_{n}$ be the sequence of tuples of resources defined recursively as follows. 
\begin{equation}\label{eq:zSequenceDefinition}
	\left\{\begin{array}{l}
	b_{d}^{n} = 0, \\
\\
	b_{v}^{0} = \infty \text{ and } b_{v}^{n+1} = \displaystyle b_{v}^{n} \meet \bigmeet_{(v,u)\in \delta^{+}(v)}(\re_{(v,u)} \rplus b_{u}^{n}) \text{ for } v\in V\backslash \{d\}.
	\end{array}
	\right.
\end{equation}

As $\rset$ is a complete lattice, we can define $b_{v}^{\infty} = \displaystyle \bigmeet_{n \in \mathbb{Z}_{+}}b_{v}^{n}$ for each vertex $v$. 
Let $\mathbf{b}^{n}$ denote the tuple $(b_{v}^{n})_{v\in V}$. Recall that we have defined $\mathbf{b^{\dagger}}= (b_{v}^{\dagger})_{v\in V}$ to be the greatest solution of the following equation.

\begin{equation}\label{eq:meetEquation}
	\left\{\begin{array}{l}
	b_{d} = 0, \\
\\
	b_{v} = \displaystyle b_{v}\meet \bigmeet_{(v,u)\in \delta^{+}(v)}(\re_{(v,u)} \rplus b_{u}) \text{ for all } v\in V\backslash \{d\}.
	\end{array}
	\right.
\end{equation}
\noindent The existence of a greatest solution of Equation \eqref{eq:meetEquation} is a direct consequence of the Knaster-Tarski fixed point theorem applied in the complete product lattice $\rset^{V}$. This theorem states that the set of fixed points of a monotone mapping in a complete lattice is a non-empty complete lattice. Details on the Knaster-Tarski fixed point theorem can be found in \cite{davey2002introduction}. We underline that both $b_{v}^{\dagger}$ and $b_{v}^{\infty}$ may be defined only in the completion of $\rset$.  

\index{Knaster-Tarski fixed point theorem}

\begin{theo}\label{theo:relationsBetweenBounds}
Let $\ell^{*}$ be the length of the longest elementary $v$-$d$ path. If $\re_{C} \rgeq 0$ for each cycle $C$ in $D$, then for each vertex and $v$-$d$ path $P$, we have
\begin{equation}\label{eq:relationsBetweenBounds}
	b_{v}^{\dagger} \rleq b_{v}^{\infty} \rleq b_{v}^{\ell^{*}} \rleq b_{v}^{\mathrm{opt}} \rleq \re_{P}
\end{equation}
If $\rplus$ distributes with respect to $\meet$,  then three first inequalities are equalities.
\end{theo}

These \emph{bounds $b_{v}^{\ell^{*}}$ are good candidates to be used as bounds $b_{v}$ on the resource $\re_{P}$ of all $v$-$d$ paths $P$ in the enumeration algorithms }of Section~\ref{sec:enumeration_algorithms}. Indeed, they can be computed in $O(|A|\ell^{*})$ operations~$\rplus$ and~$\meet$ by computing the $\ell^{*}$ first terms of sequence $\left(\mathbf{b}^{n}\right)_{n}$ using its definition in Equation~\eqref{eq:zSequenceDefinition}. Besides, the sequence $\left(\mathbf{b}^{n}\right)_{n}$ can be interpreted as a \emph{generalization of the Ford-Bellman algorithm}. Indeed, when $(\rset, \rplus, \rleq) = (\mathbb{R}, +, \leq)$, or more generally when $(\rset, \rplus, \rleq) $ is a totally ordered group, the meet~$\re_{1}\meet \re_{2}$ of two resources $\re_{1}$ and $\re_{2}$ is the minimum of $\re_{1}$ and $\re_{2}$. In that case, the sequence of Equation \eqref{eq:zSequenceDefinition} corresponds to the successive steps of the Ford-Bellman shortest path algorithm, and for each integer $k$, the bound $b_{v}^{k}$ is the value of a shortest $o$-$v$ path with at most $k$ arcs. 

The proof of Theorem \ref{theo:relationsBetweenBounds} relies on two lemmas. The mapping $F:\rset^{V} \rightarrow  \rset^{V}$ defined as follows is useful in the proof.
\begin{equation}\label{eq:meetF}
 F(\mathbf{b}) = \mathbf{b}' \text{ with }\left\{\begin{array}{l}
	b'_{d} = 0, \\
\\
	b'_{v} = \displaystyle b_{v}\meet\bigmeet_{(v,u)\in \delta^{+}(v)}(\re_{(v,u)} \rplus b_{u}) \text{ for all } v\in V\backslash \{d\},
	\end{array}
	\right.
\end{equation}
\noindent where $\rset^{V}$ denotes the Cartesian product. Note that $\mathbf{b}^{n+1} = F\left(\mathbf{b}^{n}\right)$ and that $F$ is isotone by monotonicity of the operators $\rplus$ and $\meet$.

\begin{lem}\label{lem:meetEqSolSmallerThanSeq}
For each vertex $v$ and integer $n$, we have $b_{v}^{\dagger} \rleq b_{v}^{\infty} \rleq b_{v}^{n}$.
\end{lem}
\begin{proof}
A straightforward induction on $n$ based on the isotony of mapping $F$ defined in Equation~\eqref{eq:meetF} gives $b_{v}^{\dagger} \rleq b_{v}^{n}$ for all $n$, which implies that $b_{v}^{\dagger} \rleq b_{v}^{\infty}$.
\end{proof}

\begin{lem}\label{lem:meetSeqAndPathsResources}
The resource $b_{v}^{k}$ is a lower bound on the resource $\re_{P}$ of the $v$-$d$ paths $P$ of length at most~$k$. 
When $\rplus$ distributes with respect to $\meet$, $b_{v}^{k}$ is the meet of the resources of the $v$-$d$ paths of length at most $k$. 
\end{lem}


\begin{proof}
The result is proved by induction on $k$. The result for $k=0$, i.e.\ $b_{v}^{0}$ is equal to $0$ if $v=d$ and $\infty$ otherwise, follows from the fact that the only path of length $0$ is the trivial path. Let $k > 0$ be an integer and suppose the result is true up to $k-1$, let $v$ be a vertex, and let $P$ be a $v$-$d$ path with $\ell(P) \leq k$. If $\ell(P) = 0$ then $v = d$ and $\re_{P} \rgeq b_{d}^{k} = 0$. Otherwise let $(v,u)$ be the first arc of $P$ and $Q$ be the subpath of $P$ obtained by removing $(v,u)$ from $P$. Then $\ell(Q)\leq k-1$, thus $b_{u}^{k-1} \rleq \re_{Q}$ which implies $\re_{(v,u)} \rplus b_{u}^{k-1} \rleq \re_{(v,u)} \rplus \re_{Q}$ and finally $b_{v}^{k} \rleq \re_{P}$, which gives that $b_{v}^{k}$ is a lower bound on the resource $\re_{P}$ of the $v$-$d$ paths $P$ of length at most~$k$.

When $\rplus$ distributes with respect to $\meet$, we have $\re_{1} \rplus (\re_{2} \meet \re_{3}) = (\re_{1} \rplus \re_{2}) \meet (\re_{1} \rplus \re_{3})$. Suppose that $b_{u}^{k-1}$ is the meet of the resource of all the $v$-$d$ paths of length at most $k$ for each vertex $u$. Then $\re_{(v,u)} \rplus b_{u}^{k-1}$ is the meet of the resources $\re_{P}$ of all the $v$-$d$ paths $P$ starting by $(v,u)$ such that $\ell(P)\leq k$. Thus $b_{v}^{k-1} \meet \displaystyle\bigmeet_{(v,u)\in \delta^{+}(v)}\left( \re_{(v,u)} \rplus b_{u}^{k-1} \right)$ is the meet of all the $v$-$d$ paths of length at most $k$.
\end{proof}
\begin{proof}[Proof of Theorem \ref{theo:relationsBetweenBounds}]
As the resource of any cycle $C$ in $D$ satisfies $\re_{C}\geq 0$, for each $o$-$d$ path $P$, there is an elementary $o$-$d$ path $P'$ such that $\re_{P'}\rleq \re_{P}$. Hence, $b_{v}^{\mathrm{opt}}$ is the meet of the resources of all the elementary $o$-$d$ paths. As the length of any elementary $v$-$d$ path is non greater than $\ell^{*}$, Lemma \ref{lem:meetSeqAndPathsResources} implies that $b_{v}^{\ell^{*}} \rleq \re_{P}$ for all elementary $v$-$d$ paths $P$, and thus $b_{v}^{\ell^{*}} \leq b_{v}^{\mathrm{opt}}$. Lemma \ref{lem:meetEqSolSmallerThanSeq} then gives Equation~\eqref{eq:relationsBetweenBounds}.

Suppose now that $\rplus$ distributes with respect to $\meet$. We already mentioned that $x_{C}\rgeq 0$ for all cycles $C$ implies that $b_{v}^{\mathrm{opt}}$ is the meet of all the elementary $o$-$d$ paths. 
As the length of an elementary path is non greater than $\ell^{*}$, Lemma \ref{lem:meetSeqAndPathsResources} implies that $b_{v}^{\ell^{*}}=b_{v}^{\mathrm{opt}}$. This implies that $F(\mathbf{b}^{\ell^{*}}) = \mathbf{b}^{\ell^{*}}$ and $b_{v}^{\ell^{*}}$ is a solution of Equation~\eqref{eq:meetEquation}. Thus, $b_{v}^{\ell^{*}} = b_{v}^{\dagger}$, which gives the result.
\end{proof}

\begin{rem}
If there are cycles with negative resources in $D$, then Theorem \ref{theo:relationsBetweenBounds} remains true provided that, first, $P$ is an elementary $v$-$d$ path, and second, $b_{v}^{\mathrm{opt}}$ is defined as the meet of the resources of all the elementary $v$-$d$ paths.
Besides, Lemma \ref{lem:meetSeqAndPathsResources} implies that $b_{v}^{\dagger} = b_{v}^{\infty} = - \infty$.
\end{rem}

\begin{rem}\label{rem:cousotSequence}
The sequence $\left(\mathbf{b}^{n}\right)_{n}$ is the sequence used in the constructive proof by \citet{cousot1979constructive} of the Knaster-Tarski fixed point theorem for mapping the $F$ defined in Equation \eqref{eq:meetF}. Given a topology and some weak assumptions on $\rset$, it can be proved that $b_{v}^{n}$ converges to $b_{v}^{\infty}$ and $b_{v}^{\infty} = b_{v}^{\dagger}$. The inequality $\bigmeet_{(v,u)\in \delta^{+}(v)}(\re_{(v,u)} \rplus b_{u}^{\infty}) \rleq b_{v}^{\infty}$ is easy to prove: indeed, as $\re_{(v,u)} \rplus b_{u}^{\infty} \rleq \re_{(v,u)} \rplus b_{u}^{n}$ for each arc $(v,u) $ in $ \delta^{+}(v)$ and for all $n$ in $\mathbb{Z}_{+}$, we have  $\bigmeet_{(v,u)\in \delta^{+}(v)}(\re_{(v,u)} \rplus b_{u}^{\infty}) \rleq \bigmeet_{(v,u)\in \delta^{+}(v)}(\re_{(v,u)} \rplus b_{u}^{n}) = b_{v}^{n+1}$ for all $n \in \mathbb{Z}_{+}$, which gives the result. The inequality $\bigmeet_{(v,u)\in \delta^{+}(v)}(\re_{(v,u)} \rplus b_{u}^{\infty}) \rgeq b_{v}^{\infty}$ requires a transfinite induction.
\end{rem}

\subsection{Generalized Dijkstra algorithm for faster bound computations} 
\label{sub:generalizedDijkstra}



In all our numerical experiments, we have observed that practically, $b_{v}^{\dagger} = b_{v}^{\infty} = b_{v}^{\ell^{*}}$. Therefore, the bounds we obtain if we compute $b_{v}^{\dagger}$ are as good as those we obtain if we compute $b_{v}^{\ell^{*}}$. 
When $\rplus$ distributes with respect to $\meet$, computing $b_{v}^{\dagger}$ amounts to solving the algebraic path problem. There are algorithms for the algebraic path problem that are practically much more efficient than the generalized Ford-Bellman algorithm of the previous section. We now adapt one of these algorithms, which generalizes Dijkstra algorithm, to our lattice ordered monoid setting. 


A resource $\tilde{b}_{v}$ and an integer $\tilde{n}_{v}$ are attached to each vertex $v$ and updated during the algorithm. Initially, $\tilde{b}_v=\bigjoin \rset$, which may be defined only in the completion of $\rset$, and $\tilde{n}_v=+\infty$ for each vertex $v\neq d$, and $\tilde{b}_d=0$.
During the algorithm, a queue $L$ of vertices ``to be extended'' is maintained. Initially, the queue $L$ contains only $d$. The algorithm ends when $L$ is empty. While $L$ is not empty and the minimum $\tilde{n}_{v}$ over all vertices $v$ is non greater than $\ell^{*}$, where $\ell^{*}$ is the maximum length of an elementary path ending in $d$, the following operations are repeated:
\begin{itemize}
\item Extract from $L$ a vertex $v$ with minimum $\tilde{n}_v$. 
\item For each arc $(u,v)$ in $\delta^-(v)$, {\em extend} $v$ along $(u,v)$: if $\tilde{b}_u \nrleq \re_{(u,v)}\rplus \tilde{b}_{v}$, then
	\begin{itemize}
	\item {\em Update} $\tilde{n}_{u} $ to $ \min(\tilde{n}_{u}, 1 + \tilde{n}_{v})$.
	\item {\em Update} $\tilde{b}_{u}$ to $\tilde{b}_{u} \meet (\re_{(u,v)}\rplus \tilde{b}_{v})$.
	\item Add $u$ to $L$ (if it is not already present).
	\end{itemize}
\item Set $\tilde{n}_{v} = +\infty$.
\end{itemize}

\index{algorithm!generalized Dijkstra}

\begin{prop}\label{prop:MeetLbFastComputation}
This algorithm terminates after at most $\ell^{*}|V|$ iterations, where $\ell^{*}$ is the maximum length of an elementary path ending in $d$. The value $b_{v}$ of $\tilde{b}_v$ at the end of the algorithm is equal to $b_{v}^{\ell^{*}}$ for each $v\in V$. If $L$ is empty at the end of the algorithm, then $b_{v} = b_{v}^{\dagger}$ for all vertices $v$.

When  $\rplus$ distributes with respect to $\meet$, at the end of the algorithm, we have $L = \emptyset$ and $b_{v} = b_{v}^{\dagger}$ for all vertices $v$.
\end{prop}

The proof of Proposition \ref{prop:MeetLbFastComputation} is technical but not difficult and relies on Theorem \ref{theo:relationsBetweenBounds}. The interested reader can find it in Proposition 4.15 in \cite{parmentier2016thesis}. If the complexity bounds we obtain are identical to those of the algorithm of the previous section, this algorithm is practically faster in practice, and should be preferred. We use it in our numerical experiments.

In the first step, we can extract a vertex $v$ that minimizes a key function $\phi(\tilde{b}_{v})$ instead of one that minimizes $\tilde{n}_{v}$. Using the key $\phi(x) = x$, our algorithm corresponds to Dijkstra algorithm 
when $(\rset, \rplus,\rleq) = (\mathbb{R}_{+},+,\leq)$. When an alternative $\phi(\tilde{b}_{v})$ is used instead of $\tilde{n}_{v}$, the algorithm ends only when $L$ is empty, and we loose the convergence of Proposition \ref{prop:MeetLbFastComputation}. However, in practice, the list $L$ is always empty after a number of iterations that is not much larger than $|V|$. In our numerical experiments, we use the ratio
\begin{equation}\label{eq:gamma}
	\gamma = \frac{\text{number of vertices extended before convergence}}{|V|}
\end{equation}
\noindent to evaluate how fast the algorithm converges. As each vertex $v$ such that there exists a $v$-$d$ path is extended at least once, if there is a $v$-$d$ path $P$ for each vertex $v$ in $D$, the ratio $\gamma$ is necessarily non smaller than $1$. With a carefully chosen $\phi$, convergence is fast: the worst $\gamma$ encountered in the numerical experiments is 2.6. Using $\tilde{n}_{v}$, we have obtain ration $\gamma$ up to 20 times larger on the instance.

When $\rplus$ distributes with respect to $\meet$, we have a convergence result for arbitrary $\phi$. Indeed, in that case, the only difference between our algorithm and the one proposed by \citet{mohri2002semiring} for the algebraic path problem is that the vertex picked-up in $L$ at each iteration is arbitrarily chosen. \citet{mohri2002semiring} shows that, after a finite but possibly exponential number of iterations, $L$ is empty and the algorithm terminates.

 



\begin{rem}\label{rem:acyclicBounds}
 When digraph $D$ is acylic, the bounds $b_{v}^{\dagger}$ can easily be obtained with an even faster algorithm. Indeed, using a topological ordering, they can be computed iteratively directly from the dynamic programming equation \eqref{eq:meetEquation}.
\end{rem}

\section{Numerical experiments on usual resource constrained shortest path problem} 
\label{sec:numerical_experiments_on_usual_resource_constrained_shortest_path_problem}
In this section, we test numerically our enumeration algorithms on the usual resource constrained shortest path problem with $k = 1$ and $k=10$ constraints, 
\begin{equation}\label{eq:RCSPnumericalResultsPb}
	\begin{array}{rl}
	\displaystyle\min_{P\in \mathcal{P}_{od}} & \displaystyle\sum_{a \in P}w_{a}^{0}, \\
	\text{s.t.}& \displaystyle\sum_{a \in P} w_{a}^{i} \leq W^{i}.
	\end{array}
\end{equation}
where $w_{a}^{i} \in \R$ for each arc $a$ and $i \in \{0,\ldots,k\}$, thresholds $W^{i}\in \R$ for $i \in \{1,\ldots,k\}$, and $\mathcal{P}_{od}$ is the set of $o$-$d$ paths. We model it as a \MRCSP with resources $\re \in \R^{k+1}$, 
$$\rcost(\re)= w^{0}\quad \text{and} \quad \rmeas(\re) = \max_{i}(\ind_{(W^{i},+\infty)}(w^{i})), $$
where $\re = (w^{0},\ldots,w^{k})$, and $\ind_{(W^{i},+\infty)}$ is the indicator function of the interval $(W^{i},+\infty)$. As we focus on instances with positive resources, Theorems \ref{theo:labelSettingConvergence}.(b), \ref{theo:labelCorrectingConvergence}, and \ref{theo:labelDominanceConvergence} ensure respectively the convergence of the generalized A$^{*}$, label correcting, and label dominance algorithms.

\subsection{Instances} 
\label{sub:instances}
We use four families of graphs: road networks, acyclic graphs, grids, and random graphs. The three last families of graphs are used by \citet{cherkassky1996shortest} in their experimental study of algorithms for the \SP. We have used adapted versions of their generators \texttt{spgrid}, \texttt{sprand}, and \texttt{spacyc} to produce instances of these families. The adaptation consists in the insertion of a destination vertex. Among others, these four family of graphs have been used by \citet{dumitrescu2003improved} to test the different \RCSP algorithms available in the literature. We now describe them.

\begin{table}
\begin{center}
 	\begin{tabular}{|lll|}
 		\hline
 			\emph{Name} & \emph{Generator} & \emph{Brief description} \\
 		\hline
 			road & extraction	& Road networks with a given number of vertices. \\
 			\hline
 			square & grid & Square grid of size $m \times m$ \\
 			long & grid & Long grid of size $16m \times m$ \\
 			wide & grid & Wide grid of size $m \times 16 m$ \\
 			\hline
 			acyc & acyclic & 
 			Acyclic graph with $n$ vertices $v_{1},\ldots,v_{n}$ and $5n$ arcs $(v_{i},v_{j})$ with $i<j$ \\
 			\hline
 			rand & random &  			Hamiltonian cycle with $n$ vertices and $5n - n$ chords. \\
			\hline
 	\end{tabular}
\end{center}
	\caption{Summary of the families of graphs used}
	\label{tab:FamilyGraphs}
 \end{table} 
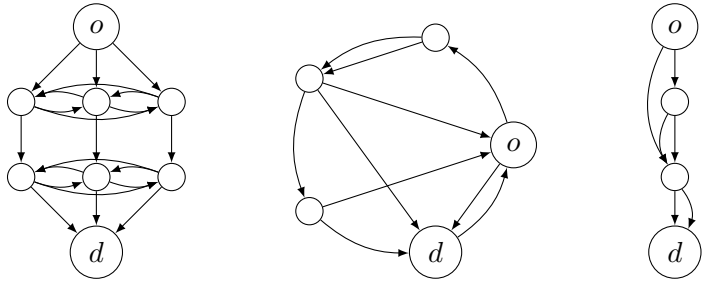
\begin{figure}[!ht]
	\begin{center} 

\begin{tikzpicture}
\tikzset{vertex/.style={circle,draw}}
\tikzset{arc/.style={->,>=latex}}
\node[vertex] (o) at (1.0,2) {$o$};
\node[vertex] (d) at (1.0,-1) {$d$};
\node[vertex] (x0y0) at (0,0) {};
\node[vertex] (x0y1) at (0,1) {};
\node[vertex] (x1y0) at (1,0) {};
\node[vertex] (x1y1) at (1,1) {};
\node[vertex] (x2y0) at (2,0) {};
\node[vertex] (x2y1) at (2,1) {};
\draw[arc] (o) -- (x0y1);
\draw[arc] (x0y0) -- (d);
\draw[arc] (x0y1) -- (x0y0);
\draw[arc] (o) -- (x1y1);
\draw[arc] (x1y0) -- (d);
\draw[arc] (x1y1) -- (x1y0);
\draw[arc] (o) -- (x2y1);
\draw[arc] (x2y0) -- (d);
\draw[arc] (x2y1) -- (x2y0);
\draw[arc] (x0y0) to[bend right = 20] (x2y0);
\draw[arc] (x2y0) to[bend right = 20] (x0y0);
\draw[arc] (x0y0) to[bend right = 20] (x1y0);
\draw[arc] (x1y0) to[bend right = 20] (x0y0);
\draw[arc] (x1y0) to[bend right = 20] (x2y0);
\draw[arc] (x2y0) to[bend right = 20] (x1y0);
\draw[arc] (x0y1) to[bend right = 20] (x2y1);
\draw[arc] (x2y1) to[bend right = 20] (x0y1);
\draw[arc] (x0y1) to[bend right = 20] (x1y1);
\draw[arc] (x1y1) to[bend right = 20] (x0y1);
\draw[arc] (x1y1) to[bend right = 20] (x2y1);
\draw[arc] (x2y1) to[bend right = 20] (x1y1);
\end{tikzpicture}
\hspace{1cm}
\begin{tikzpicture}
\tikzset{vertex/.style={circle,draw}}
\tikzset{arc/.style={->,>=latex}}
\node[vertex] (v0) at ({0.0}:1.5cm){$o$};
{};
\node[vertex] (v1) at ({72.0}:1.5cm){};
\node[vertex] (v2) at ({144.0}:1.5cm){};
\node[vertex] (v3) at ({216.0}:1.5cm){};
\node[vertex] (v4) at ({288.0}:1.5cm){$d$};
\draw[arc] (v0) to[bend right = 20] (v1);
\draw[arc] (v1) to[bend right = 20] (v2);
\draw[arc] (v2) to[bend right = 20] (v3);
\draw[arc] (v3) to[bend right = 20] (v4);
\draw[arc] (v4) to[bend right = 20] (v0);
\draw[arc] (v2) -- (v4);
\draw[arc] (v3) -- (v0);
\draw[arc] (v0) -- (v4);
\draw[arc] (v2) -- (v0);
\draw[arc] (v1) -- (v2);
\end{tikzpicture}
\hspace{1cm}
\begin{tikzpicture}
\tikzset{vertex/.style={circle,draw}}
\tikzset{arc/.style={->,>=latex}}
\node[vertex] (v0) at (0,3) {$o$};
{};
\node[vertex] (v1) at (0,2) {};
\node[vertex] (v2) at (0,1) {};
\node[vertex] (v3) at (0,0) {$d$};
\draw[arc] (v0) -- (v1);
\draw[arc] (v1) -- (v2);
\draw[arc] (v2) -- (v3);
\draw[arc] (v1) to[bend right = 30] (v2);
\draw[arc] (v0) to[bend right = 30] (v2);
\draw[arc] (v2) to[bend left = 30] (v3);
\end{tikzpicture}

		\caption[Examples of our families of graphs]{From left to right, a grid graph of width 3 and depth 2, a random graph with 5 vertices and 9 arcs, and an acyclic graph with 4 vertices and 7 arcs.}
		\label{fig:splibGraphs}
	\end{center}
\end{figure} 

The \emph{road} network graphs have been extracted from the Rome and the San Francisco Bay Area instances of the Dimacs challenge \cite{dimacsChallenge}, and the origin and the destination vertices have been chosen far way from each other. Details on how these instances have been extracted are available in \cite{parmentier2016thesis}. A \emph{grid} graph of length $\ell$ and width $m$ is composed of $\ell$ layers of $m$ vertices. Each layer $i$ is a Hamiltonian cycle $v_{i,1},\ldots, v_{i,m}$ with arcs in both direction. Each vertex $v_{i,j}$ of layer $i\in[\ell - 1]$ is connected to the corresponding vertex $v_{i+1,j}$ in the next layer. An origin vertex $o$ and a destination vertex $d$ are added. There is an arc between the origin $o$ and each vertex of the first layer, and an arc between each vertex of the last layer and the destination. A \emph{random} graph with $n$ vertices and $m \geq n$ arcs is composed of $n$ vertices on a Hamiltonian cycle and randomly generated chords on that cycle. Finally an \emph{acyclic} graph with $n$ vertices and $m\geq n$ arcs contains a path $v_{1}, \ldots, v_{n}$, and randomly generated arcs $(v_{i},v_{j})$ with $i<j$. The origin and the destination are respectively $v_{1}$ and $v_{n}$. Figure \ref{fig:splibGraphs} illustrates a grid, a random and an acyclic graph. We consider random and acyclic graphs with $n$ vertices and $5n$ arcs. We have obtained similar results are obtained when using digraphs with $hn$ arcs with $h$ between $2$ and $50$ \cite{parmentier2016thesis}. Table \ref{tab:FamilyGraphs} provides a summary of the main characteristics of these families of graph.

The weight $w_{a}^{0}$ of an arc $a$ of a road network instance is its length in kilometers. All the other weights $w_{a}^{i}$ of each type of graphs are randomly chosen using an uniform distribution on $[1,100]$. We use the same technique as \cite{dumitrescu2003improved} to ensure the existence of an optimal path: we first search an $o$-$d$ path $P_{c}$ of minimum cost, then find an $o$-$d$ path $P_{w}$ minimizing $\displaystyle\sum_{a \in P_{w}}\sum_{i \in k} w_{a}^{i}$, and finally set $ W^{i} = (1- \lambda) w_{P_{w}}^{i} + \lambda \max(w_{P}^{i},w_{P_{w}}^{i})$, where $\lambda\in[0,1]$ enables to choose the constraint strength. As the relative performances of the algorithms do not depend much of $\lambda$, we always use $\lambda = 0.5$ in this paper. A study of the influence of the constraint strength $\lambda$ is available in \cite{parmentier2016thesis}.








\subsection{Candidate paths, weaker bounds, and algorithms tested} 
 \label{sub:candidate_paths}
 
We have tested the three enumeration algorithms introduced in Section \ref{sec:enumeration_algorithms}. When bounds $b_{v}$ are needed, they are computed in a preprocessing using the generalized Dijkstra algorithm of Section \ref{sub:generalizedDijkstra}, with~$\re \mapsto \sum_{i=0}^{k}w^{i}$ as key function $\phi$ of Equation~\eqref{eq:gamma}, where $\re = (w^{0},\ldots,w^{k})$. On acyclic instances, we use instead the algorithm of Remark~\ref{rem:acyclicBounds}.  We have also tested two variants of the enumeration algorithms, that we now introduce. 

 \paragraph{Candidate paths}
 A standard technique to improve the performance of resource constrained shortest path algorithms \citep{dumitrescu2003improved} relies on the use of $v$-$d$ paths $Q_{v}$ that are likely to be subpaths of optimal $v$-$d$ paths. We find such paths $Q_{v}$ during the preprocessing: we use Dijkstra algorithm to find the $v$-$d$ path $P$ that minimizes $\sum_{a\in P}\sum_{i=0}^{k}w_{a}^{i}$. When an $o$-$v$ path $P$ satisfies the test of an enumeration algorithm, before extending it, we test if $P + Q_{v}$ is a feasible solution of cost smaller than $c_{od}^{UB}$, and update $c_{od}^{UB}$ if it is. This procedure may enable the algorithm to find faster feasible $o$-$d$ paths of good quality. The upper bound $c_{od}^{UB}$ is therefore likely to decrease faster, which enables to strengthen the lower bound test and thus reduce the total number of paths enumerated by the algorithm. 

\paragraph{Weaker bounds} To identify the respective contributions of the lower bound test (Low) and of the keys $\rcost(\re_{P} \rplus b_{v})$ to the performance of the generalized A$^{*}$ and the label correcting algorithms, we also provide numerical results for ``downgraded'' versions of the generalized A$^{*}$ and label correcting algorithms where $c(\re_{P})$ is used instead of $c(\re_{P} \rplus b_{v})$ as key.

\subsection{Experimental setting} 
\label{sub:experimental_setting}

The numerical experiments are performed on a Macbook Pro of 2012 with four 2.5 Ghz processors and 4 Gb of ram. The algorithms are not parallelized. The limiting parameter for the algorithms is the memory available. Therefore, we stop the algorithms if the list $L$ of candidate paths contains more than 1e+05 elements. We also stop the label correcting and the label dominance algorithms if the number of paths in the union of the sets of non dominated paths $M_{v}$ is larger than 1e+05. As we mention in Remark \ref{rem:gapLowerBound}, the minimum $\rcost(\re_{P} + b_{v})$ on the paths $P$ in $L$ when the algorithm is stopped provides a lower bound on the cost of an optimal solution, and $c_{od}^{UB}$ provides an upper bound. These two bounds are used to compute a gap on instances that cannot be solved to optimality. 

\subsection{Numerical results} 
\label{sub:numerical_results_with_one_constraint}

\begin{figure}
	\begin{center}
	\pgfplotsset{width=6cm}
	\begin{outdent}
	\pgfplotsset{width=6.8cm}
	\begin{tabular}{cc}
			\begin{tikzpicture}
			\pgfplotsset{
			    xmin=40, xmax=3e5,
			    legend pos=north west
			}
			\begin{axis}[
				title=road ,
			xmode=log,
			ymode=log,
			ymin=2e-4,
			ymax=1e2,
			xlabel=Vertices,
			ylabel=CPU time (s)
		]
		\addplot[mark=x, color=blue] coordinates{
(100, 0.000389)
(500, 0.010129)
(1000, 0.008211)
(2000, 0.21955)
(10000, 0.043718)
}; 
 \addlegendentry{A$^*$}
\addplot[mark=square, color = red] coordinates{
(100, 0.000619)
(500, 0.001491)
(1000, 0.002394)
(2000, 0.007248)
(3353, 0.009409)
(5000, 0.031929)
(10000, 0.03225)
(20000, 0.073206)
}; 
 \addlegendentry{cor.}
\addplot[mark=o] coordinates{
(100, 0.000735)
(500, 0.016113)
(1000, 0.071951)
(2000, 0.192499)
(3353, 0.150325)
(5000, 4.62358)
}; 
 \addlegendentry{dom.}

		\end{axis}
	\end{tikzpicture}
		 &
			\begin{tikzpicture}
			\pgfplotsset{
			    xmin=40, xmax=3e5,
			    legend pos=north west
			}
			\begin{axis}[
				title=square,
			xmode=log,
			ymode=log,
			ymin=2e-4,
			ymax=1e2,
			xlabel=Vertices,
			ylabel=CPU time (s)
		]
		\addplot[mark=x, color=blue] coordinates{
(102, 0.001451)
(402, 0.008245)
}; 
 \addlegendentry{A$^*$}
\addplot[mark=square, color = red] coordinates{
(102, 0.000999)
(402, 0.00192)
(2502, 0.085002)
(10002, 0.784424)
}; 
 \addlegendentry{cor.}
\addplot[mark=o] coordinates{
(102, 0.001757)
(402, 0.008796)
(2502, 0.744813)
}; 
 \addlegendentry{dom.}

		\end{axis}
	\end{tikzpicture}
		 \\
			\begin{tikzpicture}
			\pgfplotsset{
			    xmin=40, xmax=3e5,
			    legend pos=north west
			}
			\begin{axis}[
				title=long,
			xmode=log,
			ymode=log,
			ymin=2e-4,
			ymax=1e2,
			xlabel=Vertices,
			ylabel=CPU time (s)
		]
		\addplot[mark=x, color=blue] coordinates{
(514, 0.030449)
}; 
 \addlegendentry{A$^*$}
\addplot[mark=square, color = red] coordinates{
(514, 0.004357)
(1282, 0.074746)
(2562, 2.12957)
}; 
 \addlegendentry{cor.}
\addplot[mark=o] coordinates{
(514, 0.034257)
(1282, 0.896529)
(2562, 22.3855)
}; 
 \addlegendentry{dom.}

		\end{axis}
	\end{tikzpicture}
		 &
			\begin{tikzpicture}
			\pgfplotsset{
			    xmin=40, xmax=3e5,
			    legend pos=north west
			}
			\begin{axis}[
				title=wide,
			xmode=log,
			ymode=log,
			ymin=2e-4,
			ymax=1e2,
			xlabel=Vertices,
			ylabel=CPU time (s)
		]
		\addplot[mark=x, color=blue] coordinates{
(514, 0.001437)
(1282, 0.003189)
(2562, 0.006991)
(5122, 0.020066)
(12802, 0.03417)
(25602, 0.072345)
}; 
 \addlegendentry{A$^*$}
\addplot[mark=square, color = red] coordinates{
(514, 0.001485)
(1282, 0.00318)
(2562, 0.008197)
(5122, 0.01854)
(12802, 0.036252)
(25602, 0.073611)
}; 
 \addlegendentry{cor.}
\addplot[mark=o] coordinates{
(514, 0.009795)
(1282, 0.025864)
(2562, 0.054168)
(5122, 0.120234)
(12802, 0.273635)
(25602, 0.461938)
}; 
 \addlegendentry{dom.}

		\end{axis}
	\end{tikzpicture}
		 \\
			\begin{tikzpicture}
			\pgfplotsset{
			    xmin=40, xmax=3e5,
			    legend pos=north west
			}
			\begin{axis}[
				title=acyc,
			xmode=log,
			ymode=log,
			ymin=2e-4,
			ymax=1e2,
			xlabel=Vertices,
			ylabel=CPU time (s)
		]
		\addplot[mark=x, color=blue] coordinates{
(100, 0.000464)
(200, 0.000445)
(500, 0.000496)
(1000, 0.00043)
(2000, 0.002305)
(5000, 0.003271)
(10000, 0.009201)
}; 
 \addlegendentry{A$^*$}
\addplot[mark=square, color = red] coordinates{
(100, 0.000386)
(200, 0.000426)
(500, 0.000391)
(1000, 0.00055)
(2000, 0.0023)
(5000, 0.004111)
(10000, 0.008069)
}; 
 \addlegendentry{cor.}
\addplot[mark=o] coordinates{
(100, 0.000449)
(200, 0.000355)
(500, 0.000225)
(1000, 0.000369)
(2000, 0.000876)
(5000, 0.007376)
(10000, 0.019876)
}; 
 \addlegendentry{dom.}

		\end{axis}
	\end{tikzpicture}
		 &
			\begin{tikzpicture}
			\pgfplotsset{
			    xmin=40, xmax=3e5,
			    legend pos=north west
			}
			\begin{axis}[
				title=rand,
			xmode=log,
			ymode=log,
			ymin=2e-4,
			ymax=1e2,
			xlabel=Vertices,
			ylabel=CPU time (s)
		]
		\addplot[mark=x, color=blue] coordinates{
(100, 0.000821)
(200, 0.001487)
(500, 0.002519)
(1000, 0.002183)
(2000, 0.009682)
(5000, 0.028041)
(10000, 0.074832)
}; 
 \addlegendentry{A$^*$}
\addplot[mark=square, color = red] coordinates{
(100, 0.000825)
(200, 0.001216)
(500, 0.002524)
(1000, 0.002785)
(2000, 0.010273)
(5000, 0.026467)
(10000, 0.08017)
}; 
 \addlegendentry{cor.}
\addplot[mark=o] coordinates{
(100, 0.000506)
(200, 0.001178)
(500, 0.000877)
(1000, 0.003646)
(2000, 0.001265)
(5000, 0.023892)
(10000, 0.225905)
}; 
 \addlegendentry{dom.}

		\end{axis}
	\end{tikzpicture}
		
	\end{tabular}
	\end{outdent}
	\end{center}
	\caption{Usual RCSP with $k=1$ constraint}
	\label{fig:usualRCSP1constraint}
\end{figure}
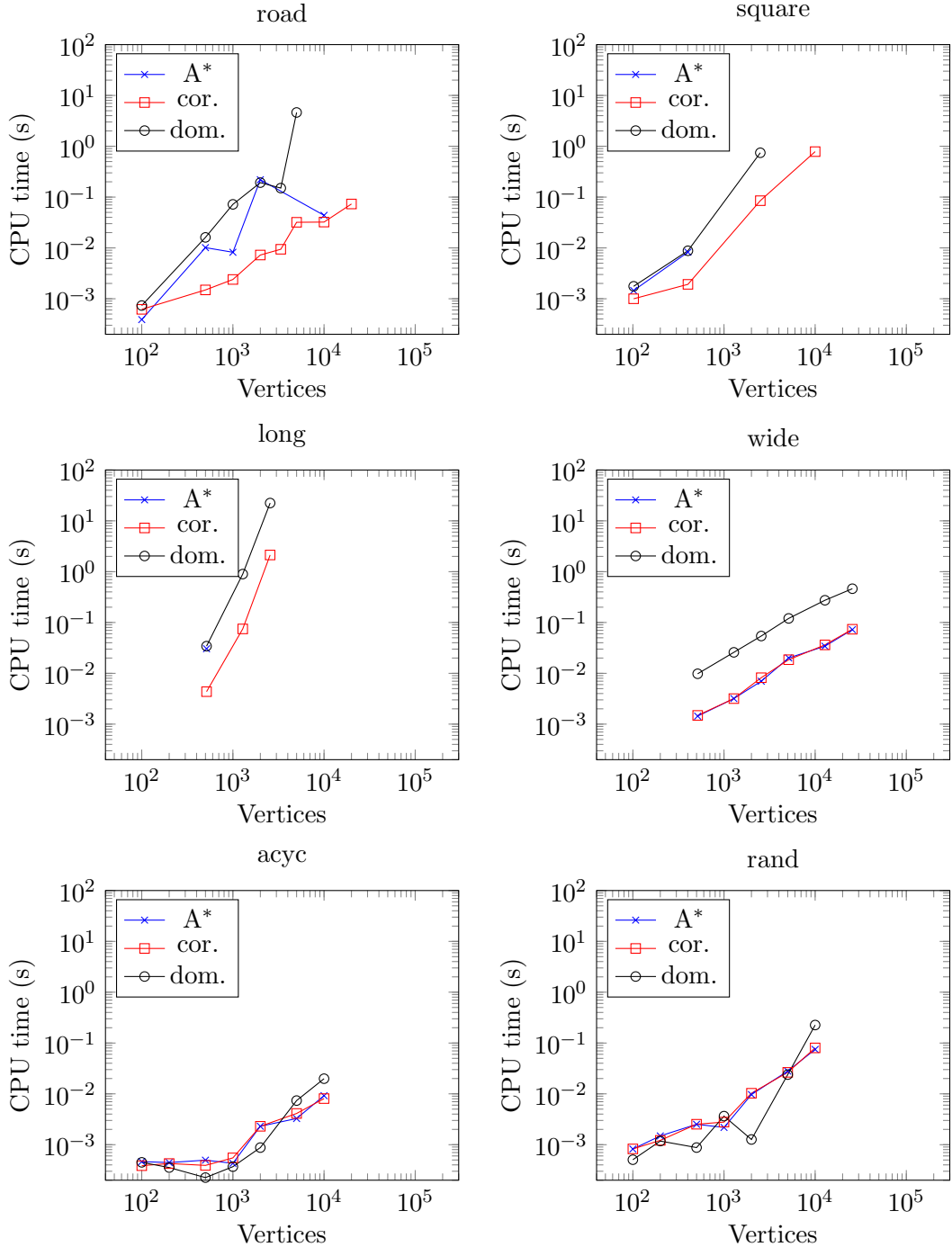



\begin{figure}
	\begin{center}
	\pgfplotsset{width=6cm}
	\begin{outdent}
	\pgfplotsset{width=6.8cm}
	\begin{tabular}{cc}
			\begin{tikzpicture}
			\pgfplotsset{
			    xmin=40, xmax=3e5,
			    legend pos=south east
			}
			\begin{axis}[
				title=road ,
			xmode=log,
			ymode=log,
			ymin=2e-4,
			ymax=1e2,
			xlabel=Vertices,
			ylabel=CPU time (s)
		]
		\addplot[mark=x, color=blue] coordinates{
(100, 0.000483)
(500, 0.002118)
(1000, 0.007863)
(2000, 0.022758)
(3353, 0.085129)
}; 
 \addlegendentry{A$^*$}
\addplot[mark=square, color = red] coordinates{
(100, 0.000668)
(500, 0.00162)
(1000, 0.003521)
(2000, 0.011561)
(3353, 0.017571)
(20000, 0.302474)
}; 
 \addlegendentry{cor.}
\addplot[mark=o] coordinates{
(100, 0.003545)
(500, 0.102751)
(1000, 49.1832)
}; 
 \addlegendentry{dom.}

		\end{axis}
	\end{tikzpicture}
		 &
			\begin{tikzpicture}
			\pgfplotsset{
			    xmin=40, xmax=3e5,
			    legend pos=south east
			}
			\begin{axis}[
				title=square,
			xmode=log,
			ymode=log,
			ymin=2e-4,
			ymax=1e2,
			xlabel=Vertices,
			ylabel=CPU time (s)
		]
		\addplot[mark=x, color=blue] coordinates{
(102, 0.000715)
(402, 0.017494)
}; 
 \addlegendentry{A$^*$}
\addplot[mark=square, color = red] coordinates{
(102, 0.000742)
(402, 0.011569)
}; 
 \addlegendentry{cor.}
\addplot[mark=o] coordinates{
(102, 0.004955)
(402, 29.1471)
}; 
 \addlegendentry{dom.}

		\end{axis}
	\end{tikzpicture}
		 \\
			\begin{tikzpicture}
			\pgfplotsset{
			    xmin=40, xmax=3e5,
			    legend pos=south east
			}
			\begin{axis}[
				title=long,
			xmode=log,
			ymode=log,
			ymin=2e-4,
			ymax=1e2,
			xlabel=Vertices,
			ylabel=CPU time (s)
		]
		\addplot[mark=x, color=blue] coordinates{
(514, 0.816493)
}; 
 \addlegendentry{A$^*$}
\addplot[mark=square, color = red] coordinates{
(514, 0.230767)
}; 
 \addlegendentry{cor.}
\addplot[mark=o] coordinates{
}; 
 \addlegendentry{dom.}

		\end{axis}
	\end{tikzpicture}
		 &
			\begin{tikzpicture}
			\pgfplotsset{
			    xmin=40, xmax=3e5,
			    legend pos=south east
			}
			\begin{axis}[
				title=wide,
			xmode=log,
			ymode=log,
			ymin=2e-4,
			ymax=1e2,
			xlabel=Vertices,
			ylabel=CPU time (s)
		]
		\addplot[mark=x, color=blue] coordinates{
(514, 0.004722)
(1282, 0.006398)
(2562, 0.013052)
(5122, 0.027395)
(12802, 0.058484)
(25602, 0.121602)
}; 
 \addlegendentry{A$^*$}
\addplot[mark=square, color = red] coordinates{
(514, 0.00335)
(1282, 0.006709)
(2562, 0.013882)
(5122, 0.029149)
(12802, 0.06101)
(25602, 0.124577)
}; 
 \addlegendentry{cor.}
\addplot[mark=o] coordinates{
(514, 1.39105)
(1282, 3.12023)
(2562, 6.34309)
}; 
 \addlegendentry{dom.}

		\end{axis}
	\end{tikzpicture}
		 \\
			\begin{tikzpicture}
			\pgfplotsset{
			    xmin=40, xmax=3e5,
			    legend pos=north west
			}
			\begin{axis}[
				title=acyc,
			xmode=log,
			ymode=log,
			ymin=2e-4,
			ymax=1e2,
			xlabel=Vertices,
			ylabel=CPU time (s)
		]
		\addplot[mark=x, color=blue] coordinates{
(100, 0.000319)
(200, 0.000386)
(500, 0.00036)
(1000, 0.000339)
(2000, 0.000653)
(5000, 0.001342)
(10000, 0.002787)
}; 
 \addlegendentry{A$^*$}
\addplot[mark=square, color = red] coordinates{
(100, 0.000322)
(200, 0.000351)
(500, 0.00029)
(1000, 0.000361)
(2000, 0.000679)
(5000, 0.00149)
(10000, 0.00252)
}; 
 \addlegendentry{cor.}
\addplot[mark=o] coordinates{
(100, 0.000398)
(200, 0.000271)
(500, 0.000199)
(1000, 0.000141)
(2000, 0.000721)
(5000, 0.00098)
(10000, 0.001079)
}; 
 \addlegendentry{dom.}

		\end{axis}
	\end{tikzpicture}
		 &
			\begin{tikzpicture}
			\pgfplotsset{
			    xmin=40, xmax=3e5,
			    legend pos=south east
			}
			\begin{axis}[
				title=rand,
			xmode=log,
			ymode=log,
			ymin=2e-4,
			ymax=1e2,
			xlabel=Vertices,
			ylabel=CPU time (s)
		]
		\addplot[mark=x, color=blue] coordinates{
(100, 0.000384)
(200, 0.001973)
(500, 0.006193)
(1000, 0.000399)
(2000, 0.000908)
(5000, 0.096801)
(10000, 0.241285)
}; 
 \addlegendentry{A$^*$}
\addplot[mark=square, color = red] coordinates{
(100, 0.000367)
(200, 0.002229)
(500, 0.004806)
(1000, 0.000316)
(2000, 0.001133)
(5000, 0.097646)
(10000, 0.232376)
}; 
 \addlegendentry{cor.}
\addplot[mark=o] coordinates{
(100, 0.000278)
(200, 0.00105)
(500, 0.000643)
(1000, 0.00023)
(2000, 0.000814)
(5000, 0.009883)
(10000, 0.041806)
}; 
 \addlegendentry{dom.}

		\end{axis}
	\end{tikzpicture}
	\end{tabular}
	\end{outdent}
	\end{center}
	\caption{Usual RCSP with $k=10$ constraints}
	\label{fig:usualRCSP10constraints}
\end{figure}
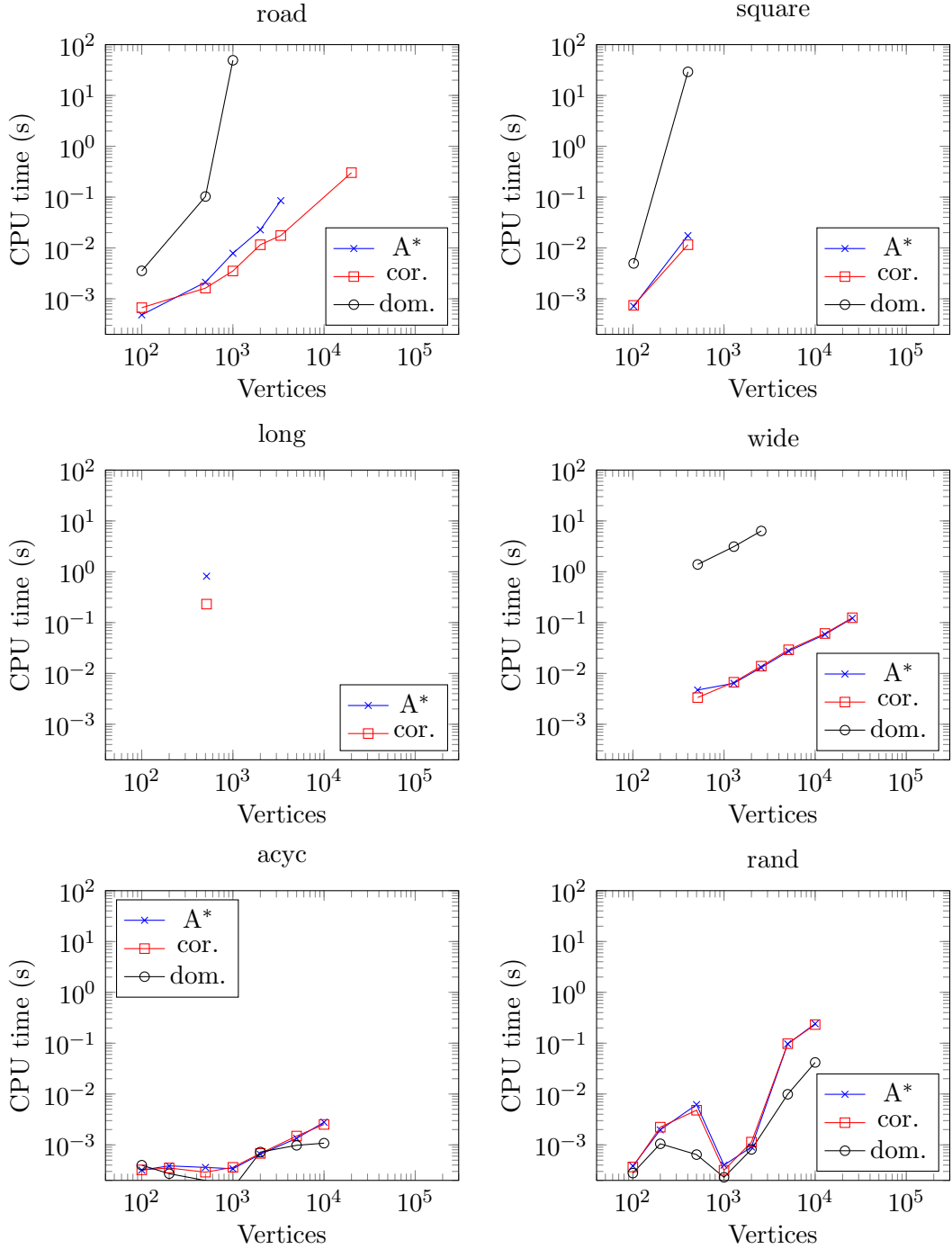 

We now come to numerical results on Problem \eqref{eq:RCSPnumericalResultsPb}. Figure \ref{fig:usualRCSP1constraint} plots the performance of our algorithms for each family of instances. As there is only one degree of freedom in the definition of instances of each family of graphs, we can identify an instance of a given family by its number of vertices. For each of our enumeration algorithms in Table \ref{tab:mrcspAlgo}, there is a curve  that gives the total CPU time as a function of the number of vertices. This total CPU time includes the computation of bounds in the preprocessing. Curves $A^{*}$, cor., and dom. correspond respectively to the generalized A$^{*}$, the label correcting and the label dominance algorithm. We only add points corresponding to instance solved to optimality. Therefore, the longer and the lower a curve, the better the performance of the corresponding algorithm. Figure \ref{fig:usualRCSP10constraints} provides the same results for the problem with $k=10$ constraints. 

\begin{table}
\begin{outdent}
\begin{footnotesize}
\begin{tabular}{|l|rrc|rr|rrr|rr|r|}
\hline
Instance 
& \multicolumn{1}{c}{$|V|$} 
& \multicolumn{1}{c}{$|A|$} 
& \multicolumn{1}{c|}{Alg.} 
& \multicolumn{1}{c}{$\gamma$} 
& \multicolumn{1}{c|}{Preproc.} 
& \multicolumn{1}{c}{Ext.} 
& \multicolumn{1}{c}{Cut.} 
& \multicolumn{1}{c|}{Dom.} 
& \multicolumn{1}{c}{$\ell$} 
& \multicolumn{1}{c|}{Gap} 
& \multicolumn{1}{c|}{CPU (s)} \\
\hline
road20 & 20000 & 55180 & A$^*$ & 1.6 & 15\% &54131 & 14121 & -- & -- &$\infty$ & 3.08e-01 \\
&&&A$^*$ CP & 1.6 & 9\%+10\% &55762 & 17579 & -- &252 & 2.5\% & 4.91e-01 \\
&&&A$^*$~K & 1.6 & 16\% &71402 & 4 & -- & -- &$\infty$ & 3.21e-01 \\
&&&cor. & 1.6 & 64\% &7899 & 8306 & 44\% &243 & opt & 7.32e-02 \\
&&&cor. CP & 1.6 & 36\%+37\% &7852 & 8228 & 44\% &243 & opt & 1.30e-01 \\
&&&cor.~K & 1.6 & 1\% &1128194 & 793042 & 92\% &243 & opt & 7.17e+00 \\
&&&dom. & -- & --  &1505043 & 995349 & -- & -- &$\infty$ & 1.06e+01 \\
&&&dom. CP & -- & -- +0\% &1505043 & 995349 & -- &243 & 21.0\% & 1.20e+01 \\
\hline
road50 & 50000 & 138112 & A$^*$ & 2.7 & 43\% &44613 & 3710 & -- & -- &$\infty$ & 4.42e-01 \\
&&&A$^*$ CP & 2.7 & 25\%+21\% &44613 & 3710 & -- &494 & 13.6\% & 7.21e-01 \\
&&&A$^*$~K & 2.7 & 42\% &71420 & 0 & -- & -- &$\infty$ & 4.71e-01 \\
&&&cor. & 2.7 & 17\% &153191 & 80428 & 96\% & -- &$\infty$ & 1.12e+00 \\
&&&cor. CP & 2.7 & 12\%+10\% &153221 & 80488 & 96\% &478 & 5.5\% & 1.60e+00 \\
&&&cor.~K & 2.7 & 1\% &1570406 & 1047263 & 99\% & -- &$\infty$ & 1.37e+01 \\
&&&dom. & -- & --  &1557409 & 1029848 & -- & -- &$\infty$ & 1.21e+01 \\
&&&dom. CP & -- & -- +1\% &1557409 & 1029848 & -- &462 & 431.2\% & 1.44e+01 \\
\hline
square100 & 10002 & 30100 & A$^*$ & 1.6 & 9\% &58533 & 17162 & -- & -- &$\infty$ & 2.69e-01 \\
&&&A$^*$ CP & 1.6 & 7\%+6\% &58533 & 17162 & -- &131 & 30.4\% & 4.36e-01 \\
&&&A$^*$~K & 1.6 & 10\% &50298 & 690 & -- & -- &$\infty$ & 2.62e-01 \\
&&&cor. & 1.6 & 3\% &168036 & 155429 & 58\% &118 & opt & 7.84e-01 \\
&&&cor. CP & 1.6 & 2\%+2\% &168016 & 155396 & 58\% &118 & opt & 1.01e+00 \\
&&&cor.~K & 1.6 & 0\% &1488625 & 981538 & 99\% & -- &$\infty$ & 1.17e+01 \\
&&&dom. & -- & --  &1493529 & 973040 & -- & -- &$\infty$ & 1.05e+01 \\
&&&dom. CP & -- & -- +0\% &1493529 & 973040 & -- &131 & 25.1\% & 1.21e+01 \\
\hline
long20 & 5122 & 15376 & A$^*$ & 1.6 & 5\% &51961 & 3933 & -- & -- &$\infty$ & 2.66e-01 \\
&&&A$^*$ CP & 1.6 & 3\%+4\% &51961 & 3933 & -- &438 & 58.6\% & 4.17e-01 \\
&&&A$^*$~K & 1.6 & 6\% &49996 & 3 & -- & -- &$\infty$ & 2.36e-01 \\
&&&cor. & 1.6 & 1\% &186590 & 93798 & 96\% & -- &$\infty$ & 1.12e+00 \\
&&&cor. CP & 1.6 & 1\%+1\% &186590 & 93798 & 96\% &419 & 33.9\% & 1.45e+00 \\
&&&cor.~K & 1.6 & 0\% &1558600 & 1025370 & 100\% & -- &$\infty$ & 2.56e+01 \\
&&&dom. & -- & --  &1558884 & 1024684 & -- & -- &$\infty$ & 2.36e+01 \\
&&&dom. CP & -- & -- +0\% &1558884 & 1024684 & -- &437 & 202.3\% & 2.62e+01 \\
\hline
wide100 & 25602 & 78400 & A$^*$ & 1.5 & 97\% &107 & 1809 & -- &21 & opt & 7.23e-02 \\
&&&A$^*$ CP & 1.5 & 51\%+47\% &104 & 1806 & -- &21 & opt & 1.42e-01 \\
&&&A$^*$~K & 1.5 & 19\% &67108 & 35812 & -- & -- &$\infty$ & 4.22e-01 \\
&&&cor. & 1.5 & 97\% &98 & 1785 & 0\% &21 & opt & 7.36e-02 \\
&&&cor. CP & 1.5 & 50\%+48\% &95 & 1782 & 0\% &21 & opt & 1.40e-01 \\
&&&cor.~K & 1.5 & 17\% &105678 & 99041 & 58\% &21 & opt & 4.51e-01 \\
&&&dom. & -- & --  &148903 & 83968 & -- &21 & opt & 4.62e-01 \\
&&&dom. CP & -- & -- +10\% &148564 & 83772 & -- &21 & opt & 6.79e-01 \\
\hline
acyc10000 & 10000 & 50000 & A$^*$ & 1 & 97\% &12 & 63 & -- &10 & opt & 9.20e-03 \\
&&&A$^*$ CP & 1 & 32\%+66\% &7 & 51 & -- &10 & opt & 2.59e-02 \\
&&&A$^*$~K & 1 & 96\% &22 & 127 & -- &10 & opt & 9.14e-03 \\
&&&cor. & 1 & 95\% &12 & 63 & 0\% &10 & opt & 8.07e-03 \\
&&&cor. CP & 1 & 29\%+70\% &7 & 51 & 0\% &10 & opt & 2.61e-02 \\
&&&cor.~K & 1 & 95\% &22 & 127 & 0\% &10 & opt & 1.01e-02 \\
&&&dom. & -- & --  &5918 & 1413 & -- &10 & opt & 1.99e-02 \\
&&&dom. CP & -- & -- +50\% &5909 & 1407 & -- &10 & opt & 3.44e-02 \\
\hline
rand10000 & 10000 & 50000 & A$^*$ & 2.6 & 99\% &41 & 161 & -- &10 & opt & 7.48e-02 \\
&&&A$^*$ CP & 2.6 & 68\%+32\% &37 & 144 & -- &10 & opt & 1.10e-01 \\
&&&A$^*$~K & 2.6 & 89\% &2078 & 9393 & -- &10 & opt & 1.05e-01 \\
&&&cor. & 2.6 & 99\% &41 & 161 & 0\% &10 & opt & 8.02e-02 \\
&&&cor. CP & 2.6 & 65\%+35\% &37 & 144 & 0\% &10 & opt & 1.13e-01 \\
&&&cor.~K & 2.6 & 89\% &1957 & 7889 & 2\% &10 & opt & 1.17e-01 \\
&&&dom. & -- & --  &67626 & 41871 & -- &10 & opt & 2.26e-01 \\
&&&dom. CP & -- & -- +12\% &67378 & 41653 & -- &10 & opt & 3.31e-01 \\
\hline
\end{tabular}
\end{footnotesize}
\end{outdent}
\caption{Standard resource constrained shortest path with $k=1$ resource constraint}
\label{tab:RCSP1constraint}
\end{table}

\begin{table}
\begin{outdent}
\begin{footnotesize}
\begin{tabular}{|l|rrc|rr|rrr|rr|r|}
\hline
Instance 
& \multicolumn{1}{c}{$|V|$} 
& \multicolumn{1}{c}{$|A|$} 
& \multicolumn{1}{c|}{Alg.} 
& \multicolumn{1}{c}{$\gamma$} 
& \multicolumn{1}{c|}{Preproc.} 
& \multicolumn{1}{c}{Ext.} 
& \multicolumn{1}{c}{Cut.} 
& \multicolumn{1}{c|}{Dom.} 
& \multicolumn{1}{c}{$\ell$} 
& \multicolumn{1}{c|}{Gap} 
& \multicolumn{1}{c|}{CPU (s)} \\
\hline
road20 & 20000 & 55180 & A$^*$ & 2.1 & 7\% &253214 & 323551 & -- & -- &$\infty$ & 9.75e-01 \\
&&&A$^*$ CP & 2.1 & 6\%+4\% &253214 & 323551 & -- &221 & 55.3\% & 1.23e+00 \\
&&&A$^*$~K & 2.1 & 5\% &340302 & 556262 & -- & -- &$\infty$ & 1.50e+00 \\
&&&cor. & 2.1 & 24\% &45066 & 66374 & 21\% &221 & opt & 3.02e-01 \\
&&&cor. CP & 2.1 & 18\%+13\% &45034 & 66317 & 21\% &221 & opt & 3.82e-01 \\
&&&cor.~K & 2.1 & 25\% &45177 & 66416 & 21\% &221 & opt & 2.96e-01 \\
&&&dom. & -- & --  &1103713 & 603707 & -- & -- &$\infty$ & 3.48e+01 \\
&&&dom. CP & -- & -- +0\% &1103713 & 603707 & -- &221 & 373.3\% & 3.91e+01 \\
\hline
road50 & 50000 & 138112 & A$^*$ & 2.5 & 47\% &56159 & 9481 & -- & -- &$\infty$ & 4.59e-01 \\
&&&A$^*$ CP & 2.5 & 30\%+19\% &56159 & 9481 & -- &400 & 18.1\% & 7.12e-01 \\
&&&A$^*$~K & 2.5 & 43\% &71421 & 1 & -- & -- &$\infty$ & 5.02e-01 \\
&&&cor. & 2.5 & 12\% &64093 & 17783 & 53\% & -- &$\infty$ & 1.80e+00 \\
&&&cor. CP & 2.5 & 10\%+7\% &64093 & 17783 & 53\% &400 & 18.0\% & 2.10e+00 \\
&&&cor.~K & 2.5 & 1\% &466563 & 303206 & 99\% & -- &$\infty$ & 2.91e+01 \\
&&&dom. & -- & --  &477467 & 307503 & -- & -- &$\infty$ & 2.67e+01 \\
&&&dom. CP & -- & -- +0\% &477467 & 307503 & -- &400 & 1321.4\% & 2.76e+01 \\
\hline
square50 & 2502 & 7550 & A$^*$ & 2.2 & 3\% &81247 & 62539 & -- & -- &$\infty$ & 3.46e-01 \\
&&&A$^*$ CP & 2.2 & 2\%+2\% &81247 & 62539 & -- &52 & 57.5\% & 4.73e-01 \\
&&&A$^*$~K & 2.2 & 3\% &56775 & 13594 & -- & -- &$\infty$ & 2.85e-01 \\
&&&cor. & 2.2 & 0\% &141815 & 103501 & 39\% & -- &$\infty$ & 2.41e+00 \\
&&&cor. CP & 2.2 & 0\%+0\% &141815 & 103501 & 39\% &52 & 49.3\% & 2.64e+00 \\
&&&cor.~K & 2.2 & 1\% &179777 & 93497 & 89\% & -- &$\infty$ & 2.00e+00 \\
&&&dom. & -- & --  &168274 & 78864 & -- & -- &$\infty$ & 1.64e+00 \\
&&&dom. CP & -- & -- +0\% &168274 & 78864 & -- &52 & 376.1\% & 1.84e+00 \\
\hline
long5 & 1282 & 3856 & A$^*$ & 2.2 & 2\% &64642 & 29295 & -- & -- &$\infty$ & 2.79e-01 \\
&&&A$^*$ CP & 2.2 & 1\%+1\% &64642 & 29295 & -- &81 & 67.9\% & 4.02e-01 \\
&&&A$^*$~K & 2.2 & 2\% &52538 & 5087 & -- & -- &$\infty$ & 2.39e-01 \\
&&&cor. & 2.2 & 0\% &101197 & 52388 & 48\% & -- &$\infty$ & 1.70e+00 \\
&&&cor. CP & 2.2 & 0\%+0\% &101197 & 52388 & 48\% &81 & 62.9\% & 1.90e+00 \\
&&&cor.~K & 2.2 & 0\% &196476 & 98269 & 99\% & -- &$\infty$ & 3.06e+00 \\
&&&dom. & -- & --  &195012 & 96678 & -- & -- &$\infty$ & 2.81e+00 \\
&&&dom. CP & -- & -- +0\% &195012 & 96678 & -- &81 & 599.7\% & 3.03e+00 \\
\hline
wide100 & 25602 & 78400 & A$^*$ & 2.1 & 85\% &6702 & 14999 & -- &17 & opt & 1.22e-01 \\
&&&A$^*$ CP & 2.1 & 59\%+27\% &6698 & 14994 & -- &17 & opt & 1.81e-01 \\
&&&A$^*$~K & 2.1 & 83\% &6711 & 15017 & -- &17 & opt & 1.38e-01 \\
&&&cor. & 2.1 & 82\% &6413 & 13731 & 3\% &17 & opt & 1.25e-01 \\
&&&cor. CP & 2.1 & 60\%+27\% &6409 & 13726 & 3\% &17 & opt & 1.74e-01 \\
&&&cor.~K & 2.1 & 83\% &6421 & 13745 & 3\% &17 & opt & 1.40e-01 \\
&&&dom. & -- & --  &108128 & 38179 & -- & -- &$\infty$ & 4.92e-01 \\
&&&dom. CP & -- & -- +7\% &108128 & 38179 & -- &17 & 314.6\% & 6.69e-01 \\
\hline
acyc10000 & 10000 & 50000 & A$^*$ & 1 & 91\% &6 & 24 & -- &5 & opt & 2.79e-03 \\
&&&A$^*$ CP & 1 & 47\%+51\% &2 & 12 & -- &5 & opt & 4.97e-03 \\
&&&A$^*$~K & 1 & 89\% &6 & 24 & -- &5 & opt & 2.46e-03 \\
&&&cor. & 1 & 83\% &6 & 24 & 0\% &5 & opt & 2.52e-03 \\
&&&cor. CP & 1 & 42\%+56\% &2 & 12 & 0\% &5 & opt & 4.93e-03 \\
&&&cor.~K & 1 & 84\% &6 & 24 & 0\% &5 & opt & 2.63e-03 \\
&&&dom. & -- & --  &211 & 0 & -- &5 & opt & 1.08e-03 \\
&&&dom. CP & -- & -- +77\% &210 & 0 & -- &5 & opt & 3.95e-03 \\
\hline
rand10000 & 10000 & 50000 & A$^*$ & 6.3 & 100\% &13 & 56 & -- &7 & opt & 2.41e-01 \\
&&&A$^*$ CP & 6.3 & 88\%+12\% &9 & 37 & -- &7 & opt & 2.58e-01 \\
&&&A$^*$~K & 6.3 & 100\% &14 & 57 & -- &7 & opt & 2.47e-01 \\
&&&cor. & 6.3 & 100\% &13 & 56 & 0\% &7 & opt & 2.32e-01 \\
&&&cor. CP & 6.3 & 88\%+12\% &9 & 37 & 0\% &7 & opt & 2.67e-01 \\
&&&cor.~K & 6.3 & 100\% &14 & 57 & 0\% &7 & opt & 2.50e-01 \\
&&&dom. & -- & --  &7183 & 160 & -- &7 & opt & 4.18e-02 \\
&&&dom. CP & -- & -- +40\% &7177 & 159 & -- &7 & opt & 8.55e-02 \\
\hline
\end{tabular}
\end{footnotesize}
\end{outdent}
\caption{Standard resource constrained shortest path with $k=10$ resource constraint}
\label{tab:RCSP10constraints}
\end{table}

Tables \ref{tab:RCSP1constraint} and \ref{tab:RCSP10constraints} provide detailed results on some difficult instances with respectively $k=1$ and $k=10$ constraints. The first columns provide the instance considered, the number of vertices, and the number of arcs in the instance. The next column provides the enumeration algorithm tested. Again, A$^{*}$ corresponds to the generalized A$^{*}$ algorithm, cor.~to the label correcting algorithm, and dom.~to the label dominance algorithm. The suffix CP indicates that the variant of the algorithm with candidate paths is considered, and the suffix K that $c(\re_{P})$ is used instead of $c(\re_{P} \rplus b_{v})$. The ratio $\gamma$ of the next column is the one of Equation~\eqref{eq:gamma}, and indicates the performance of the generalized Dijkstra algorithm. The lower $\gamma$, the better is the performance of this algorithm.  The next column provides the percentage of the total CPU time spent in the preprocessing. For the standard version of the label dominance algorithm, no preprocessing is needed, while for the generalized A$^{*}$ and the label correcting algorithm, this preprocessing time indicates the time needed to compute lower bounds. When candidate paths are computed, the percentage after the symbol $+$ indicates the percentage of the total CPU time spent computing them. The two next columns provide the number of paths extended and the number of paths cut. The next column indicates the percentage of paths cut by the dominance test, the remaining being cut by the lower bound test. The other algorithm are not concerned as they use a single test to cut paths. The column $\ell$ provides the number of arcs in the solution returned. The next column provides the gap between the lower and the upper bound when the algorithm is stopped before convergence, and the last column provides the total CPU time.

\paragraph{Difficulty of the instances.} We first, remark that the difficulty of the instances is mainly linked to the number of arcs $\ell$ in an optimal solution. The smaller this number of arcs, the easier the instances. The acyclic and the random graph instances are therefore the easiest to solve, and any version of our algorithms can solve them well. Then comes the road network instances and the wide grids. The most difficult instances are the square and the long grids. 

\paragraph{Relative performance of the algorithms.} The label dominance algorithm has the best performances on the easy instances, i.e.~the acyclic and the random instances. If we look at Tables \ref{tab:RCSP1constraint} and \ref{tab:RCSP10constraints}, we can see that on these instances, most of the computation time of the generalized A$^{*}$ and of the label correcting algorithms is spent in the preprocessing, gathering information that is not needed as the problem is easy to solve. On the other instances, the label correcting algorithm has the best performance. Using lower bounds in addition to the dominance test is always interesting on difficult instances. The relative performance of the generalized A$^{*}$ algorithm and of the label correcting algorithm depends on the family of instances considered.

\paragraph{Influence of dimension and relative performance of the algorithms.} Instance with ten constraints are more difficult to solve than instances with one constraint. We also underline the fact that the relative performance of the algorithms on instances with ten constraints is different from the performance in the case of instances with one constraint. Indeed, the label dominance algorithm exhibits poor performances in that setting, while the generalized A$^{*}$ performs almost as well as the label correcting algorithm. We can also see in the tables that the percentage of paths cut by the dominance test in the label correcting algorithm decreases with the dimension. Section \ref{sec:what_to_do_on_difficult_problems} explains why the performance of the dominance test decreases with the dimension, and provides techniques to increase the performance of our algorithms on problems in large dimension. 

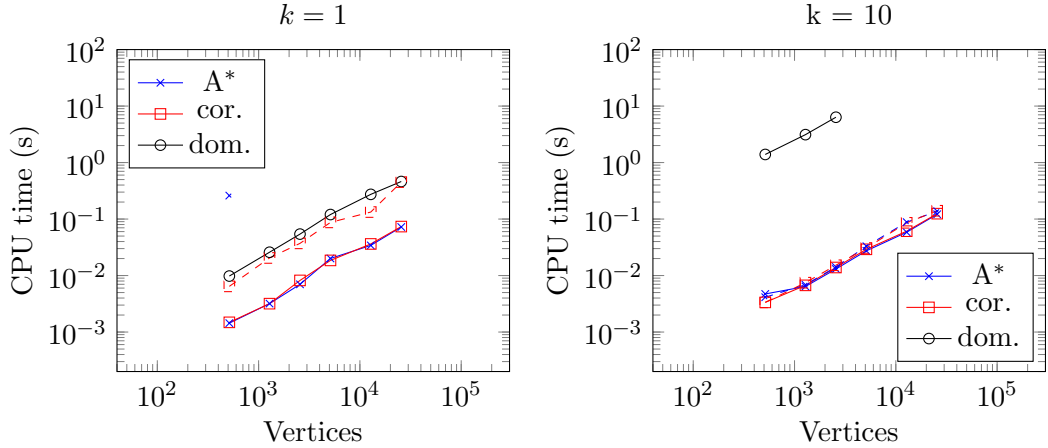
\begin{figure}
	\begin{center}
	\pgfplotsset{width=6cm}
	\begin{outdent}
	\pgfplotsset{width=6.8cm}
	\begin{tabular}{cc}
			\begin{tikzpicture}
			\pgfplotsset{
			    xmin=40, xmax=3e5,
			    legend pos=north west
			}
			\begin{axis}[
				title={$k=1$},
			xmode=log,
			ymode=log,
			ymin=2e-4,
			ymax=1e2,
			xlabel=Vertices,
			ylabel=CPU time (s)
		]
		\addplot[mark=x, color=blue] coordinates{
(514, 0.001437)
(1282, 0.003189)
(2562, 0.006991)
(5122, 0.020066)
(12802, 0.03417)
(25602, 0.072345)
}; 
 \addlegendentry{A$^*$}
\addplot[mark=square, color = red] coordinates{
(514, 0.001485)
(1282, 0.00318)
(2562, 0.008197)
(5122, 0.01854)
(12802, 0.036252)
(25602, 0.073611)
}; 
 \addlegendentry{cor.}
\addplot[mark=o] coordinates{
(514, 0.009795)
(1282, 0.025864)
(2562, 0.054168)
(5122, 0.120234)
(12802, 0.273635)
(25602, 0.461938)
}; 
 \addlegendentry{dom.}
\addplot[mark=x, dashed, color=blue] coordinates{
(514, 0.260207)
};
\addplot[mark=square, dashed, color = red] coordinates{
(514, 0.006434)
(1282, 0.020472)
(2562, 0.037129)
(5122, 0.087285)
(12802, 0.133063)
(25602, 0.450641)
};

		\end{axis}
	\end{tikzpicture}
		&
			\begin{tikzpicture}
			\pgfplotsset{
			    xmin=40, xmax=3e5,
			    legend pos=south east
			}
			\begin{axis}[
				title={k = 10},
			xmode=log,
			ymode=log,
			ymin=2e-4,
			ymax=1e2,
			xlabel=Vertices,
			ylabel=CPU time (s)
		]
		\addplot[mark=x, color=blue] coordinates{
(514, 0.004722)
(1282, 0.006398)
(2562, 0.013052)
(5122, 0.027395)
(12802, 0.058484)
(25602, 0.121602)
}; 
 \addlegendentry{A$^*$}
\addplot[mark=square, color = red] coordinates{
(514, 0.00335)
(1282, 0.006709)
(2562, 0.013882)
(5122, 0.029149)
(12802, 0.06101)
(25602, 0.124577)
}; 
 \addlegendentry{cor.}
\addplot[mark=o] coordinates{
(514, 1.39105)
(1282, 3.12023)
(2562, 6.34309)
}; 
 \addlegendentry{dom.}
\addplot[mark=x, dashed, color=blue] coordinates{
(514, 0.004134)
(1282, 0.006865)
(2562, 0.014162)
(5122, 0.033011)
(12802, 0.088141)
(25602, 0.137566)
};
\addplot[mark=square, dashed, color = red] coordinates{
(514, 0.003354)
(1282, 0.007476)
(2562, 0.015222)
(5122, 0.030777)
(12802, 0.085834)
(25602, 0.140288)
};

		\end{axis}
	\end{tikzpicture}
		
	\end{tabular}
	\end{outdent}
	\end{center}
	\caption{Influence of keys on usual RCSP with $k=1$ and $k=10$ constraint}
	\label{fig:usualRCSP1constraint-NoBiKandCP}
\end{figure}

\paragraph{Influence of keys.} Figure \ref{fig:usualRCSP1constraint-NoBiKandCP} provides the performances of the generalized A$^{*}$ and of the label correcting algorithms with different keys to solve Problem \eqref{eq:RCSPnumericalResultsPb} on wide grids with $k=1$ constraint.  Plain lines correspond to the key $c(\re_{P} \rplus b_{v})$ and dashed lines to the key $c(\re_{P})$. On instances with $k=1$ constraint, There is no-dashed line corresponding to the generalized A$^{*}$ because only one instance was solved by the algorithm with key $c(\re_{P})$. When bounds $b_{v}$ have been computed for the lower bound test, is is always more interesting to use the key $c(\re_{P} \rplus b_{v})$ than the key $c(\re_{P})$. In practice, the key has an influence only on weakly constrained problems with $k=1$, where the keys  identify which partial solutions are the most promising. On more constrained problems with $k=10$, we are not able to guess which partial solution will lead to a feasible solution, and the importance of keys decreases.



\begin{figure}
	\begin{center}
	\pgfplotsset{width=6cm}
	\begin{outdent}
	\pgfplotsset{width=6.8cm}
	\begin{tabular}{cc}
			\begin{tikzpicture}
			\pgfplotsset{
			    xmin=40, xmax=3e5,
			    legend pos=north west
			}
			\begin{axis}[
				title={k = 1},
			xmode=log,
			ymode=log,
			ymin=2e-4,
			ymax=1e2,
			xlabel=Vertices,
			ylabel=CPU time (s)
		]
		\addplot[mark=x, color=blue] coordinates{
(514, 0.001437)
(1282, 0.003189)
(2562, 0.006991)
(5122, 0.020066)
(12802, 0.03417)
(25602, 0.072345)
}; 
 \addlegendentry{A$^*$}
\addplot[mark=square, color = red] coordinates{
(514, 0.001485)
(1282, 0.00318)
(2562, 0.008197)
(5122, 0.01854)
(12802, 0.036252)
(25602, 0.073611)
}; 
 \addlegendentry{cor.}
\addplot[mark=o] coordinates{
(514, 0.009795)
(1282, 0.025864)
(2562, 0.054168)
(5122, 0.120234)
(12802, 0.273635)
(25602, 0.461938)
}; 
 \addlegendentry{dom.}
\addplot[mark=x, dashed, color=blue] coordinates{
(514, 0.002612)
(1282, 0.007271)
(2562, 0.012441)
(5122, 0.035295)
(12802, 0.067074)
(25602, 0.142346)
};
\addplot[mark=square, dashed, color = red] coordinates{
(514, 0.002878)
(1282, 0.006271)
(2562, 0.014017)
(5122, 0.0291)
(12802, 0.06482)
(25602, 0.139674)
};
\addplot[mark=o, dashed] coordinates{
(514, 0.014435)
(1282, 0.041292)
(2562, 0.077636)
(5122, 0.176324)
(12802, 0.394782)
(25602, 0.678791)
};

		\end{axis}
	\end{tikzpicture}
		&
			\begin{tikzpicture}
			\pgfplotsset{
			    xmin=40, xmax=3e5,
			    legend pos=south east
			}
			\begin{axis}[
				title={k = 10},
			xmode=log,
			ymode=log,
			ymin=2e-4,
			ymax=1e2,
			xlabel=Vertices,
			ylabel=CPU time (s)
		]
		\addplot[mark=x, color=blue] coordinates{
(514, 0.004722)
(1282, 0.006398)
(2562, 0.013052)
(5122, 0.027395)
(12802, 0.058484)
(25602, 0.121602)
}; 
 \addlegendentry{A$^*$}
\addplot[mark=square, color = red] coordinates{
(514, 0.00335)
(1282, 0.006709)
(2562, 0.013882)
(5122, 0.029149)
(12802, 0.06101)
(25602, 0.124577)
}; 
 \addlegendentry{cor.}
\addplot[mark=o] coordinates{
(514, 1.39105)
(1282, 3.12023)
(2562, 6.34309)
}; 
 \addlegendentry{dom.}
\addplot[mark=x, dashed, color=blue] coordinates{
(514, 0.005035)
(1282, 0.008804)
(2562, 0.017298)
(5122, 0.037513)
(12802, 0.087256)
(25602, 0.181193)
};
\addplot[mark=square, dashed, color = red] coordinates{
(514, 0.004722)
(1282, 0.008896)
(2562, 0.017942)
(5122, 0.041012)
(12802, 0.085367)
(25602, 0.174259)
};
\addplot[mark=o, dashed] coordinates{
(514, 1.52978)
(1282, 3.38565)
(2562, 6.86693)
};

		\end{axis}
	\end{tikzpicture}
		
	\end{tabular}
	\end{outdent}
	\end{center}
	\caption{Influence of candidate paths on usual RCSP with $k=1$ and $k=10$ constraint}
	\label{fig:usualRCSP1constraint-CP}
\end{figure}
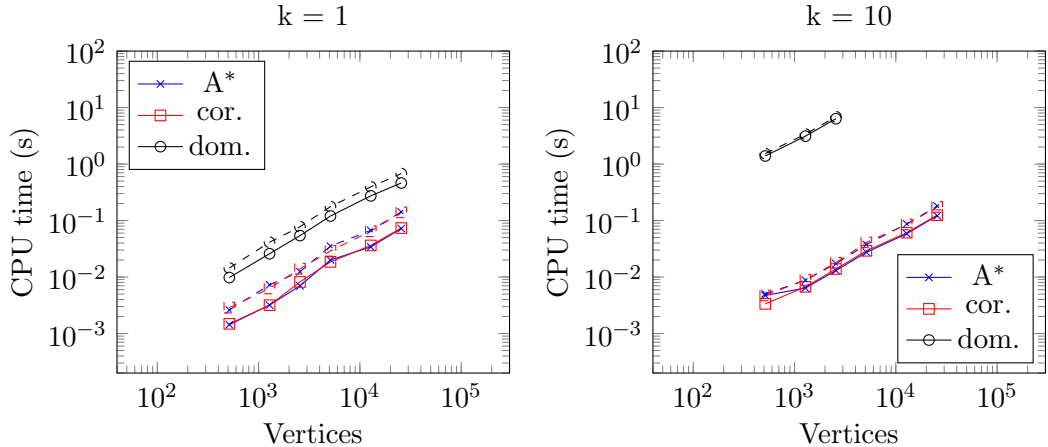 

\paragraph{Influence of candidate paths.}
Figure \ref{fig:usualRCSP1constraint-CP} shows the influence of candidate paths on the performances of the algorithms on the wide grid instances with $k=1$ and $k=10$ constraints. Plain lines correspond to algorithms without candidate paths, and dashed lines to algorithms with candidate paths. Candidate paths are not interesting on instances that can be solved to optimality: the decrease in $c_{od}^{UB}$ is not large enough to enable to reduce significantly the number of paths enumerated. Therefore, the use of candidate paths only slow the algorithm by requiring a preprocessing and slowing each step of the enumeration. On the contrary, we can see in Tables \ref{tab:RCSP1constraint} and \ref{tab:RCSP10constraints} that candidate paths become interesting on difficult instances: they enable to find feasible $o$-$d$ paths of small cost along the algorithm, and thus to obtain a smaller gap.




\section{Stochastic path problems: models and numerical experiments} 
\label{sec:stochastic_path_problems}

We now come to the stochastic path problems mentioned in the introduction. We suppose to have a random variable $\xi_{a}$ for each arc $a$, and denote $\xi_{P} = \sum_{a \in P}\xi_{a}$ for each path $P$. Given an origin $o$ and a destination $v$, a stochastic path problem typically seeks an $o$-$d$ path $P$ minimizing a probability functional~$\mu$
$$\min_{P\in \mathcal{P}_{od}} \mu\left(\sum_{a \in P}\xi_{a}\right).$$
Three types of probability functionals $\mu$ are of specific interest. First, given~$\tau \in \R$, the functional $\P(\cdot >\tau)$ enables to model the \emph{probability of arriving on time }at an event starting at $\tau$. Second, utility theory considers the expectation $\E(f(\cdot))$ of a non-decreasing cost function $f$. Third, stochastic optimization often deals with \emph{risk measures} \cite{artzner1999coherent}, i.e. probability functionals $\mu$ that are normalized, $\mu(0) = 0$, monotone with respect to the almost sure order, i.e.~$\mu(\xi) \leq \mu(\tilde{\xi})$ if $\xi\leq \tilde{\xi}$ almost surely, and invariant by translation $\mu(\xi+c) = \mu(\xi) + c$ for all $c\in \mathbb{R}$. Alternative definitions of risk measures exist in the literature, but all assume the monotonicity with respect to the almost sure order.  
Stochastic constraints are typically of the form
$$\mu'(\xi_{P}) \leq \alpha,$$
where $\mu'$ is a probability functional. The most common stochastic constraints are probability constraints of the type 
$$\P(\xi_{P} > \tau) \leq \alpha.$$

We now provide techniques to model stochastic path problems within the \MRCSP framework. To that purpose, with introduce several lattice ordered monoids of random variables. On a given lattice ordered monoid $(\rset,\rplus,\rleq)$, we can model any stochastic path problem such that the functions
$$\rcost(\xi) = \mu(\xi) \quad \text{and} \quad \rmeas(\xi) = \ind_{(\alpha,\infty]}(\mu'(\xi))$$ 
are isotone with respect to $\rleq$. On the lattice ordered monoids we consider, this property is satisfied by any probability functional $\mu$ or $\mu'$ that is monotone with respect to the almost sure order. This is notably the case of all the probability functionals $\mu$ and $\mu'$ of interest aforementioned.



\subsection{General case} 
\label{sub:general_case}
Given arbitrary random variables $\xi_{a}$, the vector space of the random variables $\xi_{a}$ endowed with the addition and the almost sure order is a lattice ordered monoid. Indeed, the addition is compatible with the almost sure order, as given three random variables $\xi,\tilde{\xi},$ and $\xi'$ such that $\xi \leq \tilde{\xi}$ a.s., we have $\xi + \xi' \leq \tilde{\xi} + \xi'$ a.s.. Besides, the almost sure order induces a lattice structure and the meet of two random variables is their essential-infimum, which is unique up to a.s.~equality.

Practically, to be able to make the computations, we sample from the initial random variables a finite number $N$ of scenarios $\omega_{1},\ldots,\omega_{N}$. The sampled distribution can therefore be encoded as elements of $\R^{N}$. The a.s.~order between the sampled distributions becomes the component by component order on $\R^{N}$, and the essential infimum becomes the minimum:
$$(\xi \meet \tilde{\xi})(\omega_{i}) = \min(\xi(\omega_{i}),\tilde{\xi}(\omega_{i})).$$
The lattice ordered monoid we use is therefore $(\R^{N}, +, \leq)$. We provide in \cite{parmentier2016thesis} upper bounds on the error committed due to sampling that are exponential in the number of samples.

\subsection{Independent random variables} 
\label{sub:independent_random_variables}

When the random variables $\xi_{a}$ are independent, we can work on their distributions. Indeed, the distribution of the sum of two random variables is their convolution product $\ast$. In that case, we can use the \emph{usual stochastic order} $\leqst$ which is such that
$$ \xi \leqst \tilde{\xi} \quad \text{if} \quad \P(\xi \leq t) \geq \P(\tilde{\xi} \leq t) \text{ for all }t\in \R. $$ 
The usual stochastic order is compatible with the convolution product. Practically, we work on the set $\bbM$ of distributions of random variables with finite support on $\Z$. We denote $F_{\xi}$ the cumulative distribution of $\xi$. The usual stochastic order induces a lattice structure on $\bbM$, and the meet $\xi \meetst \tilde{\xi}$ of two random variables $\xi$ and $\tilde{\xi}$ is defined through its cumulative distribution:
$$ F_{\xi \meetst \tilde{\xi}} = \max(F_{\xi},F_{\tilde{\xi}}).$$
We can therefore work with resources in the lattice ordered monoid $(\bbM,\ast,\leqst)$. A probability functional $\mu$ is version independent if $\mu(\xi)$ depends only on the distribution of $\xi$. As we work on the random variables distributions, we can use only version-independent probability functionals. 

All the probability functionals of interest mentioned at the beginning of the section are version independent. And as they are also monotone with respect to the almost sure order, the following proposition ensures that we can deal with them using the \MRCSP framework with resource in $(\bbM,\ast,\leqst)$.

\begin{prop}\label{prop:MonotonicityLeqSt}
A version independent probability functional that is monotone with respect to the almost sure order is monotone with respect to the usual stochastic order $\leqst$.
\end{prop}
The proof of this proposition is straightforward and can be found in \cite{parmentier2016thesis}. The converse statement holds even for probability functionals that are not version independent as the usual stochastic order is \emph{coarser} than the almost sure order, i.e.
$$\xi \leq \tilde{\xi} \quad \text{ a.s.\ implies } \quad\xi \leqst \tilde{\xi}.$$

As $\leqst$ is coarser than the almost sure order, the lower bounds we obtain using our algorithms with resources in $(\bbM,\ast,\leqst)$ are tighter than the one we obtain using the sampling approach of the previous section with resources in $(R^{N},+,\leqst)$. Therefore, when the $\xi_{a}$ are independent, it is always more interesting from an algorithmic point of view to use resources in $(\bbM,\ast,\leqst)$ than to use the sampling approach.

We can also use parametric families of distributions that are stable by convolution product instead of discrete distributions. We notably introduce in \cite{parmentier2016thesis} lattice ordered monoid structures to model the families of distributions considered in the literature: normal distributions, gamma distributions, and Cauchy distributions.

\begin{rem}
$(\bbM,\ast,\leqst)$ is an example of lattice ordered monoid that is not an indempotent semiring, as $\ast$ does not distribute with respect to $\meetst$.
\end{rem}

\subsection{Numerical results} 

We now benchmark our algorithms on two stochastic path problems. We consider here independent random variables $\xi_{a}$. We provide in \cite{parmentier2016thesis} numerical results on the same problems and graph instances, but with sampled distributions of non-independent random variables $\xi_{a}$.

The \emph{Condition Value-at-Risk} \cite{rockafellar2000optimization}
 of level $\beta \in [0,1)$ is
\begin{equation*}
	\cvar_{\beta}(\xi) = \frac{1}{1-\beta}\int_{\beta}^{1}\mathrm{VaR}_{\alpha}(\xi)d\alpha,
\end{equation*}
where $\mathrm{VaR}_{\alpha}(\xi) = \inf\left\{t |\P(\xi\leq t) \geq \alpha \right\}$. Intuitively, the Conditional Value at Risk of level $\beta$ can be interpreted as the expectation in the $\beta$ worst case. Parameter $\beta$ enables to choose a level of risk awareness. 
As the conditional value at risk is one of the most popular risk measures in stochastic optimization, the first problem on which we test our algorithms is
\begin{equation}\label{eq:stoShortestPathCvar}
	\min_{P\in \mathcal{P}_{od}} \cvar_{\beta}\left(\sum_{a\in P} \xi_{a}\right).
\end{equation}
For instance, the optimal solution with $\beta=0.05$ of this problem is the best itinerary for a commuter going to work: it gives an itinerary that is still good the most congested day of the month.

The second problem we consider is a shortest path problem with deterministic cost and probability constraints:
\begin{equation}\label{eq:stoConstraintedShortestPath}
	\begin{array}{rl}
		\displaystyle\min_{P\in \mathcal{P}_{od}}& \displaystyle \sum_{a \in P}w_{a}, \\
		\text{s.t.}& \displaystyle\P\left(\sum_{a \in P}\xi_{a} > \tau\right) \leq \alpha.
	\end{array}
\end{equation}
where $\tau \in \R$ and $w_{a}$ are weights in $\R$. This problem can be interpreted as a truck delivery problem. The objective is to find a path of minimum cost among those that guarantee a certain probability of arriving on time. 

On both problems, we use independent distributions $\xi_{a}$, and model them as elements of the lattice ordered monoid $(\bbM,\ast,\leqst)$. The resources of Problem \eqref{eq:stoShortestPathCvar} therefore belong to $\bbM$, and those of Problem \eqref{eq:stoConstraintedShortestPath} to $\R\times\bbM$ endowed with the product sum and order. As we focus on instances with positive resources, Theorems \ref{theo:labelSettingConvergence}.(b), \ref{theo:labelCorrectingConvergence}, and \ref{theo:labelDominanceConvergence} ensure respectively the convergence of the generalized A$^{*}$, label correcting, and label dominance algorithm.

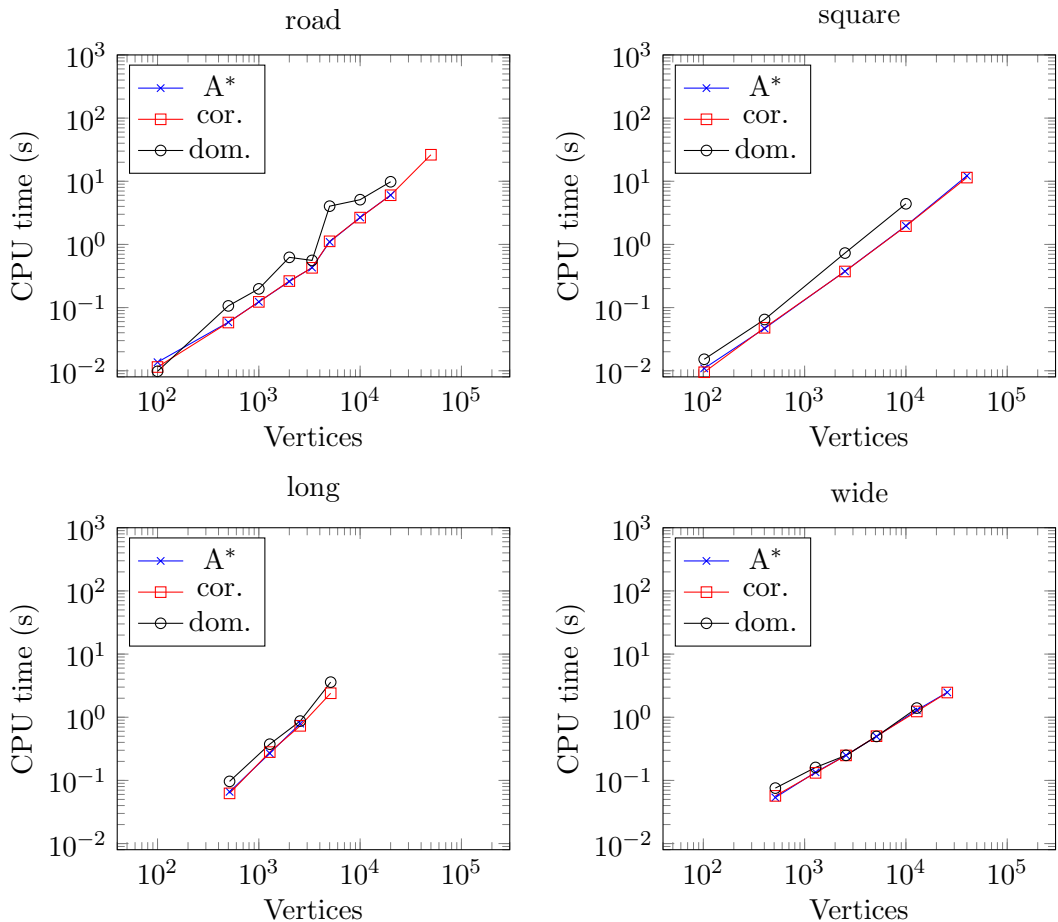
\begin{figure}[!ht]
	\begin{center}
	\pgfplotsset{width=6cm}
	\begin{outdent}
	\pgfplotsset{width=6.8cm}
	\begin{tabular}{cc}
			\begin{tikzpicture}
			\pgfplotsset{
			    xmin=40, xmax=3e5,
			    legend pos=north west
			}
			\begin{axis}[
				title=road ,
			xmode=log,
			ymode=log,
			ymin=8e-3,
			ymax=1e3,
			xlabel=Vertices,
			ylabel=CPU time (s)
		]
		\addplot[mark=x, color=blue] coordinates{
(100, 0.013664)
(500, 0.058606)
(1000, 0.123415)
(2000, 0.258685)
(3353, 0.430084)
(5000, 1.09567)
(10000, 2.67523)
(20000, 6.03718)
}; 
 \addlegendentry{A$^*$}
\addplot[mark=square, color = red] coordinates{
(100, 0.011458)
(500, 0.058021)
(1000, 0.123153)
(2000, 0.262931)
(3353, 0.426142)
(5000, 1.11986)
(10000, 2.65882)
(20000, 6.00871)
(50000, 26.2838)
}; 
 \addlegendentry{cor.}
\addplot[mark=o] coordinates{
(100, 0.009793)
(500, 0.106376)
(1000, 0.197945)
(2000, 0.624449)
(3353, 0.558449)
(5000, 4.03461)
(10000, 5.10854)
(20000, 9.81511)
}; 
 \addlegendentry{dom.}

		\end{axis}
	\end{tikzpicture}
		 &
			\begin{tikzpicture}
			\pgfplotsset{
			    xmin=40, xmax=3e5,
			    legend pos=north west
			}
			\begin{axis}[
				title=square,
			xmode=log,
			ymode=log,
			ymin=8e-3,
			ymax=1e3,
			xlabel=Vertices,
			ylabel=CPU time (s)
		]
		\addplot[mark=x, color=blue] coordinates{
(102, 0.01091)
(402, 0.046675)
(2502, 0.37531)
(10002, 1.97274)
(40002, 12.197)
}; 
 \addlegendentry{A$^*$}
\addplot[mark=square, color = red] coordinates{
(102, 0.009481)
(402, 0.047906)
(2502, 0.373112)
(10002, 1.94999)
(40002, 11.4339)
}; 
 \addlegendentry{cor.}
\addplot[mark=o] coordinates{
(102, 0.015202)
(402, 0.064808)
(2502, 0.729879)
(10002, 4.39842)
}; 
 \addlegendentry{dom.}

		\end{axis}
	\end{tikzpicture}
		 \\
			\begin{tikzpicture}
			\pgfplotsset{
			    xmin=40, xmax=3e5,
			    legend pos=north west
			}
			\begin{axis}[
				title=long,
			xmode=log,
			ymode=log,
			ymin=8e-3,
			ymax=1e3,
			xlabel=Vertices,
			ylabel=CPU time (s)
		]
		\addplot[mark=x, color=blue] coordinates{
(514, 0.066068)
(1282, 0.268499)
(2562, 0.79418)
}; 
 \addlegendentry{A$^*$}
\addplot[mark=square, color = red] coordinates{
(514, 0.06237)
(1282, 0.280649)
(2562, 0.731268)
(5122, 2.39913)
}; 
 \addlegendentry{cor.}
\addplot[mark=o] coordinates{
(514, 0.096503)
(1282, 0.374592)
(2562, 0.869536)
(5122, 3.59673)
}; 
 \addlegendentry{dom.}

		\end{axis}
	\end{tikzpicture}
		 &
			\begin{tikzpicture}
			\pgfplotsset{
			    xmin=40, xmax=3e5,
			    legend pos=north west
			}
			\begin{axis}[
				title=wide,
			xmode=log,
			ymode=log,
			ymin=8e-3,
			ymax=1e3,
			xlabel=Vertices,
			ylabel=CPU time (s)
		]
		\addplot[mark=x, color=blue] coordinates{
(514, 0.053614)
(1282, 0.13456)
(2562, 0.249215)
(5122, 0.496487)
(12802, 1.29543)
(25602, 2.47793)
}; 
 \addlegendentry{A$^*$}
\addplot[mark=square, color = red] coordinates{
(514, 0.056878)
(1282, 0.13166)
(2562, 0.248782)
(5122, 0.501471)
(12802, 1.23456)
(25602, 2.46209)
}; 
 \addlegendentry{cor.}
\addplot[mark=o] coordinates{
(514, 0.075538)
(1282, 0.160443)
(2562, 0.249335)
(5122, 0.497101)
(12802, 1.39702)
}; 
 \addlegendentry{dom.}

		\end{axis}
	\end{tikzpicture}
		 \\
	\end{tabular}
	\end{outdent}
	\end{center}
	\caption{Stochastic Shortest Path Problem with generic distributions and $\beta = 0.05$}
	\label{fig:dctDist}
\end{figure}

\paragraph{Instances.} 
We build instances by generating resources on the road, square grid, long grid, and wide grid digraphs introduced in Section \ref{sec:numerical_experiments_on_usual_resource_constrained_shortest_path_problem}. We have generated generic discrete distributions with finite support as follows. First, the length of the arcs are rescaled to be non greater than 200. Second, their support size is chosen as 10 plus a random integer between 0 and the scaled length of the arc. For each $t$ in the support, a weight $\tilde{\P}(\xi = t)$ is randomly chosen using a uniform distribution on $[0,1]$ for each $t$. The distributions $\tilde{\P}$ is then normalized to obtain a probability distribution $\P$. The measure is then normalized to obtain a probability distribution. The probability functionals $\P(\cdot>\tau)$ and $\cvar_{\beta}$ have been tested for different values of $\tau$ and $\beta$. We have chosen $\tau$ as follows: a first parameter $\tau^{-1}$ is chosen between 0 and 1, then $\tau$ is chosen as the smallest $t$ such that $\P(b_{o}>t) \leq \tau^{-1}$ where $b_{o}$ is the bound provided by the bounding algorithm for the origin vertex $o$. It enables to obtain a threshold $\tau$ such that the $\P(\xi_{P} > \tau) \simeq \tau^{-1}$ where $\xi_{P}$ is the resource of an optimal path. To choose the parameter $\alpha$, we use the same technique as the one we used in Section \ref{sec:numerical_experiments_on_usual_resource_constrained_shortest_path_problem} to choose the parameter $\omega^{i}$: we set $\alpha = \max\left(\P\left(\sum_{a \in P_{c}}\xi_{a} > \tau\right), \left(\P\left(\sum_{a \in P_{c}}\xi_{a} > \tau\right) + \P(\sum_{a \in P_{\rmeas}}\xi_{a} > \tau)\right)/2\right)$, where $P_{c}$ is an $o$-$d$ path $P$ with minimum $\sum_{a \in P}w_{a}$, and $P_{\rmeas}$ is an $o$-$d$ path $P$ with minimum $\P\left(\sum_{a \in P}\xi_{a} > \tau\right)$. 

For both problems, numerical experiments on the same digraphs but with truncated and discretized \emph{lognormal} distributions are available in \cite{parmentier2016thesis}. The performances of the algorithms with these distributions are similar.


\paragraph{Experimental setting.} On medium and large instances, the order of magnitude of the number of non-zero terms in the support of $\xi$ is a few thousands. Storing the resources therefore requires more memory than in the usual RCSP case. We have therefore set a maximum size of $1e+04$ for the list of candidate paths $L$. To avoid spending too much time in the convolution products, we compute them using a Fast Fourier Transform. To that purpose, we use the Fast Fourier Transform \texttt{C++} library \texttt{kissFFT} \cite{kissFFT}. Concerning the computation of lower bounds, we use the generalized Dijkstra algorithm of Section \ref{sub:generalizedDijkstra}. We use $\xi \mapsto \E(\xi)$ as key function $\phi$ when solving Problem~\eqref{eq:stoShortestPathCvar}, and $(w,\xi)\mapsto w + \E(\xi)$ when solving Problem~\eqref{eq:stoConstraintedShortestPath}.



\begin{table}
\begin{outdent}
\begin{small}
\begin{tabular}{|l|rrc|rr|rrr|rr|r|}
\hline
Instance 
& \multicolumn{1}{c}{$|V|$} 
& \multicolumn{1}{c}{$|A|$} 
& \multicolumn{1}{c|}{Alg.} 
& \multicolumn{1}{c}{$\gamma$} 
& \multicolumn{1}{c|}{Preproc.} 
& \multicolumn{1}{c}{Ext.} 
& \multicolumn{1}{c}{Cut.} 
& \multicolumn{1}{c|}{Dom.} 
& \multicolumn{1}{c}{$\ell$} 
& \multicolumn{1}{c|}{Gap} 
& \multicolumn{1}{c|}{CPU (s)} \\
\hline
road50 & 50000 & 138112 & A$^*$ & 1 & 74\% &5044 & 0 & -- & -- &$\infty$ & 3.15e+01 \\
&&&A$^*$ CP & 1 & 65\%+34\% &281 & 610 & -- &468 & opt & 3.27e+01 \\
&&&A$^*$~K & 1 & 80\% &7330 & 0 & -- & -- &$\infty$ & 2.91e+01 \\
&&&cor. & 1 & 91\% &1648 & 3135 & 4\% &468 & opt & 2.63e+01 \\
&&&cor. CP & 1 & 64\%+35\% &281 & 610 & 0\% &468 & opt & 3.49e+01 \\
&&&cor.~K & 1 & 25\% &147208 & 97207 & 100\% & -- &$\infty$ & 9.69e+01 \\
&&&dom. & -- & --  &147208 & 97207 & -- & -- &$\infty$ & 1.74e+01 \\
&&&dom. CP & -- & -- +14\% &147208 & 97207 & -- &469 & 140.3\% & 7.84e+01 \\
\hline
square200 & 40002 & 120200 & A$^*$ & 1 & 82\% &2801 & 5800 & -- &261 & opt & 1.22e+01 \\
&&&A$^*$ CP & 1 & 60\%+39\% &194 & 576 & -- &261 & opt & 1.54e+01 \\
&&&A$^*$~K & 1 & 82\% &4903 & 0 & -- & -- &$\infty$ & 1.27e+01 \\
&&&cor. & 1 & 92\% &1162 & 2488 & 1\% &261 & opt & 1.14e+01 \\
&&&cor. CP & 1 & 60\%+39\% &194 & 576 & 0\% &261 & opt & 1.53e+01 \\
&&&cor.~K & 1 & 22\% &147641 & 97640 & 100\% & -- &$\infty$ & 4.64e+01 \\
&&&dom. & -- & --  &147641 & 97640 & -- & -- &$\infty$ & 1.09e+01 \\
&&&dom. CP & -- & -- +5\% &423004 & 280038 & -- &261 & opt & 1.18e+02 \\
\hline
long50 & 12802 & 38416 & A$^*$ & 1 & 36\% &4995 & 0 & -- & -- &$\infty$ & 2.07e+01 \\
&&&A$^*$ CP & 1 & 52\%+35\% &591 & 1172 & -- &1014 & opt & 1.28e+01 \\
&&&A$^*$~K & 1 & 57\% &4995 & 0 & -- & -- &$\infty$ & 1.32e+01 \\
&&&cor. & 1 & 34\% &5427 & 288 & 100\% & -- &$\infty$ & 2.18e+01 \\
&&&cor. CP & 1 & 52\%+35\% &591 & 1172 & 0\% &1014 & opt & 1.29e+01 \\
&&&cor.~K & 1 & 6\% &148493 & 98492 & 100\% & -- &$\infty$ & 1.34e+02 \\
&&&dom. & -- & --  &148493 & 98492 & -- & -- &$\infty$ & 2.28e+01 \\
&&&dom. CP & -- & -- +4\% &148493 & 98492 & -- &1018 & 112.6\% & 1.21e+02 \\
\hline
wide100 & 25602 & 78400 & A$^*$ & 1 & 98\% &20 & 1638 & -- &20 & opt & 2.48e+00 \\
&&&A$^*$ CP & 1 & 61\%+38\% &1 & 1591 & -- &20 & opt & 3.60e+00 \\
&&&A$^*$~K & 1 & 76\% &4203 & 0 & -- & -- &$\infty$ & 3.17e+00 \\
&&&cor. & 1 & 98\% &20 & 1638 & 0\% &20 & opt & 2.46e+00 \\
&&&cor. CP & 1 & 61\%+38\% &1 & 1591 & 0\% &20 & opt & 3.59e+00 \\
&&&cor.~K & 1 & 39\% &46003 & 27868 & 100\% & -- &$\infty$ & 6.26e+00 \\
&&&dom. & -- & --  &46003 & 27868 & -- & -- &$\infty$ & 1.56e+00 \\
&&&dom. CP & -- & -- +28\% &46003 & 27868 & -- &20 & 33.3\% & 5.22e+00 \\
\hline
\end{tabular}
\end{small}
\end{outdent}
\caption{Stochastic shortest path problem with generic distributions and $\beta = 0.05$.}
\label{tab:dctDist}
\end{table}

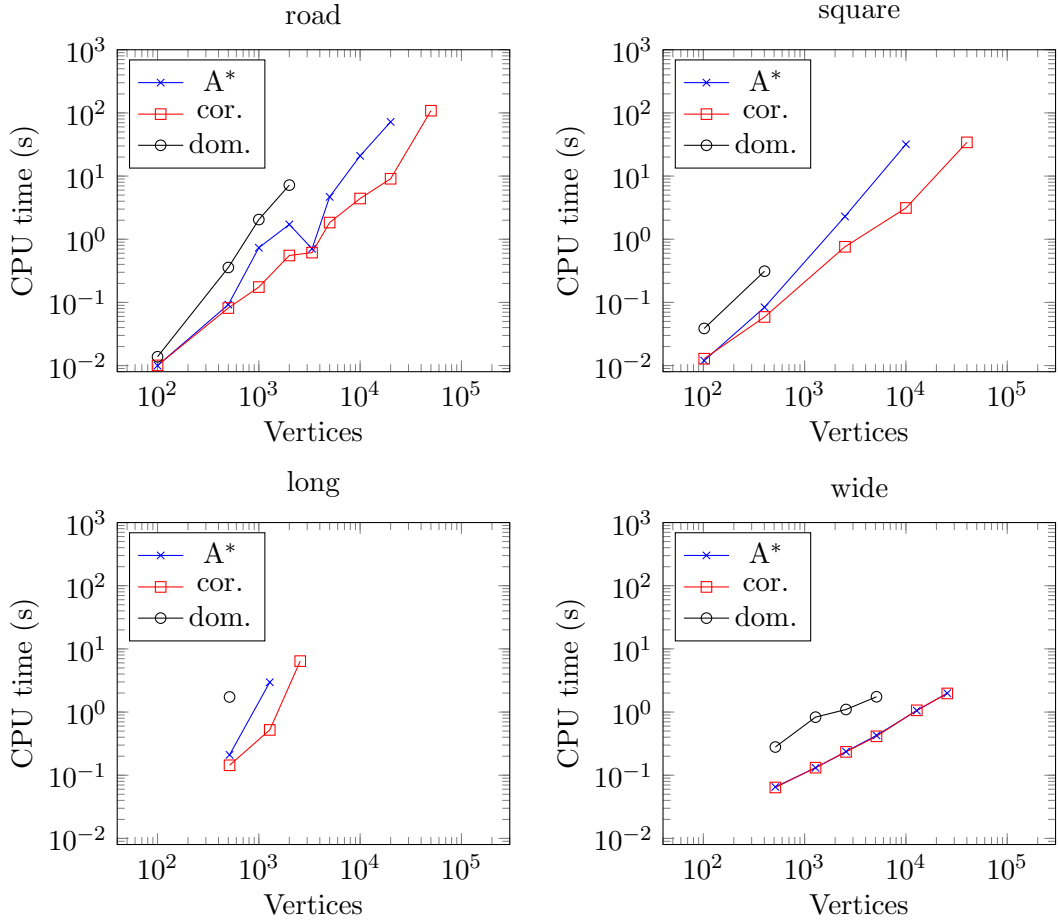
\begin{figure}[!ht]
	\begin{center}
	\pgfplotsset{width=6cm}
	\begin{outdent}
	\pgfplotsset{width=6.8cm}
	\begin{tabular}{cc}
				\begin{tikzpicture}
				\pgfplotsset{
				    xmin=40, xmax=3e5,
				    legend pos=north west
				}
				\begin{axis}[
					title=road ,
				xmode=log,
				ymode=log,
				ymin=8e-3,
				ymax=1e3,
				xlabel=Vertices,
				ylabel=CPU time (s)
			]
			\addplot[mark=x, color=blue] coordinates{
(100, 0.009984)
(500, 0.092379)
(1000, 0.734807)
(2000, 1.71642)
(3353, 0.707798)
(5000, 4.6879)
(10000, 20.9233)
(20000, 72.0884)
}; 
 \addlegendentry{A$^*$}
\addplot[mark=square, color = red] coordinates{
(100, 0.010073)
(500, 0.081532)
(1000, 0.174971)
(2000, 0.55318)
(3353, 0.614423)
(5000, 1.83645)
(10000, 4.41314)
(20000, 9.06752)
(50000, 108.418)
}; 
 \addlegendentry{cor.}
\addplot[mark=o] coordinates{
(100, 0.013797)
(500, 0.357094)
(1000, 2.04841)
(2000, 7.19862)
}; 
 \addlegendentry{dom.}

			\end{axis}
		\end{tikzpicture}
			 &
				\begin{tikzpicture}
				\pgfplotsset{
				    xmin=40, xmax=3e5,
				    legend pos=north west
				}
				\begin{axis}[
					title=square,
				xmode=log,
				ymode=log,
				ymin=8e-3,
				ymax=1e3,
				xlabel=Vertices,
				ylabel=CPU time (s)
			]
			\addplot[mark=x, color=blue] coordinates{
(102, 0.011989)
(402, 0.083663)
(2502, 2.29605)
(10002, 31.8706)
}; 
 \addlegendentry{A$^*$}
\addplot[mark=square, color = red] coordinates{
(102, 0.01286)
(402, 0.058833)
(2502, 0.760062)
(10002, 3.12513)
(40002, 34.1762)
}; 
 \addlegendentry{cor.}
\addplot[mark=o] coordinates{
(102, 0.038699)
(402, 0.311776)
}; 
 \addlegendentry{dom.}

			\end{axis}
		\end{tikzpicture}
			 \\
				\begin{tikzpicture}
				\pgfplotsset{
				    xmin=40, xmax=3e5,
				    legend pos=north west
				}
				\begin{axis}[
					title=long,
				xmode=log,
				ymode=log,
				ymin=8e-3,
				ymax=1e3,
				xlabel=Vertices,
				ylabel=CPU time (s)
			]
			\addplot[mark=x, color=blue] coordinates{
(514, 0.210276)
(1282, 2.96654)
}; 
 \addlegendentry{A$^*$}
\addplot[mark=square, color = red] coordinates{
(514, 0.143797)
(1282, 0.521486)
(2562, 6.4092)
}; 
 \addlegendentry{cor.}
\addplot[mark=o] coordinates{
(514, 1.73434)
}; 
 \addlegendentry{dom.}

			\end{axis}
		\end{tikzpicture}
			 &
				\begin{tikzpicture}
				\pgfplotsset{
				    xmin=40, xmax=3e5,
				    legend pos=north west
				}
				\begin{axis}[
					title=wide,
				xmode=log,
				ymode=log,
				ymin=8e-3,
				ymax=1e3,
				xlabel=Vertices,
				ylabel=CPU time (s)
			]
			\addplot[mark=x, color=blue] coordinates{
(514, 0.065261)
(1282, 0.131058)
(2562, 0.237071)
(5122, 0.427055)
(12802, 1.05459)
(25602, 1.98974)
}; 
 \addlegendentry{A$^*$}
\addplot[mark=square, color = red] coordinates{
(514, 0.063975)
(1282, 0.131802)
(2562, 0.234212)
(5122, 0.413164)
(12802, 1.0655)
(25602, 1.97162)
}; 
 \addlegendentry{cor.}
\addplot[mark=o] coordinates{
(514, 0.279557)
(1282, 0.830834)
(2562, 1.09825)
(5122, 1.74581)
}; 
 \addlegendentry{dom.}

			\end{axis}
		\end{tikzpicture}
			 \\	
	\end{tabular}
	\end{outdent}
	\end{center}
	\caption{Stochastic Resource Constrained Shortest Path Problem with generic distributions}
	\label{fig:consDctDist}
\end{figure}

\paragraph{Non-constrained problem.} 
Figure \ref{fig:dctDist} and Table \ref{tab:dctDist} provide numerical results for Problem \eqref{eq:stoShortestPathCvar} with $\beta = 0.05$. They can be read like the figures and tables of Section \ref{sec:numerical_experiments_on_usual_resource_constrained_shortest_path_problem}. This non-constrained stochastic problem is fairly easy to solve: large instances are solved in reasonable time. The main limit on the size of the instances we can solve is the memory needed to solve the instance. On Figure \ref{fig:dctDist}, we can see that our three algorithms exhibit similar performances in terms of computing time needed. In Table \ref{tab:dctDist}, we can see that these similar CPU time recover totally different realities. The label dominance algorithm enumerates hundreds of thousands of paths when the two other algorithms enumerate at most a few thousands. We can see in the proportion of paths cut by the dominance test in the label correcting algorithm the explanation: the lower bound test cut paths much better than the dominance test in that setting. This enables the generalized A$^{*}$ and the label correcting algorithm to tackle with larger instances than the label dominance algorithm. The similar CPU time we obtain at the end come from the fact that the preprocessing time needed to compute the lower bounds is rather long. We also note that, on this non-constrained problem, the key plays an important role in the performance of the generalized A$^{*}$ and the label correcting algorithm. Indeed, the versions of the algorithms with the key $c(\re_{P})$ instead of $c(\re_{P} \rplus b_{v})$, identified by the suffix K in Table~\ref{tab:dctDist} exhibit much poorer performances. On difficult instances, candidate paths are an important element of the performance of the generalized A$^{*}$ algorithms. Finally, the numerical results confirm that the choice of the expectation of $\xi$ as key in the bounding algorithm is relevant: the parameter $\gamma$ of Equation~\eqref{eq:gamma} remains small.


\begin{table}
\begin{outdent}
\begin{small}
\begin{tabular}{|l|rrc|rr|rrr|rr|r|}
\hline
Instance 
& \multicolumn{1}{c}{$|V|$} 
& \multicolumn{1}{c}{$|A|$} 
& \multicolumn{1}{c|}{Alg.} 
& \multicolumn{1}{c}{$\gamma$} 
& \multicolumn{1}{c|}{Preproc.} 
& \multicolumn{1}{c}{Ext.} 
& \multicolumn{1}{c}{Cut.} 
& \multicolumn{1}{c|}{Dom.} 
& \multicolumn{1}{c}{$\ell$} 
& \multicolumn{1}{c|}{Gap} 
& \multicolumn{1}{c|}{CPU (s)} \\
\hline
road20 & 20000 & 55180 & A$^*$ & 1.4 & 11\% &79249 & 166646 & -- &240 & opt & 7.21e+01 \\
&&&A$^*$ CP & 1.4 & 9\%+3\% &79207 & 166575 & -- &240 & opt & 8.54e+01 \\
&&&A$^*$~K & 1.4 & 3\% &314771 & 669005 & -- & -- &$\infty$ & 2.87e+02 \\
&&&cor. & 1.4 & 88\% &1544 & 2647 & 9\% &240 & opt & 9.07e+00 \\
&&&cor. CP & 1.4 & 66\%+24\% &1502 & 2576 & 9\% &240 & opt & 1.15e+01 \\
&&&cor.~K & 1.4 & 76\% &3538 & 6105 & 9\% &240 & opt & 1.07e+01 \\
&&&dom. & -- & --  &113537 & 63535 & -- & -- &$\infty$ & 1.26e+01 \\
&&&dom. CP & -- & -- +7\% &113537 & 63535 & -- & -- &$\infty$ & 3.28e+01 \\
\hline
square200 & 40002 & 120200 & A$^*$ & 1.5 & 20\% &87822 & 165841 & -- & -- &$\infty$ & 7.10e+01 \\
&&&A$^*$ CP & 1.5 & 15\%+7\% &87822 & 165841 & -- &255 & 51.3\% & 8.91e+01 \\
&&&A$^*$~K & 1.5 & 19\% &95221 & 180639 & -- & -- &$\infty$ & 7.65e+01 \\
&&&cor. & 1.5 & 40\% &30302 & 50233 & 11\% &255 & opt & 3.42e+01 \\
&&&cor. CP & 1.5 & 31\%+14\% &30301 & 50234 & 11\% &255 & opt & 4.31e+01 \\
&&&cor.~K & 1.5 & 40\% &32288 & 53435 & 11\% &255 & opt & 3.65e+01 \\
&&&dom. & -- & --  &37691 & 21859 & -- & -- &$\infty$ & 1.92e+00 \\
&&&dom. CP & -- & -- +72\% &37691 & 21859 & -- & -- &$\infty$ & 6.17e+00 \\
\hline
long20 & 5122 & 15376 & A$^*$ & 1.5 & 3\% &115855 & 221722 & -- & -- &$\infty$ & 8.42e+01 \\
&&&A$^*$ CP & 1.5 & 2\%+1\% &115855 & 221722 & -- &395 & 67.0\% & 1.04e+02 \\
&&&A$^*$~K & 1.5 & 2\% &128449 & 246911 & -- & -- &$\infty$ & 9.43e+01 \\
&&&cor. & 1.5 & 3\% &62412 & 99413 & 12\% & -- &$\infty$ & 7.27e+01 \\
&&&cor. CP & 1.5 & 2\%+1\% &62412 & 99413 & 12\% &395 & 7.4\% & 9.00e+01 \\
&&&cor.~K & 1.5 & 3\% &62800 & 99675 & 13\% & -- &$\infty$ & 7.49e+01 \\
&&&dom. & -- & --  &140783 & 90781 & -- & -- &$\infty$ & 1.80e+01 \\
&&&dom. CP & -- & -- +4\% &140783 & 90781 & -- & -- &$\infty$ & 1.79e+01 \\
\hline
wide100 & 25602 & 78400 & A$^*$ & 1 & 99\% &55 & 1705 & -- &21 & opt & 1.99e+00 \\
&&&A$^*$ CP & 1 & 65\%+34\% &47 & 1692 & -- &21 & opt & 2.83e+00 \\
&&&A$^*$~K & 1 & 99\% &68 & 1731 & -- &21 & opt & 2.02e+00 \\
&&&cor. & 1 & 99\% &45 & 1683 & 0\% &21 & opt & 1.97e+00 \\
&&&cor. CP & 1 & 66\%+34\% &37 & 1670 & 0\% &21 & opt & 2.83e+00 \\
&&&cor.~K & 1 & 99\% &54 & 1701 & 0\% &21 & opt & 2.01e+00 \\
&&&dom. & -- & --  &8631 & 2939 & -- & -- &$\infty$ & 3.34e-01 \\
&&&dom. CP & -- & -- +71\% &8631 & 2939 & -- & -- &$\infty$ & 1.18e+00 \\
\hline
\end{tabular}
\end{small}
\end{outdent}
\caption{Stochastic resource constrained shortest path problem with generic distributions.}
\label{tab:consDctDist}
\end{table}

\paragraph{Constrained problem.}
Figure~\ref{fig:consDctDist} and Table \ref{tab:consDctDist} are the analogues of Figure~\ref{fig:dctDist} and Table~\ref{tab:dctDist} for Problem \eqref{eq:stoConstraintedShortestPath}. This constrained problem is much more difficult than the non constrained one. The relative behaviour of the algorithms is quite similar to the one we observe on the usual resource constrained shortest path problem \eqref{eq:RCSPnumericalResultsPb} with $k=10$ constraints. The label correcting and the generalized A$^{*}$ algorithms behave much better than the label dominance algorithm. Using candidate paths enable to speed-up the algorithms on difficult instances. The key in the enumeration algorithm is less important than for the non-constrained problem.





\section{What to do on difficult problems} 
\label{sec:what_to_do_on_difficult_problems}

\begin{figure}[!ht]
	\begin{center}
		\begin{tikzpicture}[scale=0.8]
			\tikzset{axe/.style={->}}
			\draw[axe] (0,0) -- (0,3.6) node[below left]{$w^{1}$} ;
			\draw[axe] (0,0) -- (3.6,0) node[right]{$w^{0}$};
			\node (m) at (.8,.8) {};
			\node[color=red] (x) at (.8,1.2) {$\times$};
			\node[above] at (.8,1.2) {$\re_{P}$};
			\node[color=blue] (y) at (3.1,.8) {$\times$};
			\node[right] at (3.1,.8) {$\re_{\tilde{P}}$};
			\draw[dashed] (m.center) -- (x.center);
			\draw[dashed] (m.center) -- (y.center);
			\node at (2.9,2.9) {(a)};
		\end{tikzpicture}
		\begin{tikzpicture}[scale=0.8]
			\tikzset{axe/.style={->}}
			\draw[axe] (0,0) -- (0,3.6) node[below left]{$w^{1}$} ;
			\draw[axe] (0,0) -- (3.6,0) node[right]{$w^{0}$};
			\node (m) at (0.8,0.8) {\textbullet};
			\node at (.7,.5) {$\bigmeet \re_{P}$};
			\node (x1) at (0.8,3.3) {$\times$};
			\node (x2) at (1.2,2.9) {$\times$};
			\node (z) at (1.7,1.9) {$\times$};
			\node (y1) at (2.7,1.2) {$\times$};
			\node (y2) at (3.2,0.8) {$\times$};
			\node[right] at (2.8,.5) {};
			\draw[dashed] (m.center) -- (x1.center);
			\draw[dashed] (m.center) -- (y2.center);
			\node at (2.9,2.9) {(b)};
		\end{tikzpicture}
		\caption[Dominance and lower bound tests when dimension increases.]{Dominance and lower bound tests when dimension increases. On Figure (b), each symbol~$\times$ corresponds to the resource of a $v$-$d$ path.}
		\label{fig:DominanceAndBoundsLimits}
	\end{center}
\end{figure}
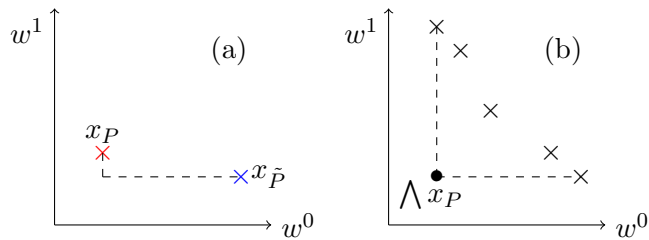

We have noted in Section \ref{sec:numerical_experiments_on_usual_resource_constrained_shortest_path_problem} that the performance of the algorithms decreases with the number of constraints. To illustrate what happens, we plot on Figure~\ref{fig:DominanceAndBoundsLimits} the resources $x = (w^{0},w^{1}) \in \R^{2}$ of the usual resource constrained shortest path problem \eqref{eq:RCSPnumericalResultsPb} with $k=1$ constraint. Each symbol $\times$ corresponds to the the resource $x_{P}$ of a path $P$. Figure \ref{fig:DominanceAndBoundsLimits}.(a) illustrates why the performance of the dominance test decreases with the dimension. All it takes is one coordinate such that $w_{P}^{i}\nleq w_{\tilde{P}}^{i}$ for $P$ not to dominate $\tilde{P}$. When the number of coordinates increases, it becomes rare that a path dominates another, and the dominance test does not enable to cut well paths. Figure \ref{fig:DominanceAndBoundsLimits}.(b) illustrates the fact that the lower bound $b_{v}$ on the resource of all the $v$-$d$ paths $P$ must be non-greater than the meet $\bigmeet \re_{P}$ of the resources of the $v$-$d$ paths, illustrated by the red diamond on the figure. When the dimension increases, the gap between $\bigmeet \re_{P}$ and the resources $\re_{P}$ tends to increase, and the quality of the bounds $b_{v}$ decreases. However, a bound $b_{v}$ can still be computed, which explains the reasonably good performance of the algorithms using the lower bound test (Low) when the dimension increases.

Suppose that for each vertex $v$, we have a set $B_{v}$ of resources in $\rset$ such that, for each $v$-$d$ path $P$, there exists a resource $b \in B_{v}$ satisfying $b \rleq \re_{P}$. We can then replace the lower bound test by the following \emph{clustered lower bounds test}.

\smallskip
	(Clu) \begin{minipage}{\dimexpr\textwidth-2cm} \emph{Is there a bound $b$ in $B_{v}$ such that $\rmeas(\re_{P}\rplus b) =0$ and $\rcost(\re_{P}\rplus b) \leq c_{od}^{UB}$?}
\end{minipage}
\smallskip

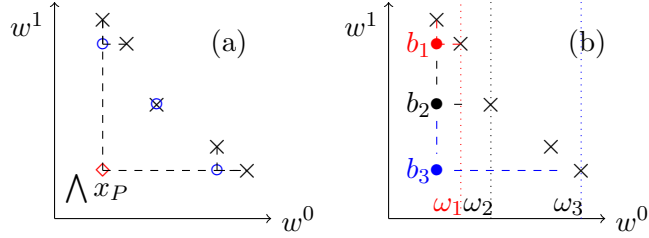
\begin{figure}[!ht]
	\begin{center}
		\begin{tikzpicture}[scale=0.8]
			\tikzset{axe/.style={->}}
			\draw[axe] (0,0) -- (0,3.6) node[below left]{$w^{1}$} ;
			\draw[axe] (0,0) -- (3.6,0) node[right]{$w^{0}$};
			\node[color = red] (m) at (0.8,0.8) {$\diamond$};
			\node at (.7,.5) {$\bigmeet \re_{P}$};
			\node (x1) at (0.8,3.3) {$\times$};
			\node (x2) at (1.2,2.9) {$\times$};
			\node (z) at (1.7,1.9) {$\times$};
			\node (y1) at (2.7,1.2) {$\times$};
			\node (y2) at (3.2,0.8) {$\times$};

			\node[color = blue] (x) at (0.8,2.9) {$\circ$};
			\node[color = blue] (z1) at (1.7,1.9) {$\circ$};
			\node[color = blue] (y) at (2.7,0.8) {$\circ$};
			\draw[dashed] (x.center) -- (x1.center);
			\draw[dashed] (x.center) -- (x2.center);
			\draw[dashed] (y.center) -- (y1.center);
			\draw[dashed] (y.center) -- (y2.center);
			\node[right] at (2.8,.5) {};
			\draw[dashed] (m.center) -- (x1.center);
			\draw[dashed] (m.center) -- (y2.center);
			\node at (2.9,2.9) {(a)};
		\end{tikzpicture}
		\begin{tikzpicture}[scale=0.8]
			\tikzset{axe/.style={->}}
			\draw[axe] (0,0) -- (0,3.6) node[below left]{$w^{1}$} ;
			\draw[axe] (0,0) -- (3.6,0) node[right]{$w^{0}$};
			\node[color=red] (x) at (0.8,2.9) {\textbullet};
			\node[color=red] at (0.5,2.9) {$b_{1}$};
			\node (x1) at (0.8,3.3) {$\times$};
			\node (x2) at (1.2,2.9) {$\times$};
			\node (w1) at (1.2,0) {};
			\node[color=red] at (1,0.2) {$\omega_{1}$};
			\draw[dotted,color=red] (w1) -- (1.2,3.6);
			\draw[dashed,color=red] (x.center) -- (x1.center);
			\draw[dashed,color=red] (x.center) -- (x2.center);

			\node at (1.5,0.2) {$\omega_{2}$};
			\node (z1) at (1.7,1.9) {$\times$};
			\node (z) at (0.8,1.9) {\textbullet};
			\node at (0.5,1.9) {$b_{2}$};
			\draw[dashed] (x) -- (z) -- (z1);
			\draw[dotted] (1.7,0) -- (1.7,3.6);

			\node (y1) at (2.7,1.2) {$\times$};
			\node (y2) at (3.2,0.8) {$\times$};
			\node[right] at (2.8,.5) {};
			\node[color=blue] (y) at (0.8,0.8) {\textbullet};
			\node[color=blue] at (0.5,0.8) {$b_{3}$};
			\draw[dotted,color=blue] (3.2,0) -- (3.2,3.6);
			\draw[dashed, color=blue] (z) -- (y) -- (y2);
			\node at (3,0.2) {$\omega_{3}$};

			\node at (3.3,2.9) {(b)};
		\end{tikzpicture}

		\caption[Clusters and conditional lower bounds]{Lower bounds on clusters of paths (a) and conditional lower bounds (b).}
		\label{fig:ClusteringAndConditionalBounds}
	\end{center}
\end{figure}

\noindent The idea behind this new test is illustrated on Figure \ref{fig:ClusteringAndConditionalBounds}.(a), where the lower bounds $b \in B_{v}$ are represented by blue circles. If we partition the $v$-$d$ paths into clusters of paths with ``similar'' resources, these lower bounds $b \in B_{v}$ are much tighter than $\bigmeet \re_{P}$, and thus enable to discard more paths. We can also replace the key $\rcost(\re_{P} \rplus b_{v})$ by $\min\{\rcost(\re_{P} \rplus b)|b\in B_{v}, \rmeas(\re_{P} \rplus b)=0\}$, which is a better approximation of the minimum cost of a feasible $o$-$d$ path starting by~$P$.

Figure \ref{fig:ClusteringAndConditionalBounds}.(b) illustrates another idea to improve the quality of the bounds. Consider the usual resource constrained shortest path problem with $k$ constraints of Equation \eqref{eq:RCSPnumericalResultsPb}. An $o$-$v$ path $P$ can be a subpath of an optimal path only if there exists a feasible $v$-$d$ path $Q$ such that $w_{Q}^{0} \leq c_{od}^{UB} - w_{P}^{0}$. Thus, instead of a bound $b_{v}$ on the resources of all the $v$-$d$ paths, we can use a bound $b'_{v}$ on the resource of the $v$-$d$ paths $Q$ such that $w_{Q}^{0} \leq c_{od}^{UB} - w_{P}^{0}$. If $c_{od}^{UB}$ is not too large, we can expect the set of such paths $Q$ to be much smaller than the complete set of $v$-$d$ paths, and thus $b'_{v}$ to be larger than $b_{v}$. 

We now formalize this idea. We assume to have a \emph{weight} morphism $\omega$ from $(\rset,\rplus,\rleq)$ to $(\mathbb{R},+,\leq)$ such that there is no cycle $C$ of negative weight $\sum_{a \in C} \omega(\re_{a})$. As $\omega$ is a morphism, we have $\omega(\re_{P}) = \sum_{a \in P}\omega(\re_{a})$. Moreover, we assume to have a scalar $\omega^{UB}$ such that $\omega(\re_{P}) > \omega^{UB}$ implies that $P$ is not an optimal feasible path. Note that such a scalar always exists if the set of optimal paths is finite, which is always the case if we consider only elementary paths. For each vertex $v$ in $V$, let $n_{v}$ be an integer, and $\omega_{v}^{1} < \omega_{v}^{2}< \ldots < \omega_{v}^{n_{v}} < \omega^{UB} \rleq \omega_{v}^{n_{v} + 1} = +\infty$ be a sequence of real numbers such that $\omega_{v}^{1}$ is the minimum weight of a $v$-$d$ path, which is well defined because there is no cycle of negative weight. Finally, for each vertex $v$, we suppose to have a set of bounds $b_{v}^{1}, \ldots, b_{v}^{n_{v}}$ such that $b_{v}^{i}$ is a lower bound on the resource of any $v$-$d$ path $P$ such that $\omega(\re_{P}) < \omega_{v}^{i+1}$. We can now define the \emph{conditional lower bounds test}.

\smallskip 
	(Con) \begin{minipage}{\dimexpr\textwidth-2cm} \emph{Do we have $\rmeas(\re_{P} \rplus b_{v}^{i})= 0$ and $\rcost(\re_{P} \rplus b_{v}^{i}) \leq \rcost_{od}^{UB}$, where $i$ is the minimum index such that $\omega_{v}^{i+1} \geq \omega^{UB}-\omega(\re_{P})$?}
	\end{minipage}
\smallskip

\noindent We can also replace the key $\rcost(\re_{P} \rplus b_{v})$ by $\rcost(\re_{P} \rplus b_{v}^{i})$, where $i$ is defined as in the test. When the cost function $c$ is a morphism, we can use it as $\omega$, and $c_{od}^{UB}$ as $\omega^{UB}$. This is for instances the case of the usual resource constrained shortest path problem \eqref{eq:RCSPnumericalResultsPb}, and of the stochastic resource constrained shortest path problem \eqref{eq:stoConstraintedShortestPath}. An alternative choice for Problem \eqref{eq:stoConstraintedShortestPath}  would be $\omega : (w,\xi) \mapsto \E(\xi)$.

We can adapt the proofs of Theorems \ref{theo:labelSettingConvergence} and \ref{theo:labelCorrectingConvergence} to show that they remain true if we replace the lower bound test (Low) by the clustered lower bounds test (Clu) or the conditional lower bounds test (Con) \cite{parmentier2016thesis}.  After a brief overview of the technique that enables to build the sets of bounds required by both tests, we detail the relative advantages of each of them, and we conclude with numerical experiments showing the gain they enable.

\begin{rem}
Provided that we are able to build the bounds $b_{v}^{i}$, any isotone function $\omega : \rset \rightarrow R$ can be used in the conditional lower bounds test. However, our technique to build the bounds  $b_{v}^{i}$ works only when $\omega$ is a morphism.
\end{rem}

\subsection{Computing the sets of bounds} 
\label{sub:computing_the_set_of_bounds}



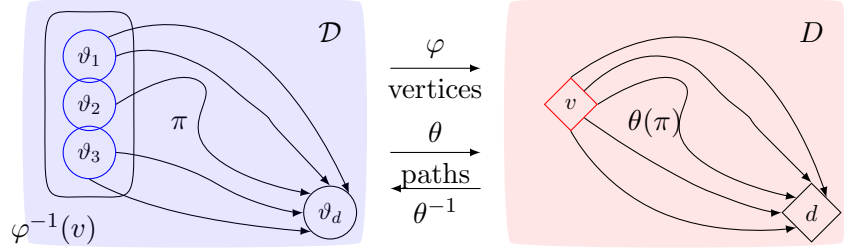
\begin{figure}[!ht]
	\begin{center}
		\begin{tikzpicture}[scale=0.8]

\tikzset{
	vert/.style={draw,circle,scale=0.8},
	edge/.style={>=latex},
	arc/.style={->,edge},
	path/.style={->,>=latex}
	}

\def\h{0.8}

\node[vert, draw = blue, text = black] (s1) at (0,2*\h) {$\vartheta_{1}$};
\node[vert, draw = blue, text = black] (s2) at (0,1*\h) {$\vartheta_{2}$};
\node[vert, draw = blue, text = black] (s3) at (0,0*\h) {$\vartheta_{3}$};

\node[above = 0.1cm of s1] (a1) {};
\node[left = 0.1cm of s2] (a2) {};
\node[below = 0.1cm of s3] (a3) {};
\node[right = 0.1cm of s2] (a4) {};
\draw plot [smooth cycle, tension = 2] coordinates {(a1) (a2) (a3) (a4)};
\node[below left = 0.4*\h and -0.3*\h of s3.south] {$\varphi^{-1}(v)$};

\node[vert] (sd) at (4,-1) {$\vartheta_{d}$};

\draw[path] plot [smooth, tension = 1] coordinates {(s1.north east) (2.5,2.2*\h) (sd.north east)};
\draw[path] plot [smooth, tension = 1] coordinates {(s1.east) (1.5,2*\h) (2.7, 0.8) (3.2,0.3) (sd.north)};

\node at (1.5,.5) {$\pi$};

\draw[path] plot [smooth, tension = 1] coordinates {(s2.east) (1.8,1.5*\h) (2, 0) (sd.north west)};

\draw[path] plot [smooth, tension = 1] coordinates {(s3.east) (1.5,-0.3*\h) (2.7, -0.8) (sd.west)};
\draw[path] plot [smooth, tension = 1] coordinates {(s3.south) (1.5,-1) (sd.south west)};

\fill[blue, opacity=0.1] plot [smooth cycle, tension = 0.1] coordinates {(-1,2.5) (-1,-1.5) (4.5,-1.5) (4.5,2.5)};
\node at (4,2) {$\mathcal{D}$};

\draw[arc] (5,1.75*\h) -- node[midway,above] {$\varphi$} node[midway,below]{vertices} (6.5,1.75*\h);
\draw[arc] (5,0.*\h) -- node[midway,above] {$\theta$} node[midway,below]{paths} (6.5,0.*\h);
\draw[arc] (6.5,-0.75*\h) -- node[midway,below] {$\theta^{-1}$} (5,-0.75*\h);

\fill[red, opacity=0.1] plot [smooth cycle, tension = 0.1] coordinates {(7,2.5) (7,-1.5) (12.5,-1.5) (12.5,2.5)};
\node at (12,2) {$D$};

\node[draw = red,diamond,scale=0.8] (v) at (8,1*\h) {$v$};

\node[draw,diamond,scale=0.8] (d) at (12,-1) {$d$};

\draw[path] plot [smooth, tension = 1] coordinates {(v.north) (10.5,2.2*\h) (d.north east)};

\draw[path] plot [smooth, tension = 1] coordinates {(v.north east) (9.5,2*\h) (10.7, 0.8) (11.2,0.3) (d.north)};
\draw[path] plot [smooth, tension = 1] coordinates {(v.east) (9.8,1.5*\h) (10, 0) (d.north west)};
\draw[path] plot [smooth, tension = 1] coordinates {(v.south east) (9.5,-0.3*\h) (10.7, -0.8) (d.west)};
\draw[path] plot [smooth, tension = 1] coordinates {(v.south) (9.5,-1) (d.south west)};

\node at (9.4,.5) {$\theta(\pi)$};


\end{tikzpicture}
		\caption{Blown-up graph. $\varphi$ is a surjective mapping between vertices, and $\theta$ a bijective mapping between paths.}
		\label{fig:stateGraph1}
	\end{center}
\end{figure}

It is not required to define new bounding algorithms to compute the lower bounds set of the clustered lower bound test. Our strategy to compute the lower bounds $B_{v}$ is to blow-up the  digraph $D = (V,A)$ in a much larger digraph $\mathcal{D} = (\mathcal{V},\mathcal{A})$, and to use the algorithm of Section \ref{sec:bounding_algorithms} in $\mathcal{D}$.  Digraphs $\mathcal{V}$ and $V$ are respectively illustrated on the left and on the right part of Figure \ref{fig:stateGraph1}. We provide here the properties of $\mathcal{D}$ that enable to retrieve the bounds $B_{v}$ from the blown-up digraph $\mathcal{D}$. These properties are enforced when $\mathcal{D}$ is built. The procedures we provide in \cite{parmentier2016thesis} to build such a graph $\mathcal{D}$ are not difficult but technical: we therefore refer the interested reader to the Chapter 6 of \cite{parmentier2016thesis}.

We denote  $\vartheta$ the vertices of $\mathcal{D}$. We have a surjective mapping $\varphi$ that associates to each vertex $\vartheta$ of $\mathcal{V}$ a vertex $\varphi(\vartheta)$ in $V$. We denote $\varphi^{-1}(v)$ the set of vertices $\vartheta$ such that $\varphi(\vartheta) = v$. There are typically many vertices in $\varphi^{-1}(v)$, the only exception being the set $\varphi^{-1}(d) $ of vertices corresponding to the destination $d$, which is the singleton $ \{\vartheta_{d}\}$. We deduce from $\varphi$ a mapping $\theta$ from the set of paths $\pi$ in $\mathcal{D}$ ending in $\vartheta_{d}$ to the set of paths in $P$ in $d$. We define the resources of the arcs in $\mathcal{A}$ is such a way that the resource $\re_{\pi}$ of a path $\pi$ is equal to the resource $\re_{P}$ of the corresponding path $P = \theta(\pi)$ in $D$. The most important assumption is the following one: \emph{the mapping $\theta$ induces a bijection between the $\varphi^{-1}(v)$-$\varphi^{-1}(d)$ paths in $\mathcal{D}$ and the $v$-$d$ paths in $D$}, where a $\varphi^{-1}(v)$-$\varphi^{-1}(d)$ path is a path between a vertex $\vartheta \in \varphi^{-1}(v)$ and $\vartheta_{d}$.

As we illustrate on Figure \ref{fig:stateGraph1}, under these assumptions, the bijection $\theta$ provides a natural partition the set $v$-$d$ paths $P$ into clusters. The cluster of a $v$-$d$ path $P$ is given by the origin vertex $\vartheta$ of the corresponding path $\pi = \theta^{-1}(P)$. To illustrate this partition in terms of lower bounds on resources, the red diamond on Figure \ref{fig:ClusteringAndConditionalBounds}.(a) is the a lower bound on the resource all the $v$-$d$ path in $\mathcal{D}$, and the blue circles are the lower bounds on the resource of all the $\vartheta$-$\vartheta_{d}$ paths in $\mathcal{D}$ for each $\vartheta$ in $\phi^{-1}(d)$. Practically, in the same way we use the algorithm of Equation \eqref{eq:zSequenceDefinition} to compute the lower bounds $b_{v}^{\ell^{*}}$ or $b_{v}^{\dagger}$ on the resource of all the $v$-$d$ paths in $D$, we can now use these algorithm in digraph $\mathcal{D}$ to obtain the lower bounds $b_{\vartheta}^{\ell^{*}}$ or $b_{\vartheta}^{\dagger}$ on the resources of all the $\vartheta$-$\vartheta_{d}$ paths. 
As $\theta$ induces a bijection, for each $v$-$d$ path $P$, there is a vertex $\vartheta \in \varphi^{-1}(v)$ and a $\vartheta$-$\vartheta_{d}$ path $\pi$ such that $b_{\vartheta}^{\dagger} \rleq b_{\vartheta}^{\ell^{*}} \rleq \re_{\pi} = \re_{P}$. We can therefore use $\{b_{\vartheta}^{\ell^{*}}, \vartheta \in \varphi^{-1}(v)\}$ or $\{b_{\vartheta}^{\dagger}, \vartheta \in \varphi^{-1}(v)\}$ as lower bounds set $B_{v}$ in the clustered lower bounds test.

The bounds of the conditional graph can also be computed using a similar ``blown-up graph'' approach. In both case, the procedure building graph $\mathcal{D}$ takes as input the maximum cardinal $\kappa$ of $\varphi^{-1}(v)$. Parameter $\kappa$ corresponds to the maximum number of bounds we obtain for a given vertex $v$:  $|B_{v}| \leq \kappa$ for the clustered lower bounds test, and the maximum $n_{v}\leq \kappa$ for the conditional lower bounds test. See Chapter 6 of \cite{parmentier2016thesis} for more details.

\subsection{When to use the clustered and the conditional lower bound test} 
\label{sub:choosing_between_the_clustered_and_the_conditional_lower_bound_test}

The clustered and conditional lower bounds tests (Clu) and (Con) are stronger versions of the lower bound test (Low) which require a longer preprocessing but enable to reduce the number of paths enumerated by the generalized A$^{*}$ and label correcting algorithms. They are therefore not interesting on easy instances, where even the preprocessing for the lower bound test (Low) is a waste of time. On the contrary, they are interesting on difficult instances where the time spent in the enumeration is much larger than the one spent in the preprocessing, and even more interesting on very difficult instances that we cannot solve to optimality because the list $L$ becomes to large. 

Increasing the size of $\mathcal{D}$ increases the preprocessing time and reduces the number of paths enumerated by and the time spent in the enumeration algorithms. On instances that cannot be solved to optimality, we advise to use the largest graph $\mathcal{D}$ that can be stored in the memory. On instances that can be solved to optimality, we advise to choose the size of $\mathcal{D}$ in order to spend about the same time in the preprocessing and in the enumeration algorithm.


Clustered and conditional lower bound tests have both their pros and cons. 
The \emph{clustered lower bound test} can in theory be applied to any instance of the \MRCSP. Indeed, the procedures that build the digraph $\mathcal{D}$ only requires a similarity measure that enable to compare two resources $\re$ and $\tilde{\re}$. Besides, as we can see on Figure \ref{fig:ClusteringAndConditionalBounds}, the bounds it produces tend to be tighter than those produced by the conditional lower bound test. Its first drawback comes from the fact that, each time a test (Clu) is performed at a vertex $v$, up to $|B_{v}|$ operations $\rplus$, $\rcost(\cdot)$, and $\rmeas(\cdot)$ must be performed, which can be time consuming. The other drawbacks are linked to the procedure we use to build the bounds $B_{v}$. Indeed, this procedure calls a clustering subroutine for each vertex $v$ of the digraph $D$ \cite{parmentier2016thesis}, which is time consuming when $\mathcal{D}$ is large. Finally, our procedure that builds $\mathcal{D}$ has been thought for acyclic digraphs. We have extended it to digraphs with cycles, but the quality of the bounds returned is mitigated.

Due to these properties, the clustered lower bound approach works particularly well in the context of column generation if the pricing subproblem is a resource constrained shortest path problem on an acyclic digraph. This is often the case when the problem solved by column generation is a vehicle or a crew scheduling problem. Along a column generation scheme, many instances of the same resource constrained shortest path problem must be solved successively, the only difference between two instances being the reduced costs. Hence, we can compute once and for all the digraph $\mathcal{D}$ in a (time-consuming) pre-processing, and then use it to compute the lower bound sets $B_{v}$ and speed-up the resolutions of all the instances of the pricing subproblem. We have applied this technique to a column generation approach to the airline crew pairing problem. As it can be seen in Table 9.2 of \cite{parmentier2016thesis}, the use of the clustered lower bound test within a generalized A$^{*}$ or a label correcting algorithm enables to divide by three the total time spent in the column generation scheme on some industrial instances.

Compared to the clustered lower bound test, the main advantage of the \emph{conditional lower bound test} is that operations $\rplus$, $\rcost(\cdot)$, and $\rmeas(\cdot)$ need to be performed only once when a test (Con) is performed. Its main drawback is that a morphism $\omega$ is required, which reduces the number of problems to which it can be applied.

The performance of the test depends on the choice of $\omega$. If the set of $o$-$d$ paths $P$ such that $\omega(\re_{P})\leq \omega^{UB}$ is much smaller than the set of $o$-$d$ paths, then the conditional lower bound test tends to exhibit good performances.

\subsection{Numerical results} 
\label{sub:numerical_results}

We now test the performance of the clustered and conditional lower bounds tests on difficult instances of the usual resource constrained shortest path problem \eqref{eq:RCSPnumericalResultsPb} with $k=10$. We use the cost as morphism $\omega$ for the conditional lower bounds test: $\omega(\re) = w^{0}$ where $\re = (w^{0},\ldots,w^{k})$. When we build the sets of bounds $B_{v}$ or the conditional bounds $b_{v}^{i}$, the number of bounds we can take for a given vertex $v$ is practically limited by the memory available. We therefore consider the instances long5 and square50 of Table \ref{tab:RCSP10constraints} because they are difficult instances of reasonable size, which enables to build large sets of bounds $B_{v}$. The numerical experiments have been performed on the same computer as those of Section \ref{sec:numerical_experiments_on_usual_resource_constrained_shortest_path_problem}. Both the generalized A$^{*}$ and the label correcting algorithms have been tested with the new tests (Clu) and (Con). We have used algorithms with candidate paths, as these versions are the most efficient ones on difficult instances.  In Section \ref{sec:numerical_experiments_on_usual_resource_constrained_shortest_path_problem}, in order to have comparable results between algorithms and instances, we have set the maximum size of $L$ to 1e+05 because it is the maximum size that could fit in the memory for all the instances and all the algorithms. Here, we want to obtain the best possible results with each algorithm. For a list $L$ of identical size, the label correcting algorithm requires more memory than then generalized A$^{*}$ as it needs to store the lists of non-dominated paths $M_{v}$. We have therefore set 2e+05 as the of maximum size of $L$ for the label correcting, and 2e+06 for the generalized A$^{*}$.

\begin{table}
\begin{outdent}
\begin{small}
\begin{tabular}{|l|ccr|rr|rrr|rr|r|}
\hline
Instance  
& \multicolumn{1}{c}{Alg.} 
& \multicolumn{1}{c}{Test} 
& \multicolumn{1}{c|}{$|\mathcal{V}|/|V|$}
& \multicolumn{1}{c}{$\gamma$} 
& \multicolumn{1}{c|}{Prep.} 
& \multicolumn{1}{c}{Ext.} 
& \multicolumn{1}{c}{Cut.} 
& \multicolumn{1}{c|}{Dom.} 
& \multicolumn{1}{c}{$\ell$} 
& \multicolumn{1}{c|}{Gap} 
& \multicolumn{1}{c|}{CPU (s)} \\
\hline
long5 & cor. CP & (Low) & --  &2.2 & 0\% & 230537 & 130784 & 50\% &81 & 60.8\% & 7.61e+00 \\
& & (Con) & 1.8e+01 & 2.2 & 6\% & 167283 & 50956 & 82\% &81 & 60.0\% & 4.91e+00 \\
& &  & 1.7e+02 & 2.6 & 41\% & 137348 & 24950 & 100\% &81 & 32.2\% & 8.14e+00 \\
& &  & 9.1e+02 & 1 & 73\% & 133873 & 22655 & 100\% &81 & 29.2\% & 2.05e+01 \\
& & (Clu) & 1.7e+01 & 1.7 & 17\% & 190559 & 88647 & 52\% &81 & 55.9\% & 1.17e+01 \\
& &  & 8.2e+01 & 1.8 & 22\% & 187568 & 83110 & 55\% &81 & 53.4\% & 2.54e+01 \\
& &  & 3.8e+02 & 1.8 & 35\% & 190368 & 85960 & 55\% &81 & 49.5\% & 9.85e+01 \\
&A$^*$ CP & (Low) & --  &2.2 & 0\% & 1424776 & 849563 & -- &81 & 64.4\% & 9.49e+00 \\
& & (Con) & 1.8e+01 & 2.2 & 2\% & 1011810 & 23631 & -- &81 & 62.3\% & 1.41e+01 \\
& &  & 1.7e+02 & 2.6 & 16\% & 1000137 & 285 & -- &81 & 31.0\% & 2.32e+01 \\
& &  & 9.1e+02 & 1 & 37\% & 1000309 & 629 & -- &81 & 27.8\% & 4.07e+01 \\
& & (Clu) & 1.7e+01 & 1.8 & 6\% & 1276845 & 553700 & -- &81 & 57.9\% & 3.45e+01 \\
& &  & 8.2e+01 & 1.8 & 5\% & 1218043 & 436097 & -- &81 & 54.5\% & 1.09e+02 \\
& &  & 3.8e+02 & 1.8 & 7\% & 1165819 & 331651 & -- &81 & 51.7\% & 4.35e+02 \\
\hline
square50 & cor. CP & (Low) & --  &2.2 & 0\% & 350553 & 273819 & 42\% &52 & 44.9\% & 1.15e+01 \\
& & (Con) & 9.7e+00 & 2.2 & 6\% & 217971 & 100571 & 67\% &52 & 41.1\% & 5.61e+00 \\
& &  & 8.6e+01 & 2.5 & 41\% & 158700 & 57177 & 53\% &52 & 15.1\% & 8.85e+00 \\
& &  & 4.7e+02 & 1 & 73\% & 165658 & 71086 & 42\% &52 & 12.3\% & 2.24e+01 \\
& & (Clu) & 1.6e+01 & 1.7 & 23\% & 299643 & 207262 & 46\% &52 & 36.5\% & 1.64e+01 \\
& &  & 7.4e+01 & 1.9 & 30\% & 285296 & 203651 & 41\% &52 & 35.4\% & 3.32e+01 \\
& &  & 3.3e+02 & 1.9 & 47\% & 307779 & 231886 & 40\% &52 & 31.6\% & 1.22e+02 \\
&A$^*$ CP & (Low) & --  &2.2 & 0\% & 9954128 & 11908301 & -- &52 & 44.9\% & 5.93e+01 \\
& & (Con) & 9.7e+00 & 2.2 & 1\% & 4627965 & 1255974 & -- &52 & 40.2\% & 5.96e+01 \\
& &  & 8.6e+01 & 2.5 & 3\% & 6772542 & 5545130 & -- &52 & 7.5\% & 1.11e+02 \\
& &  & 4.7e+02 & 1 & 3\% & 30098520 & 60197088 & -- &52 & opt & 4.73e+02 \\
& & (Clu) & 1.6e+01 & 1.7 & 2\% & 8425382 & 8850810 & -- &52 & 34.9\% & 1.61e+02 \\
& &  & 7.4e+01 & 1.8 & 2\% & 8146176 & 8292398 & -- &52 & 33.8\% & 4.73e+02 \\
& &  & 3.3e+02 & 1.9 & 3\% & 9374152 & 10748349 & -- &52 & 27.9\% & 1.71e+03 \\
\hline
\end{tabular}
\end{small}
\end{outdent}
\caption{Results of clustered and conditional lower bounds tests Problem \eqref{eq:RCSPnumericalResultsPb} with $k=10$ constraints. List $L$ maximum size is 2e+05 for label correcting algorithm and 2e+06 for A$^{*}$ algorithm}
\label{tab:clusCondRes}
\end{table}

The numerical results are available in Table \ref{tab:clusCondRes}. The column ``Test'' provides the test used, which can be the lower bound test (Low), the clustered lower bound test (Clu), or the conditional lower bounds test (Con). When the clustered or the conditional lower bound test is used, we provide the ratio $\frac{|\mathcal{V}|}{|V|}$ of the number of vertices in the blown-up digraph $\mathcal{D}$ and in the digraph $D$: it is is the average number of bounds used per vertex $v$. The percentage of time spent in the preprocessing includes the construction of the digraph $\mathcal{D}$, the bounding algorithm in $\mathcal{D}$, and the computation of candidate paths. The other columns are identical to those of Table \ref{tab:RCSP10constraints}. 

As most algorithms do not solve the instance to optimality, the interesting statistic to evaluate the performance of the algorithm is the gap. With each algorithm, both the clustered lower bound test and the conditional lower bound test enable to reduce the gap. Indeed, they enable to divide by two the gap on the instance long5, and to solve the instance square50 to optimality. Besides, the larger the number of bounds per vertex, i.e., the larger the ratio $\frac{|\mathcal{V}|}{|V|}$, the smaller is the gap at the end. Larger $\frac{|\mathcal{V}|}{|V|}$ means longer preprocessing. We have been limited by the memory of the computer on the choice of the size of $\mathcal{D}$. As, on each algorithm launched, either the instance is not solved to optimality, or most of the time is spent in the enumeration algorithm, better performances  would be obtained with larger digraphs $\mathcal{D}$.

On our two instances, the conditional lower bounds test performs better than the clustered lower bounds test. This is not surprising because instances long5 and the square50 have cycle, and, as we already mentioned, our procedure that builds the lower bounds sets $B_{v}$ does not work well on digraphs with cycles. 

 We finish with two statistics that can look surprising. First, the ratio $\gamma$ of Equation~\eqref{eq:gamma} is equal to $1$ on very large graphs $\mathcal{D}$ for the conditional lower bound test. This comes from the fact that, due to our building procedure, these graphs are acyclic. Second, on the long5 instance, the percentage of paths cut by the dominance test increases with the size of $\mathcal{D}$. This is due to the fact that, as the key used is a good approximation, only the promising paths are considered, and thus few paths are cut by the lower bound test at the beginning of the algorithm. The lower bound test enable to cut paths at the end of the algorithm. As the long5 instance is difficult, the algorithm is stopped early, and few paths are cut by this test. On the contrary, on the easier square50 instance, the algorithm is stopped later, and the proportion of paths cut by the dominance test decreases when the size of $\mathcal{D}$ increases.




\section*{Acknowledgments}
I greatly thank my PhD advisor Fr\'ed\'eric Meunier for his numerous and deep remarks on the mathematics and the way to write this article. 

\bibliographystyle{authordate1}
\bibliography{biblioShortestPath1}

\end{document}